\documentclass[conference]{IEEEtran}
\IEEEoverridecommandlockouts

\usepackage{cite}

\usepackage{array}
\usepackage{multicol}
\usepackage{multirow}

\usepackage{graphicx}
\usepackage{booktabs}
\usepackage[skip=5pt]{caption}
\usepackage{subcaption}
\usepackage{multirow}
\usepackage{weiwAlgorithm}
\usepackage{amsmath}
\usepackage{newtxmath}
\usepackage{diagbox}
\usepackage{url}
\usepackage[colorlinks,citecolor=black,linkcolor=black,urlcolor=black]{hyperref}
\usepackage{dsfont}
\usepackage{enumitem}
\usepackage{caption}
\usepackage{balance}
\usepackage{alphabeta}
\usepackage{amsmath}
\usepackage{mathtools}
\usepackage{color}
\usepackage{colortbl} 
\usepackage{xcolor} 
\usepackage{bm}
\usepackage{cuted}
\usepackage{float}
\usepackage{placeins}
\usepackage[normalem]{ulem}

\usepackage{ntheorem}

\renewcommand{\baselinestretch}{0.9965}

\def\BibTeX{{\rm B\kern-.05em{\sc i\kern-.025em b}\kern-.08em
    T\kern-.1667em\lower.7ex\hbox{E}\kern-.125emX}}

\newtheorem{example}{\textit{\textbf{Example}}}
\newtheorem{theorem}{{\textit{Theorem}}}

\newtheorem{lemma}{{\textit{Lemma}}}
\newtheorem{observation}{{\textit{Observation}}}
\newtheorem{remark}{\textit{\textit{Remark}}}
\newtheorem{definition}{{\textit{Definition}}}
\newtheorem*{proof}{{\textit{Proof}}\textit{.}}

\newcommand{\myparagraph}[1]{\vspace{1mm} \noindent \textbf{#1}.\xspace}

\newcommand{\ie}{{i.e.,}\xspace}

\newcommand{\etal}{et al.\xspace}

\newcommand{\dtTSG}{{dtTSG}\xspace}
\newcommand{\tgTSG}{{tgTSG}\xspace}
\newcommand{\esTSG}{{esTSG}\xspace}
\newcommand{\VUG}{{VUG}\xspace}

\newcommand{\stspgraph}{{tspG}\xspace}

\newcommand{\tcv}[3]{TCV_{#1}(#2,#3)\xspace}

\newcommand{\ea}[1]{\mathcal{A}(#1)\xspace}
\newcommand{\ld}[1]{\mathcal{D}(#1)\xspace}

\newcommand{\dt}{\texttt{dtTSG}\xspace}
\newcommand{\es}{\texttt{esTSG}\xspace}
\newcommand{\tg}{\texttt{tgTSG}\xspace}

\newcommand{\epdt}{\texttt{EPdtTSG}\xspace}
\newcommand{\epes}{\texttt{EPesTSG}\xspace}
\newcommand{\eptg}{\texttt{EPtgTSG}\xspace}

\newcommand{\quick}{\texttt{QuickUBG}\xspace}
\newcommand{\tight}{\texttt{TightUBG}\xspace}
\newcommand{\eev}{\texttt{EEV}\xspace}

\newcommand{\gq}{\mathcal{G}_q\xspace}
\newcommand{\gt}{\mathcal{G}_t\xspace}

\newcommand{\cblack}[1]{{\color{black}#1}}

\newcommand\blfootnote[1]{%
  \begingroup
  \renewcommand\thefootnote{}\footnote{#1}%
  \addtocounter{footnote}{-1}%
  \endgroup
}

\begin{document}

\title{Efficient Temporal Simple Path Graph Generation}

\author{
Zhiyang Tang$^{\dag}$, Yanping Wu$^{\ddag *}$, Xiangjun Zai$^{\natural}$, Chen Chen$^{\S *}$, Xiaoyang Wang$^{\natural}$, Ying Zhang$^{\dag}$\vspace{1mm}\\
\fontsize{9}{9}\selectfont\itshape
$^{\dag}$\textit{Zhejiang Gongshang University}, China~$^{\ddag}$\textit{University of Technology Sydney}, Australia\\
$^{\natural}$\textit{The University of New South Wales}, Australia
$^{\S}$\textit{University of Wollongong}, Australia\vspace{1mm}\\
\fontsize{9}{9}\selectfont\ttfamily\upshape
zhiyangtang.zjgsu@gmail.com~yanping.wu@student.uts.edu.au\\
xiangjun.zai@student.unsw.edu.au~chenc@uow.edu.au\\
xiaoyang.wang1@unsw.edu.au~ying.zhang@zjgsu.edu.cn
}

\maketitle

\begin{abstract}
Interactions between two entities often occur at specific timestamps, which can be modeled as a temporal graph.
Exploring the relationships between vertices based on temporal paths is one of the fundamental tasks.
In this paper, we conduct the first research to propose and investigate the problem of generating the temporal simple path graph ($\stspgraph$), which is the subgraph consisting of all temporal simple paths from the source vertex to the target vertex within the given time interval.
Directly enumerating all temporal simple paths and constructing the $\stspgraph$ is computationally expensive.
To accelerate the processing,
we propose an efficient method named {{V}erification~in~{U}pper-bound~{G}raph}.
It first incorporates the temporal path constraint and simple path constraint to exclude unpromising edges from the original graph, which obtains a tight upper-bound graph as a high-quality approximation of the $\stspgraph$ in polynomial time. 
Then, an Escape Edges Verification algorithm is further applied in the upper-bound graph to construct the exact $\stspgraph$ without exhaustively enumerating all temporal simple paths between given vertices.
Finally, comprehensive experiments on \cblack{10} real-world graphs are conducted to demonstrate the efficiency and effectiveness of the proposed techniques.
\blfootnote{*Corresponding author}
\end{abstract}

\section{Introduction}
\label{sec:intro}


Path enumeration is a fundamental problem in graph analysis, which aims to list all paths from one vertex to another~\cite{DBLP:journals/vldb/PengLZZQZ21,DBLP:conf/sigmod/SunCHH21,DBLP:conf/icde/YuanH0024,DBLP:conf/icde/LaiPY0021,DBLP:journals/pvldb/HaoYZ21,DBLP:conf/wise/LiHYCZYL22,DBLP:conf/icde/ZhangYO00Y23,DBLP:journals/pvldb/LiangOZYLT24,DBLP:conf/soda/BirmeleFGMPRS13,DBLP:conf/latin/GrossiMV18,DBLP:conf/iwoca/RizziSS14,DBLP:journals/kais/JinCL24,DBLP:conf/tamc/MutzelO19}.
In reality, relationships between entities are often associated with timestamps, which can be modeled as the temporal graph~\cite{DBLP:journals/pvldb/ChenWLZQZ21,DBLP:journals/pvldb/GouYZY24,DBLP:conf/icde/WuSWZ00024,DBLP:journals/pvldb/CaiKWCZLG23,DBLP:journals/pvldb/ChenWZZWL24,DBLP:conf/icde/WenHZQZ020,DBLP:conf/icde/WuHCLK16,DBLP:journals/vldb/ZhangGCGPZJ19,DBLP:journals/pvldb/WuCHKLX14,DBLP:conf/sigmod/WangLYXZ15,DBLP:journals/pvldb/WuSWWZQL24,DBLP:conf/icde/GaoCYCHD22,DBLP:conf/sigmod/GurukarRR15,DBLP:journals/tkde/GaoZQLC21}.
Due to the wide applications of temporal graphs in different domains, several path models and enumeration problems have been defined by integrating temporal constraints \cite{DBLP:journals/kais/JinCL24,DBLP:conf/tamc/MutzelO19}.
\cblack{For example, infectious disease spread can be analyzed by enumerating temporal paths between infected people based on their movement trajectories~\cite{DBLP:journals/kais/JinCL24}}.


Although path enumeration can benefit many fields, 
it still may not be adequate for studying relations in certain aspects.
\cblack{Large-scale path enumeration often results in significant overlaps in vertices and edges~\cite{DBLP:journals/vldb/PengLZZQZ21}.
Moreover, 
such an enumeration fails to globally consider and fully utilize the structural connectivity across different paths~\cite{DBLP:conf/sigmod/WangWKL21}
}.
To address the challenges, a few works propose the concept of \textit{path graph}, which consists of all paths with certain features between vertices\cite{DBLP:conf/dasfaa/LiuGP021,DBLP:journals/pacmmod/CaiLZ023,DBLP:conf/sigmod/WangWKL21}.
For example,
\cite{DBLP:conf/dasfaa/LiuGP021} defines the $k$-hop $s$-$t$ subgraph query for all paths from $s$ to $t$ within $k$ hops, and 
\cite{DBLP:journals/pacmmod/CaiLZ023} extends this concept to the hop-constraint simple path graph.
Wang \etal~\cite{DBLP:conf/sigmod/WangWKL21} define the shortest path graph, which exactly contains all the shortest paths between a pair of vertices.
Unfortunately, these path graphs disregard the temporal information, making them applicable only to static networks and unable to capture dynamic features.
Furthermore, they impose hop constraint or shortest path constraint, limiting the information they provide.

Motivated by this, in this paper, we propose a novel problem, named temporal simple path graph ($\stspgraph$) generation.
\cblack{It allows for a comprehensive and dynamic analysis of connectivity in temporal graphs, which is crucial for understanding the underlying structure and behavior of real-world systems.}
In our problem, we focus on the strict and simple temporal path, that is, timestamps of edges along the path follow a strict ascending temporal order and no repeated vertices exist in the path~\cite{DBLP:journals/tcs/CasteigtsCS24}.
Specifically, given a directed temporal graph $\mathcal{G}=(\mathcal{V},\mathcal{E})$, two vertices $s$, $t$ $\in \mathcal{V}$, and a time interval $[\tau_b,\tau_e]$, the temporal simple path graph, denoted by $\stspgraph_{[\tau_b,\tau_e]}(s,t)$, is the subgraph of $\mathcal{G}$ containing exactly all temporal simple paths from $s$ to $t$ within the time interval $[\tau_b,\tau_e]$.

\begin{figure}[t]
    \centering
    \begin{subfigure}{0.18\textwidth}
        \includegraphics[width=\textwidth]{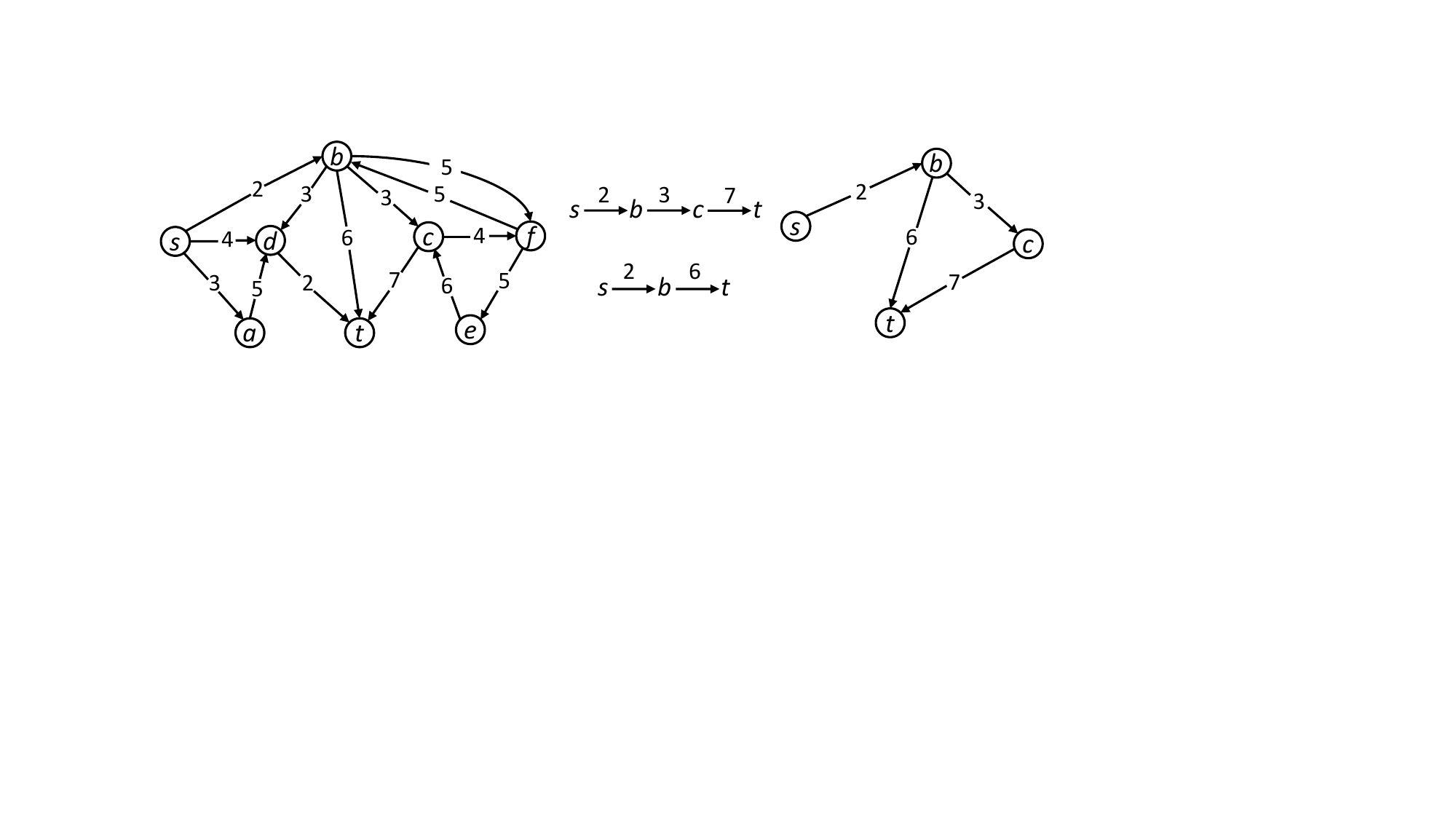}
        \caption{$\mathcal{G}=(\mathcal{V},\mathcal{E})$}
        \label{intro:tg_a}
    \end{subfigure}
    \begin{subfigure}{0.11\textwidth}
        \includegraphics[width=\textwidth]{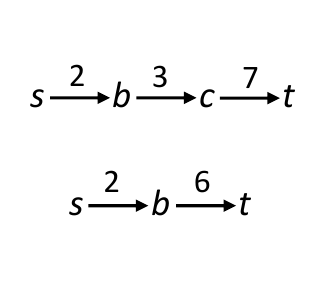}
        \caption{\vspace{-0.55ex}$\mathcal{P}^*_{[2,7]}(s,t)$}
        \label{intro:tg_b}
    \end{subfigure}
    \begin{subfigure}{0.13\textwidth}
        \includegraphics[width=\textwidth]{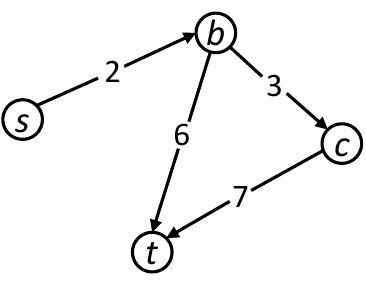}
        \caption{$\stspgraph_{[2,7]}(s,t)$}
        \label{intro:tg_c}
    \end{subfigure}
    \caption{Running example}
    \label{intro:tg}
\end{figure}

\begin{example}
    Fig.~\ref{intro:tg}(a) shows a directed temporal graph.
    Given the source vertex $s$, the target vertex $t$ and the time interval $[2,7]$, there exist two temporal simple paths from $s$ to $t$ within $[2,7]$, whose details are shown in Fig.~\ref{intro:tg}(b).
    These two paths share the same edge $e(s,b,2)$, and we illustrate the corresponding temporal simple path graph in Fig.~\ref{intro:tg}(c).
    \cblack{As observed, the path graph facilitates path enumeration and explicitly reveals structural connectivity between vertices}.
\end{example}

\myparagraph{Applications}
In the following, we illustrate \cblack{four} representative applications of temporal simple path graph generation.

\begin{itemize}[leftmargin=*]
    \item \textbf{Outbreak Control.} A disease transmission network can be modeled as a temporal graph, 
    where vertices correspond to locations of individual contact, and temporal edges indicate the movement of individuals to specific locations at particular timestamps~\cite{eubank2004modelling}. The transmission paths in such a graph reflect the spread of pathogens over time.
    Generating the temporal simple path graph from the outbreak source to the protected area, \cblack{health authorities can visualize transmission routes, identify critical nodes, and prioritize containment strategies for optimized epidemic control}. 
    
    \item \textbf{Financial Monitor.} Temporal graphs can be used to model transaction networks, where each vertex represents an account and each edge represents a transaction from one account to another at a specific time. 
    \cblack{Money laundering patterns often manifest as cyclic transaction sequences with ascending timestamps within tight temporal windows~\cite{DBLP:journals/pvldb/QiuCQPZLZ18}.
    A transaction $e(t,s,\tau)$ belongs to such a cycle if a temporal simple path exists from $s$ to $t$ within specified interval~\cite{DBLP:journals/pacmmod/CaiLZ023}.
}    
    Thus, we can visualize the flow of money from one account to another by our problem, providing a clear representation of the transaction process
    and aiding in identification of risk-associated accounts and transactions.
    \item \textbf{\cblack{Travel Planning.}} 
    \cblack{Transportation networks modeled as temporal graphs can efficiently capture dynamic traffic patterns, 
    with vertices denoting intersections or road segments and temporal edges denoting vehicle movements~\cite{DBLP:conf/sigmod/WangLYXZ15,DBLP:conf/esa/BastCEGHRV10}. 
    In a large urban area with a complex public transportation system, passengers often face challenges in planning their routes, especially when they miss their initial connections. By generating the temporal simple path graph, we can visualize all possible transfer options that allow passengers to reach their destinations within a desirable time interval.
    }
    \item \textbf{\cblack{Trend Detection.}} \cblack{In social networks, information propagation is a time-dependent process. Interactions among users, such as retweets, comments, and likes, can be modeled as temporal graphs~\cite{DBLP:conf/sigmod/GurukarRR15}. 
    A temporal simple path graph from the information source to a target user within a specified time interval enables the analysis of the information flow and the recognition of key influencers while avoiding the explicit tracing of dissemination routes. 
    }
\end{itemize}

\myparagraph{Challenges}
To the best of our knowledge,
we are the first to propose and investigate the temporal simple path graph generation problem.
Naively, we can enumerate all temporal simple paths from the source vertex to the target vertex within the query time interval, then extract distinct vertices and edges from those paths to form the $\stspgraph$.
However, this approach is far from efficient since it suffers from an exponential time complexity of $\mathcal{O}(d^{\theta}\cdot \theta \cdot m)$, where $d$ is the largest vertex out-degree or in-degree in the original graph $\mathcal{G}$, $\theta$ is the length of the query time interval and $m$ is the number of edges in $\mathcal{G}$.
The key drawbacks of the naive enumeration strategy can be observed as follows:
$i)$ The original graph $\mathcal{G}$ includes numerous edges that are definitely not part of any temporal simple paths, which leads to unnecessary large search space for the enumeration.
$ii)$ Some edges may appear in multiple enumerated temporal
simple paths from $s$ to $t$ and will be processed repeatedly during the $\stspgraph$ formation process.

\myparagraph{Our solutions}
To address the aforementioned drawbacks and 
efficiently generate the temporal simple path graph, a novel method named \textit{\underline{V}erification~in~\underline{U}pper$-$bound~\underline{G}raph} (shorted as \texttt{\VUG}) is proposed with two main components, \ie Upper-bound Graph Generation and Escaped Edges Verification.

Specifically, \texttt{\VUG} reduces the search space for temporal simple paths by constructing an upper-bound graph as a high-quality approximation of the temporal simple path graph in polynomial time through two phases.
In the first phase, it utilizes the \texttt{QuickUBG} method to efficiently filter out edges that are not contained in any temporal paths from $s$ to $t$ within $[\tau_b,\tau_e]$ and obtain a quick upper-bound graph $\gq$.
This can be achieved by comparing the timestamp of an edge with the earliest arrival time $\ea{\cdot}$ and latest departure time $\ld{\cdot}$ of its end vertices, where $\ea{u}$ (resp. $\ld{u}$) is when a temporal path from $s$ within $[\tau_b,\tau_e]$ first arrives at $u$ (resp. a temporal path to $t$ within $[\tau_b,\tau_e]$ last departs $u$).
$\ea{u}$ and $\ld{u}$ for all vertices can be computed using an extended BFS-like procedure in $\mathcal{O}(n + m)$ time, where $n$ and $m$ are the number of vertices and edges in $\mathcal{G}$, respectively.
In the second phase, \texttt{\VUG} applies the \texttt{TightUBG} method on $\gq$, which further excludes several edges that can be efficiently determined as not contained in any temporal simple path from $s$ to $t$ within $[\tau_b,\tau_e]$.
This method is powered by the introduction of \textit{time-stream common vertices}, which summarizes the vertices that commonly appear across all temporal simple paths from $s$ to $u$ (resp. from $u$ to $t$) within a time interval of interest.
The computation of time-stream common vertices utilizes a recursive method
in $\mathcal{O}(n + \theta \cdot m)$ time without explicitly listing the paths, more clearly, time-stream common vertices of paths from $s$ to vertex $u$ can be calculated from time-stream common vertices of paths from $s$ to the in-neighbors of $u$.
After that, an edge $e(u, v, \tau)$ with two disjoint sets of time-stream common vertices, one set for temporal simple paths from $s$ to $u$ arriving before $\tau$ and the other set for temporal simple paths from $v$ to $t$ departuring after $\tau$, is considered to have the potential of forming a temporal simple path from $s$ to $t$ and is retained in the tight upper-bound graph $\gt$.

Finally, we generate the exact $\stspgraph$ by verifying each edge in $\gt$. 
Instead of employing a brute-force path enumeration, we propose an Escaped Edges Verification ($\eev$) method that mitigates the repetition in edge verification.
This method iteratively selects an unverified edge, identifies a temporal simple path from $s$ to $t$ containing it through bidirectional DFS, then adds a set of vertices and edges to the $\stspgraph$. 
Two carefully designed optimization strategies regarding search direction and neighbor exploration order are integrated into bidirectional DFS to further accelerate the process of algorithm.

\myparagraph{Contributions}
We summarize the contributions in this paper.
\begin{itemize}[leftmargin=*]
    \item We conduct the first research to propose and investigate the problem of generating temporal simple path graph.
    \item To address this challenging problem, we propose an efficient method, namely \texttt{\VUG}, consisting of two components, including Upper-bound Graph Generation for effectively yielding a high-quality approximate solution and Escaped Edges Verification for efficiently achieving the exact solution.
    \item We conduct extensive experiments on \cblack{10} real-word temporal graphs to compare \texttt{\VUG} against baselines. The results demonstrate the effectiveness and efficiency of our methods.
\end{itemize}


\vspace{1mm}
\textit{Note that, due to the limited space, all omitted contents can be found in our appendix online~\cite{appendix}}.

\section{Preliminaries}
\label{sec:preli}



A directed temporal graph $\mathcal{G}=(\mathcal{V},\mathcal{E})$ consists of a set of vertices $\mathcal{V}=\mathcal{V}(\mathcal{G})$ and a set of directed temporal edges $\mathcal{E}=\mathcal{E}(\mathcal{G})$.
We use $n = |\mathcal{V}|$ and $m = |\mathcal{E}|$ to denote the number of vertices and edges, respectively.
$e(u,v,\tau) \in \mathcal{E}$ denotes a directed temporal edge from vertex $u$ to vertex $v$, where $u$, $v \in \mathcal{V}$ and $\tau$ is the interaction timestamp from $u$ to $v$. 
Without loss of generality, we use the same setting as previous studies for timestamp, which is the integer, since the UNIX timestamps are integers in practice (e.g., \cite{DBLP:journals/pvldb/WuCHKLX14,DBLP:conf/icde/WuSWZ00024}).
We use $\mathcal{T} = \{\tau|e(u, v, \tau) \in \mathcal{E}\}$ to represent the set of timestamps.
Given a directed temporal graph $\mathcal{G}=(\mathcal{V},\mathcal{E})$, a subgraph $\mathcal{S}=(\mathcal{V_S},\mathcal{E_S})$ is an induced subgraph of $\mathcal{G}$, if 
$\mathcal{E_S} \subseteq \mathcal{E}$, $\mathcal{V_S}=\{u|e(u,v,\tau) \in \mathcal{E_S} \lor e(v, u,\tau) \in \mathcal{E_S}\}$.
Given a subgraph $\mathcal{S}$, let $N_{out}(u,\mathcal{S}) = \{(v,\tau)|e(u,v,\tau) \in \mathcal{E_S}\}$ (resp. $N_{in}(u,\mathcal{S}) = \{(v,\tau)|e(v,u,\tau) \in \mathcal{E_S}\}$) be the set of out-neighbors (resp. in-neighbors) of $u$ in $\mathcal{S}$.
We use $\mathcal{T}_{out}(u,\mathcal{S})$ (resp. $\mathcal{T}_{in}(u,\mathcal{S})$) to represent all distinct timestamps in $N_{out}(u,\mathcal{S})$ (resp. $N_{in}(u,\mathcal{S})$).

Given a time interval $[\tau_b,\tau_e]$, where $\tau_b$, $\tau_e \in \mathcal{T}$ and~$\tau_b \leq \tau_e$, we use $\theta$ to denote the span of $[\tau_b,\tau_e]$, \ie $\theta=\tau_e - \tau_b + 1$.
The projected graph of $\mathcal{G}$ within $[\tau_b,\tau_e]$ is a subgraph of $\mathcal{G}$, denoted by $\mathcal{G}_{[\tau_b,\tau_e]}=(\mathcal{V}',\mathcal{E}')$, 
where $\mathcal{E}'=\{e(u,v,\tau)| e(u, v, \tau) \in \mathcal{E} \land \tau \in [\tau_b,\tau_e]\}$ and $\mathcal{V}'=\{u|e(u,v,\tau) \in \mathcal{E}' \lor e(v, u,\tau) \in \mathcal{E}'\}$.



Given two vertices $s$, $t \in \mathcal{V}$ and a time interval $[\tau_b,\tau_e]$, a temporal path $p_{[\tau_b,\tau_e]}(s,t)$ from $s$ to $t$ in $[\tau_b,\tau_e]$, simplified as $p$, is a sequence of edges $\langle e(s=v_0,v_1,\tau_1),\dots,e(v_{l-1},v_l=t,\tau_{l}) \rangle$ such that $\tau_b \leq \tau_i < \tau_{i+1} \leq \tau_e$ for all integers $1 \leq i \leq l-1$. 
$\mathcal{V}(p)$ and $\mathcal{E}(p)$ are the set of vertices and edges included in $p$, respectively.
The length of $p$ is the number of edges on the path, denoted by $l=|\mathcal{E}(p)|$.




\begin{remark}\label{rem:lengthlimit}
Let $l$ be the length of a temporal path $p_{[\tau_b,\tau_e]}(s,t)$, and $\theta$ be the span of $[\tau_b,\tau_e]$, it is easy to obtain that $l \leq \theta$.
\end{remark}

\begin{definition}[Temporal Simple Path]
Given a directed temporal graph $\mathcal{G}=(\mathcal{V},\mathcal{E})$, two vertices $s$, $t \in \mathcal{V}$, and a time interval $[\tau_b,\tau_e]$, 
a temporal simple path $p_{[\tau_b,\tau_e]}^*(s,t)=\langle e(s=v_0,v_1,\tau_1),\dots,e(v_{l-1},v_l=t,\tau_{l}) \rangle$ is a temporal path from $s$ to $t$ within $[\tau_b,\tau_e]$, without repeated vertices. That is, for any integers $0 \leq i < j \leq l$, $v_i \neq v_j$. 


\end{definition}



We use $\mathcal{P}_{[\tau_b,\tau_e]}(s,t)$ to denote the set of all temporal paths within $[\tau_b,\tau_e]$. 
The vertex set and edge set of $\mathcal{P}_{[\tau_b,\tau_e]}(s,t)$ are denoted by $\mathcal{V}(\mathcal{P}_{[\tau_b,\tau_e]}(s,t))=\cup_{p \in \mathcal{P}_{[\tau_b,\tau_e]}(s,t)} \mathcal{V}(p)$ and $\mathcal{E}(\mathcal{P}_{[\tau_b,\tau_e]}(s,t))=\cup_{p \in \mathcal{P}_{[\tau_b,\tau_e]}(s,t)} \mathcal{E}(p)$, respectively.
Similarly, we denote the temporal simple path set as $\mathcal{P}_{[\tau_b,\tau_e]}^*(s,t)$, along with its corresponding vertex set $\mathcal{V}(\mathcal{P}_{[\tau_b,\tau_e]}^*(s,t))$ and edge set $\mathcal{E}(\mathcal{P}_{[\tau_b,\tau_e]}^*(s,t))$.

\begin{definition}[Temporal Simple Path Graph]\label{the:graph}
Given a directed temporal graph $\mathcal{G}=(\mathcal{V},\mathcal{E})$, 
a source vertex $s \in \mathcal{V}$, a target vertex $t \in \mathcal{V}$ 
and a time interval $[\tau_b,\tau_e]$, the temporal simple path graph from $s$ to $t$ within $[\tau_b,\tau_e]$, denoted by $\stspgraph_{[\tau_b,\tau_e]}(s,t)$, is a subgraph $\mathcal{S}=(\mathcal{V_S},\mathcal{E_S})$ of $\mathcal{G}$ such that $\mathcal{V_S}= \mathcal{V}(\mathcal{P}_{[\tau_b,\tau_e]}^*(s,t))$ and $\mathcal{E_S}= \mathcal{E}(\mathcal{P}_{[\tau_b,\tau_e]}^*(s,t))$.


\end{definition}


\myparagraph{Problem Statement}
Given a directed temporal graph $\mathcal{G}=(\mathcal{V},\mathcal{E})$, 
a source vertex $s \in \mathcal{V}$, a target vertex $t \in \mathcal{V}$ 
and a time interval $[\tau_b,\tau_e]$, we aim to efficiently generate the temporal simple path graph $\stspgraph_{[\tau_b,\tau_e]}(s,t)$.

Note that, when the context is clear, we can omit the subscript notation $[\tau_b,\tau_e]$ and vertex pair $(s,t)$, that is, $p_{[\tau_b,\tau_e]}(s,t)$, $p_{[\tau_b,\tau_e]}^*(s,t)$, $\mathcal{P}_{[\tau_b,\tau_e]}(s,t)$, $\mathcal{P}_{[\tau_b,\tau_e]}^*(s,t)$ and $\stspgraph_{[\tau_b,\tau_e]}(s,t)$ can be simplified to $p$, $p^*$, $\mathcal{P}$, $\mathcal{P}^*$ and $\stspgraph$.
\section{Solution Overview}
\label{sec:sol_overview}
In this section, we first present three reasonable baseline methods to generate temporal simple path graphs by extending the existing studies for temporal simple path enumeration.
Then, we introduce the general framework of our solution. 

\subsection{The Baseline Methods}
\label{sec:sol_overview_a}

\cblack{A naive method for generating the $\stspgraph_{[\tau_b,\tau_e]}(s,t)$ is to first exhaustively enumerate all temporal simple paths from the source vertex $s$ to the target vertex $t$ within the time interval $[\tau_b,\tau_e]$ in $\mathcal{G}$, then construct the $\stspgraph$ by taking the union of the vertex sets and edge sets of these paths.}
The running time of this method 
is bounded by $\mathcal{O}(d^{\theta}\cdot \theta \cdot m)$,
where $d$ is the largest out-degree or in-degree of the vertices in $\mathcal{G}$, i.e., $d=\max_{u\in\mathcal{V}}\{\max(|N_{in}(u,\mathcal{G})|,|N_{out}(u,\mathcal{G})|)\}$, and $\theta$ is the span of $[\tau_b,\tau_e]$, i.e., $\theta=\tau_e-\tau_b+1$.
Although this approach can return the result, it may not be efficient due to the large search space, which is the original graph, including numerous vertices and edges that definitely cannot be involved in any temporal simple paths. This leads to unnecessary computational costs and slows down the enumeration process.

\myparagraph{Baseline methods}
To reduce the large search space of the naive method, 
our baseline methods first filter out many unpromising edges to identify a subgraph of $\mathcal{G}$, which we refer to as the \textbf{upper-bound graph} for $\stspgraph$.
Then, we enumerate temporal simple paths {in the upper-bound graph} rather than the original graph, to construct $\stspgraph$.




For the temporal simple path enumeration, a straightforward upper-bound graph would be the projected graph $\mathcal{G}_{[\tau_b,\tau_e]}$ of $\mathcal{G}$, 
whose computation method is called $\dt$ in this paper, \cblack{\ie it prunes edges $e(u,v,\tau)$ where $\tau \notin [\tau_b,\tau_e]$ in $\mathcal{O}(m)$ time}.
Recently, Jin \etal~\cite{DBLP:journals/kais/JinCL24} propose two graph reduction methods (named $\es$ and $\tg$) for temporal path enumeration, and the two reduced graphs can also serve as upper-bound graphs of our problem.
Specifically, $\es$ excludes edges that do not belong to any path from $s$ to $t$ with non-decreasing timestamps, 
\cblack{\ie it performs bidirectional temporal edge traversal (forward with non-decreasing and backward with non-ascending), uses flags to retain vertices and edges that are bidirectionally marked, achieving $\mathcal{O}(m)$ time complexity.}
$\tg$ discards edges that are not contained in any path from $s$ to $t$ with ascending timestamps,
\cblack{\ie it employs bidirectional Dijkstra with strict temporal constraints to identify reachable vertices and edges, with an overall time complexity of $\mathcal{O}((n+m) \cdot logn + m)$.}
Obviously, these two methods can construct smaller upper-bound graphs than the projected graph.

In our experiment, we design three baseline algorithms, namely $\epdt$, $\epes$ and $\eptg$, based on the above three upper-bound graph generation methods, i.e., we first generate an upper-bound graph using each method, then enumerate temporal simple paths on the generated upper-bound graph \cblack{by initiating a DFS from $s$ to explore all temporal simple paths to $t$ within $[\tau_b,\tau_e]$.
During the traversal, for each processed edge, we check its timestamp to ensure a strict temporal constraint in the formed path.
After enumeration, we obtain the $\stspgraph$ by combining all vertices and edges from these paths, where each vertex and edge is added to the $\stspgraph$ once through checking the inserted vertices and edges.
} 

\begin{figure}[t]
    \centering
    \begin{subfigure}{0.17\textwidth}
        \includegraphics[width=\textwidth]{Figure_Example/tg_a.pdf}
        \caption{\texttt{\dtTSG}}
        \label{baseline:dtTSG}
    \end{subfigure}
    \begin{subfigure}{0.13\textwidth}
        \includegraphics[width=\textwidth]{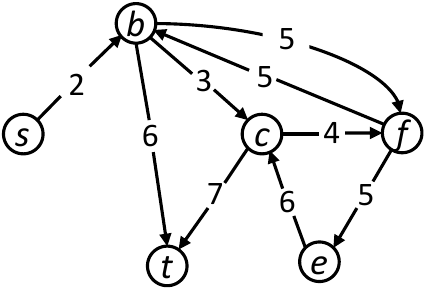}
        \caption{\texttt{\esTSG}}
        \label{baseline:esTSG}
    \end{subfigure}
    \begin{subfigure}{0.13\textwidth}
        \includegraphics[width=\textwidth]{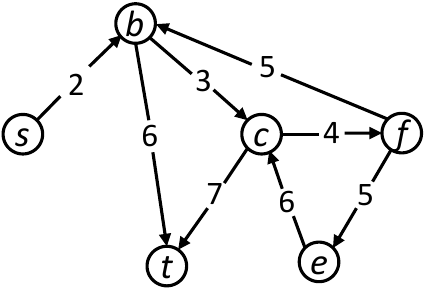}
        \caption{\texttt{\tgTSG}}
        \label{baseline:tgTSG}
    \end{subfigure}
    \caption{Upper-bound graphs of baselines within interval $[2,7]$}
    \label{baselines:upper-bound graphs}
\end{figure}

\begin{example}
\label{expample:threeupper}
    Given the directed temporal graph $\mathcal{G}$ in Fig.~\ref{intro:tg}(a).
    Suppose the source vertex is $s$, the target vertex is $t$, and the time interval is $[2,7]$.
    In Figs.~\ref{baselines:upper-bound graphs}(a)-\ref{baselines:upper-bound graphs}(c), we illustrate three upper-bound graphs obtained by three baseline algorithms.
\end{example}

\myparagraph{Limitations of baselines} 
Although the baseline methods can successfully generate temporal simple path graphs, they still have limitations that present opportunities for improvement.

\begin{itemize}[leftmargin=*]
    \item[$i)$] The upper-bound graph could be further reduced. 
    For example, edge $e(e, c, 6)$ will not be included in the resulting $\stspgraph$ as it only appears in a temporal path with a cycle. 
    However, since these upper-bound graphs ignore the constraint of the simple path, such edges are retained in these graphs. 
    \item[$ii)$] Efficiency limitation exists due to the enumeration process.
    Since some edges may appear in multiple temporal simple paths, the enumeration may lead to repeated verifications to check whether an edge is already included in the result.
    Therefore, developing efficient techniques for path enumeration can further enhance the performance of the algorithm.
\end{itemize}

\subsection{The Verification in Upper-bound Graph Framework}
\label{sec:framework}
To better approximate $\stspgraph$ by the upper-bound graph and avoid repeatedly checking edges that are already added to the result by previously discovered paths, 
we propose a novel algorithm named \textit{\underline{V}erification~in~\underline{U}pper$-$bound~\underline{G}raph} (i.e., \texttt{\VUG}) with two main components, namely Upper-bound Graph Generation and Escaped Edges Verification, 
with framework shown in Algorithm~\ref{alg:framework}.


\myparagraph{$i)$ Upper-bound Graph Generation}
We generate the upper-bound graph for $\stspgraph$ in two phases. In the first phase, the quick upper-bound graph $\gq$ is computed based on temporal constraints. In the second phase, $\gq$ is further reduced to the tight upper-bound graph $\gt$ by applying the simple path constraint (more details can be found in Section~\ref{sec:ubg}).

\myparagraph{$ii)$ Escaped Edges Verification}
We obtain exact $\stspgraph$ by extending bidirectional DFS, \ie select an unverified edge $e(u,v,\tau)$, identify a temporal simple path $p_{[\tau_b,\tau_e]}^*(s,t)$ through $e(u,v,\tau)$ and add a set of vertices and edges to $\stspgraph$ (more details can be found in Section~\ref{sec:ver}).

\begin{algorithm}[t]
\caption{Framework of VUG}\label{alg:framework}
{
    \SetVline
    \Input{a directed temporal graph $\mathcal{G}$, a source vertex $s$, a target vertex $t$ and a query time interval $[\tau_b,\tau_e]$}
    \Output{the temporal simple path graph $\stspgraph_{[\tau_b,\tau_e]}(s,t)$}
    
    \tcp{Upper-bound Graph Generation}
    \StateCmt{$\gq \gets$ \quick($\mathcal{G}$, $\mathcal{A}$, $\mathcal{D}$)}{\footnotesize Algorithm~\ref{alg:tpgconstruct}}
    \StateCmt{$\gt \gets$ \tight($\gq$, $TCV$)}{\footnotesize Algorithm~\ref{alg:ipgconstruct}}
    
    \tcp{Escaped Edges Verification}

    \StateCmt{$\stspgraph \gets$ \eev($s$, $t$, $[\tau_b,\tau_e]$, $\gt$)}{\footnotesize Algorithm~\ref{alg:tspgconstruct}}
    \State{\Return $\stspgraph$}
}
\end{algorithm}
\section{Upper-bound Graph Generation}
\label{sec:ubg}


In this section, we present two upper-bound graph computation methods, named quick upper-bound graph generation (\quick) and tight upper-bound graph generation (\tight).
\cblack{Given the potentially large size of the original graph, we start with a fast approach \quick to obtain an upper-bound graph $\gq$ by considering the temporal constraints of our problem.}
Then, we apply our stronger method \tight to obtain a more precise upper-bound graph $\gt$.


\subsection{Quick Upper-bound Graph Generation}
\label{sec:ubg_one}

Given a directed temporal graph $\mathcal{G}=(\mathcal{V},\mathcal{E})$ and a temporal simple path graph query of $s$ to $t$ in the time interval $[\tau_b,\tau_e]$, 
for an edge to be included in the $\stspgraph$,  
it must be contained in at least one temporal path in $\mathcal{P}_{[\tau_b,\tau_e]}(s,t)$.
On this basis,
we have the observation below.


\begin{observation} 
\label{obs:tpexist}
An edge $e(u,v,\tau)$ is contained in a temporal path $p_{[\tau_b,\tau_e]}(s,t) \in \mathcal{P}_{[\tau_b,\tau_e]}(s,t)$ iff any one of the following conditions holds:
\begin{enumerate}
    \item[$i)$] $u \neq s$, $v \neq t$, there exist two temporal paths $p_{[\tau_b,\tau_i]}(s,u)$ and $p_{[\tau_j,\tau_e]}(v,t)$ s.t. $\tau_i < \tau < \tau_j$.
    \item[$ii)$] $u = s$, $v \neq t$, there exist a temporal path $p_{[\tau_j,\tau_e]}(v,t)$ s.t. $\tau_b \leq \tau < \tau_j$.
    \item[$iii)$] $u \neq s$, $v = t$, there exist a temporal path $p_{[\tau_b,\tau_i]}(s,u)$ s.t. $\tau_i < \tau \leq \tau_e$.
    \item[$iv)$] $u = s$, $v = t$, $\tau_b \leq \tau \leq \tau_e$.
\end{enumerate}

\end{observation}





Using the above observation as an edge exclusion condition, 
we can effectively construct an upper-bound graph for $\stspgraph$, referred to as quick upper-bound graph $\gq$, that only includes edges that appear in at least one temporal path $p_{[\tau_b,\tau_e]}(s,t)$.
Before presenting our detailed strategy, we first formally define the concepts of \textit{arrival time} and \textit{departure time}.


\begin{definition}[Arrival \& Departure Time]
Given a temporal path $p = p_{[\tau_b,\tau_e]}(s,u)=\langle e(s,\cdot,\cdot), \cdots,e(\cdot, u, \tau_a) \rangle$,
the arrival time of $u$ regarding $s$ in this path, denoted by $a(p, u)$, is the timestamp of the last edge within the path, i.e., $a(p, u)=\tau_a$.
Similarly, given a temporal path $p=p_{[\tau_b,\tau_e]}(v,t)=\langle e(v, \cdot, \tau_d),\cdots, e(\cdot, t, \cdot)\rangle$, the departure time of $v$ regarding $t$ in this path, denoted by $d(p, v)$, is the timestamp of the first edge within the path, i.e., $d(p, v)=\tau_d$.
\end{definition}


Since each vertex may appear in multiple temporal paths from $s$ to it or from it to $t$, it can have multiple arrival and departure times. 
Therefore, we define \textit{polarity time} to store each vertex's earliest arrival time and latest departure time.


 
\begin{definition}[Polarity Time]\label{def:polarity}
For each vertex $u\in \mathcal{V} \setminus \{s\}$, the earliest arrival time of $u$, denoted by $\ea{u}$, is the smallest arrival time of $u$ regarding $s$ among all the paths from $s$ to $u$ within $[\tau_b,\tau_e]$, i.e., $\ea{u} = \min\{a(p, u) | p \in \mathcal{P}_{[\tau_b,\tau_e]}(s,u)\}$.
Similarly, for each vertex $u\in \mathcal{V} \setminus \{t\}$, the latest departure time of $u$, denoted by $\ld{u}$, is the largest departure time of $u$ regarding $t$ among all the paths from $u$ to $t$ within $[\tau_b,\tau_e]$, i.e., $\ld{u} = \max\{d(p, u)|p \in \mathcal{P}_{[\tau_b,\tau_e]}(u,t)\}$.
\end{definition}

\begin{example}
Given the directed temporal graph $\mathcal{G}=(\mathcal{V},\mathcal{E})$ in Fig.~\ref{intro:tg}(a), a source vertex $s$, a target vertex $t$ and a time interval $[2,7]$.
For the vertex $f \in \mathcal{V} \setminus \{s,t\}$,
$\mathcal{P}_{[2,7]}(s,f)=\{\langle e(s,b,2),e(b,f,5)\rangle,\langle e(s,b,2),e(b,c,3),e(c,f,4)\rangle\}$, thus $\mathcal{A}(f)=\min\{4,5\}=4$.
Similarly, $\mathcal{P}_{[2,7]}(f,t)
=\{\langle e(f,e,5),$ $e(e,c,6),e(c,t,7)\rangle,\langle e(f,b,5),e(b,t,6)\rangle\}$, thus $\mathcal{D}(f)=5$.
\end{example}








\begin{figure}[t]
    \centering
    \begin{subfigure}{0.1\textwidth}
        \includegraphics[width=\textwidth]{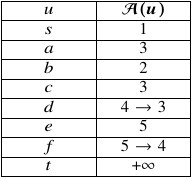}
        \caption{$\ea{u}$}
        \label{tg:time}
    \end{subfigure}
    \hspace{2mm}
    \begin{subfigure}{0.1\textwidth}
        \includegraphics[width=\textwidth]{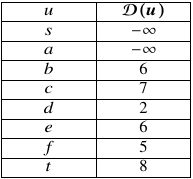}
        \caption{$\ld{u}$}
        \label{tg:v_pru}
    \end{subfigure}
    \begin{subfigure}{0.18\textwidth}
        \includegraphics[width=\textwidth]{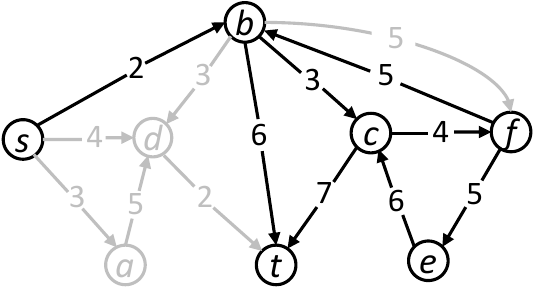}
        \caption{$\gq$}
        \label{tg:e_pru}
    \end{subfigure}
    \caption{Quick Upper-bound Graph Generation}
    \label{tg:computing tpgraph}
\end{figure}



With Observation~\ref{obs:tpexist} and Definition~\ref{def:polarity} in hand, we can~determine whether an edge is contained in any $p_{[\tau_b,\tau_e]}(s,t)$ based on the polarity time of its end vertices, as stated in Lemma \ref{lem:tpedge}.
\cblack{The proof follows directly and is omitted here.}


\begin{lemma}\label{lem:tpedge}
Given an edge $e(u,v,\tau) \in \mathcal{E}$, there exists a temporal path from the source vertex $s$ to the target vertex $t$ containing $e(u,v,\tau)$ within $[\tau_b,\tau_e]$ iff $\ea{u} < \tau < \ld{v}$. 
\end{lemma}


Based on Lemma~\ref{lem:tpedge}, we can efficiently generate $\gq$, as illustrated in Algorithm~\ref{alg:tpgconstruct}.
Note that, here we suppose the polarity time of every vertex is already given, and its computation will be discussed later.
A running example is provided as follows.

\begin{algorithm}[t]  
{
    \SetVline
    \caption{Quick Upper-bound Graph Generation}\label{alg:tpgconstruct}
        \Input{a directed temporal graph $\mathcal{G}=(\mathcal{V}, \mathcal{E})$, earliest arrival time $\ea{u}$ and latest departure time $\ld{u}$ for each vertex $u \in \mathcal{V}$}
        \Output{quick upper-bound graph $\gq$=$(\mathcal{V}_q,\mathcal{E}_q)$}
    \State{$\mathcal{V}_q=\emptyset$, $\mathcal{E}_q=\emptyset$}  
    \ForEach{$e(u, v, \tau) \in \mathcal{E}$}{
        \If{$\ea{u} < \tau \land \tau < \ld{v}$}{
            \State{$\mathcal{V}_q \gets \mathcal{V}_q \cup \{u,v\}$}
            \StateCmt{$\mathcal{E}_q \gets \mathcal{E}_q \cup \{e(u, v, \tau)\}$}{Lemma~\ref{lem:tpedge}}
        }
    }
    \State{\Return{$(\mathcal{V}_q,\mathcal{E}_q)$}}
}
\end{algorithm}

\begin{example}\label{example:gq}
Given the directed temporal graph $\mathcal{G}$ in Fig.~\ref{intro:tg}(a), 
the precomputed polarity time in 
Figs.~\ref{tg:computing tpgraph}(a)-\ref{tg:computing tpgraph}(b), 
the resulting quick upper-bound graph $\gq$
is shown in Fig.~\ref{tg:computing tpgraph}(c).  
We exclude $e(s,a,3)$ since $\ld{a}= -\infty < 3$, $e(d,t,2)$ as $\ea{d}=3>2$.

\end{example}


\begin{theorem}\label{the:gq}
The time complexity of Algorithm~\ref{alg:tpgconstruct} is $\mathcal{O}(m)$, and its space complexity is $\mathcal{O}(n+m)$.
\end{theorem}

\textit{Due to the space limitation, the proof for Theorem~\ref{the:gq} and other omitted proofs can be found in Appendix A online \cite{appendix}.}

Next, we introduce our method for calculating the polarity time for each vertex.
\cblack{With a BFS-oriented traversal starting from the source vertex $s$ (resp. the target vertex $t$), we record and update the earliest arrival time (resp. the latest departure time) for each reachable vertex $u$ when necessary, as there may exist multiple paths from $s$ to $u$ (resp. from $u$ to $t$).}
For computational simplicity, we define $\ea{s}=\tau_b - 1$, $\ld{t}=\tau_e + 1$.
The detailed pseudocode of the polarity time computation is given in Algorithm~\ref{alg:polaritytime}.
Note that, $\ea{u}=+\infty$ (resp. $\ld{u}=-\infty$) implies there is no temporal path from $s$ to $u$ (resp. from $u$ to $t$) within $[\tau_b,\tau_e]$ without passing through $t$ (resp. $s$).
\begin{algorithm}[t]
{
    \SetVline
    \caption{Polarity Time Computation}\label{alg:polaritytime}
    \Input{a directed temporal graph $\mathcal{G}=(\mathcal{V}, \mathcal{E})$, a source vertex $s \in \mathcal{V}$, a target vertex $t \in \mathcal{V}$ and a query time interval $[\tau_b,\tau_e]$}
    \Output{the earliest arrival time $\ea{u}$ and latest departure time $\ld{u}$ for each vertex $u \in \mathcal{V}$}
    \State{$\ea{s} \gets \tau_{b} - 1$, $\ea{u} \gets +\infty$ \textbf{for each} $u \in \mathcal{V} \setminus \{s\}$}
    \State{$\ld{t} \gets \tau_{e} + 1$, $\ld{u} \gets -\infty$ \textbf{for each} $u \in \mathcal{V} \setminus \{t\}$}

    \State{${Q} \gets$ a queue containing $s$}
    \While{${Q}\neq \emptyset$}{
        \State{$u \gets {Q}.pop()$}
        \ForEach{$(v, \tau) \in N_{out}(u,\mathcal{G})$ s.t. $v \neq t$}{
            \State{\textbf{if} $\tau > \tau_e \lor \ea{u} \geq \tau \lor \tau \geq \ea{v}$ \textbf{then continue}}
            
            \State{$\ea{v} \gets \tau$}
            \lIf{$\tau \ne \tau_{e} \land v \notin {Q}$}{\State{${Q}.push(v)$}}
        }
    }
    \State{repeat Lines 3-9 with required adjustment for $\ld{u}$} 
    \State{\Return $\ea{u}, \ld{u}$ for each $u \in \mathcal{V}$}
}
\end{algorithm}

\myparagraph{\underline{Polarity Time Computation}}
In Algorithm~\ref{alg:polaritytime}, $\mathcal{G}$ is stored such that, within the neighbor list $N_{out}(u,\mathcal{G})$, $N_{in}(u,\mathcal{G})$ of each vertex $u \in \mathcal{V}(\mathcal{G})$, neighbors are sorted based on their timestamps.
In Lines 4-9, for each popped vertex $u$ from $Q$, we iteratively explore its out-neighbors.
It is worth mentioning that, a pointer in $N_{out}(u,\mathcal{G})$ is maintained for each vertex $u \in \mathcal{V}(\mathcal{G})$ to help filter out $(v,\tau)$ that has been processed during a previous visit to $u$, which avoids scanning the entire $N_{out}(u,\mathcal{G})$ each time $u$ is visited.
In Line 7, an out-neighbor $(v, \tau)$ with $\ea{u} \geq \tau$ or $\tau \geq \ea{v}$ is ignored since a new temporal path $p_{[\tau_b,\tau]}(s,v)$ cannot be formed or $p_{[\tau_b,\tau]}(s,v)$ has already been explored previously. 
We update $\ea{v}$ in Line 8 as we find a new temporal path $p_{[\tau_b,\tau]}(s,v)$ with an earlier arrival time for $v$, and in Line 9 we add $v$ to ${Q}$ for further exploration if it is not yet in ${Q}$. 
We also repeat the process in Lines 3-9 by toggling between $s$ and $t$, ``out'' and ``in'', $\mathcal{A}$ and $\mathcal{D}$, $\tau_e$ and $\tau_b$, $>$ and $<$, $\geq$ and $\leq$ to obtain $\mathcal{D}(\cdot)$.

\begin{example}
Given the directed temporal graph $\mathcal{G}$ in Fig.~\ref{intro:tg}(a). For the query of $s$ to $t$ within $[2,7]$, the earliest arrival time of each vertex is shown in Fig.~\ref{tg:computing tpgraph}(a).
As we process $s$, its out-neighbors $\{(b,2),(a,3),(d,4)\}$ are checked. 
We add $b,a,d$ to ${Q}$ and update their earliest arrival time to $2,3,4$, respectively.
Next, $\ea{d}$ is updated to $3$ when $b$ is being processed.
Note that, we do not add $d$ to ${Q}$ since it is currently in ${Q}$. 
Then, $a$ is processed without changes to $\ea{d}$, since $5 > 3 = \ea{d}$.
\end{example}

\begin{theorem}\label{the:polarity}
The time complexity of Algorithm~\ref{alg:polaritytime} is $\mathcal{O}(n+m)$, and its space complexity is $\mathcal{O}(n)$.
\end{theorem}

\cblack{
\myparagraph{Discussions between \tg and \quick}
Note that, although both \tg and \quick adopt a strict temporal path constraint and achieve the same effect in terms of graph reduction, \quick reduces the time complexity of \tg by a factor of $\mathcal{O}(\log n)$.
This is because, \quick iteratively updates the polarity time for each vertex, avoiding the overhead of a priority queue used in the Dijkstra-based method \tg.
Experiments also show that \quick achieves a speedup of two orders of magnitude compared to \tg
(more details can be found in Section~\ref{sec:exp}).
}
\subsection{Tight Upper-bound Graph Generation}
\label{sec:ubg_two}
\cblack{
In this subsection, we present the tight upper-bound graph $\gt$, which further shrink 
the quick upper-bound graph $\gq$ based on the simple path constraint.
Our method employs a necessary but not sufficient condition for the inclusion of an edge in the $\stspgraph$.
Before introducing the details of this condition, we first give the analysis in the following.
Given the quick upper-bound graph $\gq$ of a directed temporal graph $\mathcal{G}$ for the query of $s$ to $t$ within $[\tau_b,\tau_e]$, following Algorithms~\ref{alg:tpgconstruct} and~\ref{alg:polaritytime}, it is straightforward to obtain Lemma~\ref{lem:tpgtotspexistamended} below.}


\begin{lemma}
\label{lem:tpgtotspexistamended}
If an edge $e(u,v,\tau) \in \mathcal{E}(\gq)$, then one of the following two conditions must be satisfied:
\begin{enumerate}
    \item[$i)$] $u \neq s$, $v \neq t$, there exist two temporal simple paths $p_{[\tau_b,\tau_i]}^*(s,u)$ and $p_{[\tau_j,\tau_e]}^*(v,t)$ s.t. $ \tau_i < \tau <\tau_j $, $t \notin p_{[\tau_b,\tau_i]}^*(s,u)$, $s \notin p_{[\tau_j,\tau_e]}^*(v,t)$.
    \item[$ii)$] $u = s$ or $v = t$, there exists a temporal simple path $p_{[\tau_b,\tau_e]}^*(s,t)$ s.t. $e(u,v,\tau) \in p_{[\tau_b,\tau_e]}^*(s,t)$.
\end{enumerate}
\end{lemma}

\cblack{According to Lemma~\ref{lem:tpgtotspexistamended}, 
given an edge $e(u,v,\tau) \in \mathcal{E}(\gq)$, with $u \neq s$ and $v \neq t$, we can always find a pair of simple paths $p_{[\tau_b,\tau_i]}^*(s,u)$ and $p_{[\tau_j,\tau_e]}^*(v,t)$.
However, those two paths may share the same vertex which prevents the formation of a simple path $p_{[\tau_b,\tau_e]}^*(s,t)$, as shown in the observation below.
}

\begin{observation}\label{obs:tspexist}
An edge $e(u,v,\tau)$, with $u \neq s$ and $v \ne t$, belongs to a temporal simple path $p_{[\tau_b,\tau_e]}^*(s,t)$ iff there exist two temporal simple paths $p_{[\tau_b,\tau_i]}^*(s,u)$ and $p_{[\tau_j,\tau_e]}^*(v,t)$ s.t. $i)$ $\tau_i < \tau <\tau_j$ and $ii)$ $\mathcal{V}(p_{[\tau_b,\tau_i]}^*(s,u)) \cap \mathcal{V}(p_{[\tau_j,\tau_e]}^*(v,t)) = \emptyset$.
\end{observation}

\cblack{
Following Observation~\ref{obs:tspexist}, to determine whether $e(u,v,\tau)$ is contained in any $p_{[\tau_b,\tau_e]}^*(s,t)$, we would need to exhaustively evaluate condition $ii)$ for each combination of $p_{[\tau_b,\tau_i]}^*(s,u)$ and $p_{[\tau_j,\tau_e]}^*(v,t)$, which is computationally expensive.
Thus, we aim to efficiently identify and remove edges that clearly violate condition $ii)$, rather than all such edges.
The high-level idea is that, if there exists a vertex $w$ appearing in all $p_{[\tau_b,\tau_i]}^*(s,u)$ and $p_{[\tau_j,\tau_e]}^*(v,t)$, condition $ii)$ will not hold.
Therefore, we will elaborate on identifying such common vertex $w$ ($w \ne s$ and $w \ne t$ as discussed in Lemma~\ref{lem:tpgtotspexistamended}), and formally define the \textit{time-stream common vertices} as follows.
}

\begin{definition}[Time-stream Common Vertices]
\label{def:common}
Given $\mathcal{G}=(\mathcal{V},\mathcal{E})$, two vertices $s$, $t$ $\in \mathcal{V}$, a time interval $[\tau_b,\tau_e]$ and a timestamp~$\tau \in [\tau_b,\tau_e]$, 
the time-stream common vertices of $\mathcal{P}^*_{[\tau_b, \tau]}(s, u)$, denoted by $\tcv{\tau}{s}{u}$, is the set of vertices (except $s$) appearing in all temporal simple paths from $s$ to $u$ not containing $t$ within time interval $[\tau_b,\tau]$,
i.e.,
\begin{equation}  
	\tcv{\tau}{s}{u} = \underset{p^* \in \mathcal{P}^*_{[\tau_b,\tau]}(s,u) ~s.t.~t \notin \mathcal{V}(p^*)}{\bigcap}\mathcal{V}(p^*)\setminus \{s\}
\end{equation}
Similarly, given $\tau \in [\tau_b,\tau_e]$, $\tcv{\tau}{u}{t}$ is the set of vertices (except $t$) appearing in all $p^*_{[\tau,\tau_e]}(u,t)$ not containing $s$,
i.e.,
\begin{equation}  
	\tcv{\tau}{u}{t}=\underset{p^* \in \mathcal{P}^*_{[\tau, \tau_e]}(u, t) ~s.t.~s \notin \mathcal{V}(p^*)}{\bigcap}\mathcal{V}(p^*)\setminus \{t\}
\end{equation}
\end{definition}


\begin{lemma}\label{lem:suf}
Given an edge $e(u,v,\tau) \in \mathcal{E}(\gq)$ with $u \neq s$, $v \neq t$, 
if it is contained in a temporal simple path $p_{[\tau_b,\tau_e]}^*(s,t)$,
then $\exists \tau_i,\tau_j \in [\tau_b,\tau_e]$ s.t. $i)$ $\tau_i<\tau<\tau_j$, and $ii)$ $\tcv{\tau_i}{s}{u} \cap \tcv{\tau_j}{v}{t} = \emptyset$; 
but not necessarily the reverse.
\end{lemma}

Lemma~\ref{lem:suf} reveals that,
\cblack{for the edge $e(u,v,\tau)$, if it satisfies the condition that $\exists$ $\tau_i,\tau_j \in [\tau_b,\tau_e]$ s.t. $\tau_i<\tau<\tau_j$, $\tcv{\tau_i}{s}{u} \cap \tcv{\tau_j}{v}{t} = \emptyset$, then it has the potential to form a $p_{[\tau_b,\tau_e]}^*(s,t)$, and we include it in $\gt$ for further verification; if not, 
we can directly conclude this edge is unpromising and exclude it from $\gt$.}
However, it is non-trivial to efficiently compute and store time-stream common vertices $\tcv{\tau}{s}{u}$ and $\tcv{\tau}{u}{t}$ for all vertices $u \in \mathcal{V}(\gq)$ and all timestamps $\tau \in [\tau_b,\tau_e]$, 
with the main challenges being two-fold.

\begin{itemize}[leftmargin=1.5em]
    \item[$i)$] Storing time-stream common vertices for all $\tau \in [\tau_b,\tau_e]$ requires significant space when the span of $[\tau_b,\tau_e]$ is large.
    \item[$ii)$] Explicitly enumerating temporal simple paths to obtain time-stream common vertices is computation-intensive.
\end{itemize}



{To address the first challenge,} for each vertex $u \in \mathcal{V}(\gq) \setminus \{s,t\}$, instead of computing and storing time-stream common vertices $\tcv{\tau}{s}{u}$ (resp. $\tcv{\tau}{u}{t}$) for every possible timestamp $\tau \in [\tau_b,\tau_e]$, we can compute and store those for only a finite subset of timestamps.
Time-stream common vertices for other timestamps can be inferred from stored information.
This idea is supported by the following lemmas.


\begin{lemma}\label{lem:tpgraphvertextopath}
For each in-coming edge $e(v,u,\tau)$ (resp. out-going edge $e(u,v,\tau)$) of $u$ in $\gq$, there exists a temporal simple path $p_{[\tau_b,\tau]}^*(s,u)$ (resp. $p_{[\tau,\tau_e]}^*(u,t)$) such that $e(v,u,\tau) \in \mathcal{E}(p_{[\tau_b,\tau]}^*(s,u))$ (resp. $e(u,v,\tau) \in \mathcal{E}(p_{[\tau,\tau_e]}^*(u,t))$) and $t \notin \mathcal{V}(p_{[\tau_b,\tau]}^*(s,u))$ (resp. $s \notin \mathcal{V}(p_{[\tau,\tau_e]}^*(u,t))$).
\end{lemma}

\begin{lemma}\label{lem:timestamps}
Given $\tau$ and $\tau_{l}=\max\{\tau_{i} | \tau_{i} \in \mathcal{T}_{in}(u,\gq), \tau_{i} \leq \tau\}$, we have $\tcv{\tau}{s}{u}=\tcv{\tau_{l}}{s}{u}$. Similarly, given $\tau_{r}=\min\{\tau_{j} | \tau_{j} \in \mathcal{T}_{out}(u,\gq), \tau_{j} \geq \tau\}$,  $\tcv{\tau}{u}{t}=\tcv{\tau_{r}}{u}{t}$.
\end{lemma}

Recall that, $\mathcal{T}_{in}(u,\gq)$ (resp. $\mathcal{T}_{out}(u,\gq)$) represents all distinct timestamps in $N_{in}(u,\gq)$ (resp. $N_{out}(u,\gq)$). 
\cblack{We only need to compute and store $\tcv{\tau}{s}{u}$ (resp. $\tcv{\tau}{u}{t}$) for every timestamp in $\mathcal{T}_{in}(u,\gq)$ (resp. $\mathcal{T}_{out}(u,\gq)$) based on Lemmas~\ref{lem:tpgraphvertextopath} and~\ref{lem:timestamps}, and then organize them in ascending (resp. descending) order based on $\tau$}.
Note that, we refer to the stored time-stream common vertices for a timestamp as an \textit{entry}.


\begin{figure}[t]
    \centering
    \begin{subfigure}{0.14\textwidth}
        \includegraphics[width=\textwidth]{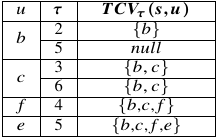}
        \caption{$TCV_\tau(s,u)$}
        \label{ac:TCVs}
    \end{subfigure}
    \begin{subfigure}{0.14\textwidth}
        \includegraphics[width=\textwidth]{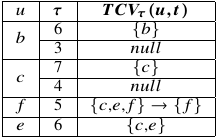}
        \caption{$TCV_\tau(u,t)$}
        \label{ac:TCVt}
    \end{subfigure}
    \begin{subfigure}{0.16\textwidth}
        \includegraphics[width=\textwidth]{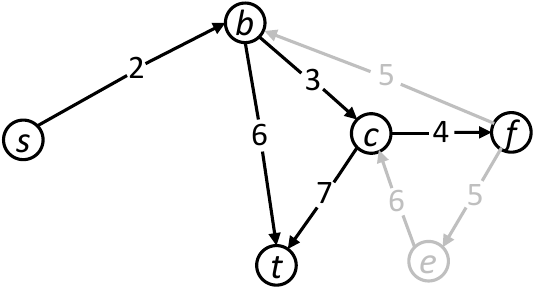}
        \caption{$\gt$}
        \label{ac:e_pru}
    \end{subfigure}
    \caption{Tight Upper-bound Graph Generation}
    \label{ac:computing acgraph}
\end{figure}


\begin{example}
Following Example~\ref{example:gq} and given the quick upper-bound graph $\gq$ shown in Fig.~\ref{tg:computing tpgraph}(c), Figs.~\ref{ac:computing acgraph}(a)-\ref{ac:computing acgraph}(b) show the stored $\tcv{\cdot}{s}{\cdot}$ and $\tcv{\cdot}{\cdot}{t}$ for all vertices in~$\gq$. 
 Take $\tcv{\cdot}{f}{t}$ as an example, 
 $N_{out}(f, \gq) = \{(b,5), (e,5)\}$,
 therefore $\mathcal{T}_{out}(f, \gq) = \{5\}$.
  As depicted in Fig.~\ref{ac:computing acgraph}(b), there is only one entry in $\tcv{\cdot}{f}{t}$, which is $\tcv{5}{f}{t}$.

\end{example}

Regarding the second challenge, according to Lemma~\ref{lem:tpgraphvertextopath}, a temporal simple path $p_{[\tau_b,\tau]}^*(s,u)$ is composed of two parts, $\langle e(v,u,\tau') \rangle$ and $p_{[\tau_b,\tau'-1]}^*(s,v)$, where $(v,\tau') \in N_{in}(u,\gq)$, $\tau' \leq \tau$ and $u \notin \mathcal{V}(p_{[\tau_b,\tau'-1]}^*(s,v))$, this inspires us to compute time-stream common vertices of a vertex based on those of its neighbors.
However, the main obstacle is the need to check for the simple path constraint during the computation.
To overcome this obstacle, we propose the following lemma.

\begin{lemma}\label{lem:tcvpath}
Intersecting the vertex set of each temporal simple path is equivalent to intersecting that of each temporal path, \ie
$\tcv{\tau}{s}{u}$ = $\bigcap_{p \in \mathcal{P}_{[\tau_b,\tau]}(s,u) ~s.t.~t \notin \mathcal{V}(p)}\mathcal{V}(p) \setminus \{s\}$; similarly, $\tcv{\tau}{u}{t}$ = $\bigcap_{p \in \mathcal{P}_{[\tau,\tau_e]}(u,t) ~s.t.~s \notin \mathcal{V}(p)}\mathcal{V}(p) \setminus \{t\}$.
\end{lemma}

According to Lemma~\ref{lem:tcvpath}, we can compute $\tcv{\tau}{s}{u}$ (resp. $\tcv{\tau}{u}{t}$) based on the temporal paths $\mathcal{P}_{[\tau_b,\tau]}(s,u)$ (resp. $\mathcal{P}_{[\tau,\tau_e]}(u,t)$) without considering the simple path constraint.
Given a vertex $w \neq s$, $w \in \tcv{\tau}{s}{u}$ iff $\forall$ $(v,\tau') \in N_{in}(u,\gq)$ s.t. $\tau' \leq \tau$ and $w \in (\tcv{\tau'-1}{s}{v} \cup \{u\})$.
Therefore, intersecting $(\tcv{\tau'-1}{s}{v} \cup \{u\})$ for each $(v,\tau') \in N_{in}(u,\gq)$ s.t. $\tau' \leq \tau$ leads to $\tcv{\tau}{s}{u}$.
Based on these, we design the computation of $\tcv{\cdot}{s}{\cdot}$ and $\tcv{\cdot}{\cdot}{t}$ in a recursive way with the formulas formally defined below.

\myparagraph{Recursive Computation}
Given the source vertex and target vertex $s,t \in \mathcal{V}(\gq)$, the time interval $[\tau_b,\tau_e]$, for each vertex $u \in \mathcal{V}(\gq) \setminus \{s,t\}$ and timestamp $\tau \in [\tau_b,\tau_e]$, 
$\tcv{\tau}{s}{u}$ is computed based on the time-stream common vertices of $u$'s in-neighbors.
For ease of computation, we define the base case to be:
for $\forall \tau \in [\tau_b-1,\tau_e-1]$, $\tcv{\tau}{s}{s} = \emptyset$.

\begin{equation}\label{equ:tcvs}
\tcv{\tau}{s}{u} = \bigcap_{\substack{
  (v,\tau') \in N_{in}(u,\gq) \\
  \tau'\leq \tau
}} (\tcv{\tau'-1}{s}{v} \cup \{u\})
\end{equation}

Similarly, 
$\tcv{\tau}{u}{t}$ is computed based on the time-stream common vertices of $u$'s out-neighbors.
And we define the base case to be:
for $\forall \tau \in [\tau_b+1,\tau_e+1]$, $\tcv{\tau}{t}{t} = \emptyset$.

\begin{equation}\label{equ:tcvt}
\tcv{\tau}{u}{t} = \bigcap_{\substack{
  (v,\tau') \in N_{out}(u,\gq) \\
  \tau'\geq \tau
}} (\tcv{\tau'+1}{v}{t} \cup \{u\})
\end{equation}

The computation of $\tcv{\tau}{s}{u}$ can be interpreted as two steps: 
$i)$ intersect $(\tcv{\tau'-1}{s}{v}\cup \{u\})$ for all $(v,\tau') \in N_{in}(u,\gq) ~$s.t.$~\tau' \leq \tau-1$;
$ii)$ intersect $(\tcv{\tau-1}{s}{v}\cup \{u\})$ for all $(v,\tau) \in N_{in}(u,\gq)$. 
The computation of $\tcv{\tau}{u}{t}$ can be interpreted in a similar manner, and we omit the details here.
\cblack{Clearly, the result of step $i)$ is $\tcv{\tau-1}{s}{u}$ if $\tcv{\tau-1}{s}{u}$ exists, therefore,
$\tcv{\tau}{s}{u} \subseteq \tcv{\tau-1}{s}{u}$, which allows the computation of $\tcv{\tau}{s}{u}$ to inherit from $\tcv{\tau-1}{s}{u}$}
and we also have the following lemma that serves as a pruning rule in the computation of time-stream common vertices.

\begin{lemma}\label{lem:prune}
If $\tcv{\tau_m}{s}{u}=\{u\}$, then $\forall$ $\tau_n>\tau_m$, $\tcv{\tau_n}{s}{u}$ $=\{u\}$. Similarly, if $\tcv{\tau_m}{u}{t}=\{u\}$, then $\forall$ $\tau_n<\tau_m$, $\tcv{\tau_n}{u}{t}=\{u\}$.
\end{lemma}

\begin{algorithm}[t]  
{
    \SetVline
    \caption{Time-stream Common Vertices Computation}\label{alg:time-stream}
    \Input{quick upper-bound graph $\gq$ of a temporal graph $\mathcal{G}=(\mathcal{V}, \mathcal{E})$, a source vertex $s \in \mathcal{V}$, a target vertex $t \in \mathcal{V}$ and a time interval $[\tau_b,\tau_e]$}
    \Output{the time-stream common vertices $\tcv{\cdot}{s}{\cdot}$ and $\tcv{\cdot}{\cdot}{t}$}


    \State{$\tcv{\tau_b-1}{s}{s} \gets \emptyset$, $P(s) \gets 1$}
    \State{$\tcv{\tau_e+1}{t}{t} \gets \emptyset$, $P(t) \gets 1$}
 
    \For{$u \in \mathcal{V}(\gq) \setminus \{s,t\}$}{
        \State{$\tcv{\tau}{s}{u} \gets null$ for each $\tau \in \mathcal{T}_{in}(u,\gq)$}
        \State{mark $u$ as uncompleted}
        \State{$P(u) \gets 1$}
    }
    \For{$e(v,u,\tau) \in \mathcal{E}(\gq)$}{
        \lIf{$u=t \lor u$ is completed}{\State{\textbf{continue}}}
        \State{$i \gets P(u)$, $\tau_i \gets$ $i$-th timestamp in $\tcv{\cdot}{s}{u}$}
        \State{$\tcv{\tau_i}{s}{u} \gets$ $i$-th entry in $\tcv{\cdot}{s}{u}$}

        \State{$j \gets P(v)$, $\tau_j \gets$ $j$-th timestamp in $\tcv{\cdot}{s}{v}$}

        \If{$\tau_j=\tau$}{
            \State{$j \gets j-1$, $\tau_j \gets$ $j$-th timestamp in $\tcv{\cdot}{s}{v}$}
        }

        \State{$\tcv{\tau_j}{s}{v} \gets$ $j$-th entry in $\tcv{\cdot}{s}{v}$}

        \lIf{$\tcv{\tau_j}{s}{v}=null$}{\State{$\tcv{\tau_j}{s}{v} \gets \{v\}$}}

        \If{$\tcv{\tau_i}{s}{u}=null$}{
            \State{$\tcv{\tau}{s}{u} \gets \tcv{\tau_j}{s}{v} \cup \{u\}$}
        }
        \Else{
            \State{$\tcv{\tau}{s}{u} \gets \tcv{\tau_i}{s}{u} \cap (\tcv{\tau_j}{s}{v} \cup \{u\})$}
            \lIf{$\tau > \tau_i$}{\State{$i \gets i+1$, $P(u) \gets i$}}
        }
        \State{$i$-th entry in $\tcv{\cdot}{s}{u} \gets \tcv{\tau}{s}{u}$}
        \If{$\tcv{\tau}{s}{u}=\{u\}$}{
            \StateCmt{mark $u$ as completed}{Lemma~\ref{lem:prune}}
        }
        
    }
    \State{repeat Lines 3-23 with required adjustment for $\tcv{\cdot}{\cdot}{t}$}
    \State{\Return{$\tcv{\cdot}{s}{\cdot}$, $\tcv{\cdot}{\cdot}{t}$}}
}
\end{algorithm}

\myparagraph{\underline{Time-stream Common Vertices Computation}} The detailed pseudocode is shown in Algorithm~\ref{alg:time-stream}. 
Lines 3-23 compute $\tcv{\tau}{s}{u}$ for each vertex $u$ $\in \mathcal{V}(\gq) \setminus \{s,t\}$ where $\tau$ $\in \mathcal{T}_{in}(u,\gq)$. 
All entries in $\tcv{\cdot}{s}{u}$ are initially set to $null$ in Line 4.
A pointer $P(u)$ in $\tcv{\cdot}{s}{u}$ is maintained for each vertex $u$ to enable constant-time access to the entry currently being processed. $P(u)$ is initialized to $1$ in Line 6,
indicating the $first$ entry (alternatively, the $first$ timestamp) in $\tcv{\cdot}{s}{u}$ is initially in processing.

Edges in $\mathcal{E}(\gq)$ are arranged in non-descending temporal order when $\gq$ is generated, and 
Line 7 scans them forward to compute $\tcv{\cdot}{s}{\cdot}$. 
For each scanned edge $e(v,u,\tau)$, 
we process $\tcv{\tau}{s}{u}$ accordingly, 
\cblack{its iterative intersection computation will be completed when all in-coming edges of $u$ with timestamps no greater than $\tau$ have been scanned.}
This ensures the processing of $\tcv{\cdot}{s}{\cdot}$ with a smaller timestamp is finalized before the processing of $\tcv{\cdot}{s}{\cdot}$ with a larger timestamp begins.
\cblack{For a scanned edge $e(v,u,\tau)$, Lines 11-13 obtain $\tau_j$ as either the currently processed timestamp or the immediately preceding timestamp
in $\tcv{\cdot}{s}{v}$, whichever is closer to but smaller than $\tau$}.
Then, $\tcv{\tau_j}{s}{v}$ is accessed in Lines 14-15 through the pointer corresponding to $\tau_j$ in $\tcv{\cdot}{s}{v}$, and $\tcv{\tau_j}{s}{v}$ is equal to $\tcv{\tau-1}{s}{v}$ according to Lemma~\ref{lem:timestamps}.
\cblack{$\tcv{\tau_i}{s}{u}=null$ in Line 16 suggests  
we are at the beginning of the process for the $first$ entry in $\tcv{\cdot}{s}{u}$ and we compute this entry for the first time in Line 17.
} 
\cblack{If $\tcv{\tau_i}{s}{u} \neq null$, we are facing two circumstances: 
$i)$ when $\tau=\tau_i$, indicating the $i$-th entry in $\tcv{\cdot}{s}{u}$ has been computed using other in-coming edges of $u$ with the same timestamp $\tau$, and we continue the iterative intersection process for the $i$-th entry;
$ii)$ when $\tau>\tau_i$, it indicates that all in-coming edges of $u$ with the timestamp $\tau_i$ has been scanned, we can conclude the $i$-th entry in $\tcv{\cdot}{s}{u}$ is finalized and start the processing of the $(i+1)$-th entry.
}
In both cases, 
Line 19 fulfills the purpose, 
and $\tcv{\tau}{s}{u}$ is efficiently stored at its position following the pointer $P(u)$ in Line 21.

Lines 22-23 and Line 8 implement the pruning strategy based on Lemma~\ref{lem:prune}, specifically, if $\tcv{\tau}{s}{u}=\{u\}$, any entry in $\tcv{\cdot}{s}{u}$ with a larger timestamp than $\tau$ will be $\{u\}$, thus, we can omit the further computation of $\tcv{\cdot}{s}{u}$ and mark $u$ as completed. 
As a result, in Line 15, since $\tau_j<\tau$, $\tcv{\tau_j}{s}{v}$ should already been finalized and $\tcv{\tau_j}{s}{v}=null$ if and only if $v$ was completed previously, therefore we can conclude $\tcv{\tau_j}{s}{v}=\{v\}$.

\cblack{We compute $\tcv{\cdot}{\cdot}{t}$ by repeating the process in Lines 3-23 but scanning edges backward in Line 7 and toggling between $s$ and $t$, ``in'' and ``out''.}

\begin{example}
Consider $\gq$ in Fig.~\ref{tg:computing tpgraph}(c), the time-stream common vertices $\tcv{\cdot}{s}{\cdot}$ and $\tcv{\cdot}{\cdot}{t}$ are shown in Figs.~\ref{ac:computing acgraph}(a)-\ref{ac:computing acgraph}(b).
Take the computation of $\tcv{5}{f}{t}$ as an example.
\cblack{When $e(f,e,5)$ is scanned, we have $i=1$, $\tau_i=5$, $\tau_j=6$.}
Since $\tcv{5}{f}{t}=null$, we are starting to process the first entry in $\tcv{\cdot}{f}{t}$ and we obtain $\tcv{5}{f}{t}=\tcv{6}{e}{t} \cup \{f\}=\{c,e\} \cup \{f\}=\{c,e,f\}$ in Line 17.
\cblack{After that, we process $e(f,b,5)$ with $i=1$, $\tau_i=5$, $\tau_j=6$.
Since $\tcv{5}{f}{t} \ne null$ and $\tau=5=\tau_i$, we continue the iterative intersection process for the first entry in $\tcv{5}{\cdot}{t}$ and update $\tcv{5}{f}{t}$ to $\{c,e,f\} \cap (\tcv{6}{b}{t} \cup \{f\}) = \{f\}$.}
Then, the pruning condition in Line 22 is satisfied, and we mark $f$ as completed.
\end{example}

\begin{theorem}\label{the:time-stream}
The time complexity of Algorithm~\ref{alg:time-stream} is 
$\mathcal{O}(n + \theta \cdot m)$ where $\theta$ is the span of $[\tau_b,\tau_e]$, i.e., $\theta=\tau_e-\tau_b+1$. 
The space complexity of Algorithm~\ref{alg:time-stream} is $\mathcal{O}(n + \theta \cdot m)$.
\end{theorem}

With the precomputed time-stream common vertices, we can exclude unpromising edges from $\gq$ to generate $\gt$ following the contrapositive of Lemma~\ref{lem:suf}, \ie for each $e(u,v,\tau) \in \mathcal{E}(\gq)$, check whether $\forall$ $\tau_i,\tau_j \in [\tau_b,\tau_e]$ s.t. $\tau_i<\tau<\tau_j$, $\tcv{\tau_i}{s}{u} \cap \tcv{\tau_j}{v}{t} \ne \emptyset$.
During the process, the following two optimization techniques can be used to accelerate computations.
First, based on Lemma~\ref{lem:tpgtotspexistamended}, all the out-going edges from $s$ and in-coming edges to $t$ can be directly added to $\gt$.
Second, instead of exhausting all possible combinations of $\tau_i$ and $\tau_j$ and intersecting the corresponding time-stream common vertices, we focus on the maximum timestamp $\tau_i$ in $\mathcal{T}_{in}(u,\gq)$ and minimum timestamp $\tau_j$ in $\mathcal{T}_{out}(v,\gq)$.


\begin{lemma}\label{lem:certain_timestamp}
Given an edge $e(u,v,\tau) \in \mathcal{E}(\gq)$, $u \neq s$, $v \neq t$, let 
$\tau_l=\max\{\tau_i|\tau_i \in \mathcal{T}_{in}(u,\gq) \land \tau_b \leq \tau_i <\tau\}$ and $\tau_r=\min\{\tau_j|\tau_j \in \mathcal{T}_{out}(v,\gq) \land \tau < \tau_j \leq \tau_e\}$, if $\tcv{\tau_l}{s}{u} \cap \tcv{\tau_r}{v}{t} \ne \emptyset$, then 
for all $\tau_b \leq \tau_i<\tau_l$ and $\tau_r < \tau_j \leq \tau_e$, $\tcv{\tau_i}{s}{u} \cap \tcv{\tau_j}{v}{t} \ne \emptyset$.
\end{lemma}

\begin{algorithm}[t]  
{
    \SetVline
    \caption{Tight Upper-bound Graph Generation}\label{alg:ipgconstruct}
    \Input{quick upper-bound graph $\gq$=$(\mathcal{V}_q,\mathcal{E}_q)$, the time-stream common vertices $\tcv{\cdot}{s}{\cdot}$ and $\tcv{\cdot}{\cdot}{t}$}
    \Output{tight upper-bound graph $\gt$=$(\mathcal{V}_t,\mathcal{E}_t)$}
    \State{$\mathcal{V}_t=\emptyset,\mathcal{E}_t=\emptyset$}
    \State{$P_{s}(u) \gets 1$, $P_{t}(u) \gets |\mathcal{T}_{out}(u,\gq)|$ \textbf{for each} $u \in \mathcal{V}_q$}
        
    \For{$e(u, v, \tau) \in \mathcal{E}(\gq)$}{
        \If{$u = s$ or $v = t$}{
            \State{$\mathcal{V}_t \gets \mathcal{V}_t \cup \{u,v\}$, $\mathcal{E}_t \gets \mathcal{E}_t \cup \{e(u,v,\tau)\}$}
            \StateCmt{\textbf{continue}}{Lemma~\ref{lem:tpgtotspexistamended}}
        }
        \State{$i \gets P_s(u)$, $j \gets P_t(v)$}
        \State{\textbf{while} $i$-th timestamp in $\tcv{\cdot}{s}{u} < \tau$ \textbf{do} $i \gets i+1$}
        \State{$i \gets i-1$, $\tau_i \gets$ $i$-th timestamp in $\tcv{\cdot}{s}{u}$}

        \State{\textbf{while} $j$-th timestamp in $\tcv{\cdot}{v}{t} \leq \tau$ \textbf{do} $j \gets j-1$}
        \State{$\tau_j \gets$ $j$-th timestamp in $\tcv{\cdot}{v}{t}$}

        \State{$P_s(u) \gets i$, $P_t(v) \gets j$}

        \State{$\tcv{\tau_i}{s}{u} \gets$ $i$-th entry in $\tcv{\cdot}{s}{u}$}
        \lIf{$\tcv{\tau_i}{s}{u}=null$}{\State{$\tcv{\tau_i}{s}{u}=\{u\}$}}

        \State{$\tcv{\tau_j}{v}{t} \gets$ $j$-th entry in $\tcv{\cdot}{v}{t}$}
        \lIf{$\tcv{\tau_j}{v}{t}=null$}{\State{$\tcv{\tau_j}{v}{t}=\{v\}$}}

        \If{$\tcv{\tau_i}{s}{u} \cap \tcv{\tau_j}{v}{t}=\emptyset$}{
            \State{$\mathcal{V}_t \gets \mathcal{V}_t \cup \{u,v\}$}
            \StateCmt{$\mathcal{E}_t \gets \mathcal{E}_t \cup \{e(u,v,\tau)\}$}{{Lemma~\ref{lem:suf}}}
        }

    }
    \State{\Return{$(\mathcal{V}_t,\mathcal{E}_t)$}}
}
\end{algorithm}

With Lemma~\ref{lem:certain_timestamp}, for each $e(u,v,\tau) \in \mathcal{E}(\gq)$ with $u \neq s$, $v \neq t$, we only need to perform the intersection operation once to determine if it should be included in $\gt$.
The detailed process of generating $\gt$ is illustrated in Algorithm~\ref{alg:ipgconstruct}.
Note that, edges are processed in a non-descending temporal order, and we utilize pointers in $\tcv{\cdot}{s}{\cdot}$ and $\tcv{\cdot}{\cdot}{t}$ to achieve constant-time access to entries in a similar way as Algorithm~\ref{alg:time-stream}.
A running example of the process is provided as follows.


\begin{example}
Given $\gq$ in Fig.~\ref{tg:computing tpgraph}(c), the precomputed $\tcv{\cdot}{s}{\cdot}$ and $\tcv{\cdot}{\cdot}{t}$ in Figs.~\ref{ac:computing acgraph}(a)-\ref{ac:computing acgraph}(b). The resulting tight upper-bound graph $\gt$ is shown in Fig.~\ref{ac:computing acgraph}(c).
\cblack{Take $e(c,f,4)$ as an example, it is contained in $\gt$ as $\tcv{3}{s}{c} \cap \tcv{5}{f}{t} = \emptyset$}.

\end{example}


\begin{lemma}\label{the:ipgcorrectness}
An edge $e(u,v,\tau) \in \mathcal{E}(\gq)$ is in $\mathcal{E}(\gt)$ iff one of the following two conditions is satisfied:
\begin{enumerate}
    \item[$i)$] $u \neq s$, $v \neq t$, $\tcv{\tau_l}{s}{u} \cap \tcv{\tau_r}{v}{t} = \emptyset$, where $\tau_l=\max\{\tau_i|\tau_i \in \mathcal{T}_{in}(u,\gq) \land \tau_b \leq \tau_i <\tau\}$ and $\tau_r=\min\{\tau_j|\tau_j \in \mathcal{T}_{out}(v,\gq) \land \tau < \tau_j \leq \tau_e\}$.
    \item[$ii)$] $u = s$ or $v = t$.
\end{enumerate}
\end{lemma}


\begin{theorem}\label{the:ipgconstruct}
The time complexity of Algorithm~\ref{alg:ipgconstruct} is $\mathcal{O}(n + \theta \cdot m)$, and its space complexity is $\mathcal{O}(n+m)$.
\end{theorem}


\section{Escaped Edges Verification}
\label{sec:ver}



In this section, we propose an Escaped Edges Verification (\eev) method \cblack{which strictly imposes the simple path constraint on $\gt$ and generates the exact $\stspgraph$}.
The main idea of \eev is that we iteratively select an unverified edge $e \in \gt$ and identify a $p_{[\tau_b,\tau_e]}^*(s,t)$ through $e$ by applying the bidirectional DFS.
All the edges along this path belong to $\stspgraph$, and we add them to the result.
The process stops until all the edges in $\gt$ have been verified. 
Before discussing the details of our method, we first introduce the following ideas.

\begin{lemma}\label{lem:2hopedges}
    Given $e(u,v,\tau) \in \mathcal{E}(\gt)$ with $u \neq s$, $v \neq t$, if $i)$ $\exists$ $e(s, u, \tau') \in \mathcal{E}(\gt)$ s.t. $\tau_b \leq \tau' < \tau$, or $ii)$ $\exists$ $e(v, t, \tau') \in \mathcal{E}(\gt)$ s.t. $\tau < \tau' \leq \tau_e$,
    then there is a temporal simple path $p_{[\tau_b,\tau_e]}^*(s,t)$ containing the edge $e(u,v,\tau)$.
\end{lemma}

By applying Lemmas~\ref{lem:tpgtotspexistamended} and~\ref{lem:2hopedges}, we can confirm certain edges belong to $\stspgraph$ without searching for the exact paths containing them, which reduces the number of edges to verify. 
Moreover, we can accelerate the edge verification process by adding edges across a batch of paths simultaneously based on an identified temporal simple path using the following lemma.



\begin{lemma}\label{lem:batchver}
    Given a temporal simple path $p^* = $
    $\langle e(u_0,u_1,\tau_1)$ $,\dots,e(u_{l-1},u_l,\tau_l) \rangle$,
    for all integers $1 \leq i \leq l$, $e(u_{i-1},u_i,\tau_i) \in \mathcal{E}(p^*)$ can be replaced by any $e(u_{i-1},u_i,\tau)$ s.t. $i)$ if $i=1$, $\tau \in [\tau_b, \tau_{i+1})$, $ii)$ if $1 < i < l-1$, $\tau \in (\tau_{i-1}, \tau_{i+1})$, $iii)$ if $i=l$, $\tau \in (\tau_i-1, \tau_b]$
   , and the resulting path remains a temporal simple path from $s$ to $t$ within the time interval $[\tau_b,\tau_e]$.
\end{lemma}

Following Lemma~\ref{lem:batchver}, when a path $p_{[\tau_b,\tau_e]}^*(s,t)$ is identified during the bidirectional DFS, 
in addition to all the edges on this path itself, edges that are able to replace each of them to form another temporal simple path from $s$ to $t$ within $[\tau_b,\tau_e]$ can be confirmed at the same time. 
Based on the above analysis,
we give the pseudocode of our Escaped Edges Verification method for producing exact $\stspgraph$ in Algorithm~\ref{alg:tspgconstruct}.

\begin{algorithm}[t]  
{
    \SetVline
    \caption{Escaped Edges Verification}\label{alg:tspgconstruct}
    \Input{tight upper-bound graph $\gt$ of a temporal graph $\mathcal{G}=(\mathcal{V}, \mathcal{E})$, a source vertex $s \in \mathcal{V}$, a target vertex $t \in \mathcal{V}$ and a time interval $[\tau_b,\tau_e]$}
    \Output{temporal simple path graph $\stspgraph=(\mathcal{V^*},\mathcal{E^*})$}
    
    \State{$\mathcal{V^*}=\emptyset,\mathcal{E^*}=\emptyset$}
    
    \State{\textbf{for each} $e \in \mathcal{E}(\gt)$ mark $e$ as unverified}
    
    
    \ForEach{$e(u, v, \tau)$ satisfies Lemma~\ref{lem:tpgtotspexistamended} or~\ref{lem:2hopedges}}{
        \State{mark $e(u, v, \tau)$ as verified}
        \State{$\mathcal{V^*} \gets \mathcal{V^*} \cup \{u, v\}$; $\mathcal{E^*} \gets \mathcal{E^*} \cup \{e(u, v, \tau)\}$}
    }
    
    \ForEach{$e(u, v, \tau) \in \mathcal{E}(\gt)$}{
        \lIf{$e(u, v, \tau)$ is verified}{\State{\textbf{continue}}}
        \State{$S_v \gets$ an empty stack, $S_e \gets$ an empty stack}
        \lIf{$\neg BiDirSearch(e(u, v, \tau))$}{\State{\textbf{continue}}}
        \State{build path $p^*$ from $S_e$}
        \For{$3 \leq i \leq l(p^*)-2$}{
        \State{$e(u_{i-2},u_{i-1},\tau_{i-1}) \gets$ the ($i$-1)-th edge in $p^*$}
        \State{$e(u_{i-1},u_{i},\tau_{i}) \gets$ the $i$-th edge in $p^*$}
        \State{$e(u_{i},u_{i+1},\tau_{i+1}) \gets$ the ($i$+1)-th edge in $p^*$}
        \State{$\mathcal{V^*} \gets \mathcal{V^*} \cup \{u_{i-1}, u_{i}\}$}
        \ForEach{$(w_j,\tau_j) \in N_{out}(u_{i-1}, \gt)$}{
        \If{$w_j = u_{i}$ and $\tau_j \in (\tau_{i-1},\tau_{i+1})$}{
        \State{mark $e(u_{i-1}, u_{i}, \tau_j)$ as verified}
        \StateCmt{$\mathcal{E^*} \gets \mathcal{E^*} \cup \{e(u_{i-1}, u_{i}, \tau_j)\}$}
        {Lemma~\ref{lem:batchver}}
        }
        }
        }
    }
    \State{\Return{$(\mathcal{V^*},\mathcal{E^*})$}}
}
\end{algorithm}

\myparagraph{\underline{Escaped Edges Verification}}
In Algorithm~\ref{alg:tspgconstruct},
Lines 2-5 initialize the verification status for each edge in $\gt$. For each 
out-neighbor $(u, \tau')$ of $s$ (resp. in-neighbor of $t$), we mark the edge $e(s, u, \tau')$ (resp. $e(u, t, \tau')$) as verified and add it to $\stspgraph$ based on Lemma~\ref{lem:tpgtotspexistamended}; we also check the out-neighbors (resp. in-neighbors) of $u$, if a neighbor $(v, \tau)$ with a timestamp $\tau > \tau'$ (resp. $\tau < \tau'$) is found, we mark $e(u, v, \tau)$ (resp. $e(v, u, \tau)$) as verified and add it to $\stspgraph$ according to Lemma~\ref{lem:2hopedges}.
In $\gt$, edges are stored in a non-descending temporal order, Lines 6-7 iterate through those edges and select an unverified one to start a bidirectional DFS for a temporal simple path. 
From Line 8 to Line 10, global stacks $S_v$, $S_e$ are maintained throughout the bidirectional DFS. 
When a temporal simple path is found, we can build it from $S_e$.
The edge confirmation process is illustrated in Lines 11-19. For each edge in this identified path, we confirm it together with all edges that can replace it to form another temporal simple path, add those edges to $\stspgraph$, and mark them as verified. Note that, we omit the confirmation for $i$-th edge in the path when $i=1,2,l(p^*)-1, l(p^*)$ since it clearly satisfies Lemma~\ref{lem:tpgtotspexistamended} or~\ref{lem:2hopedges} and has been verified in~Line~4.

\myparagraph{Optimized Bidirectional DFS}
To further speed up the path identification process, we propose two optimization techniques utilized in our implementation of bidirectional DFS. 



\myparagraph{$i)$ Prioritization of Search Direction}
When an unverified edge $e(u, v, \tau)$ is selected, the bidirectional DFS for identifying a path through $e(u, v, \tau)$ includes a forward search for a forward path $p_{[\tau,\tau_e]}^*(u,t)$ and a backward search for a backward path $p_{[\tau_b,\tau]}^*(s,v)$.
Assume $\tau-\tau_b > \tau_e-\tau$, 
based on Remark~\ref{rem:lengthlimit}, the backward path is potentially longer than the forward path. 
If we perform the backward search prior to the forward search, we may not find a valid forward path since the vertices have been possessed by the longer backward path, and the DFS will need to backtrack extensively due to the failure to form a temporal simple path. 
Following this idea, if $\tau-\tau_b > \tau_e-\tau$, we perform the forward search followed by the backward search; otherwise, the reverse.


\myparagraph{$ii)$ Neighbor Exploration Order}
Intuitively, a shorter temporal path has a lower probability of revisiting a vertex than a longer one.
To guide the DFS in favor of shorter paths, at each vertex, we traverse its neighbors following a temporal order.
\cblack{Specifically, in the forward search, 
we explore out-neighbors in a non-ascending temporal order;
while in the backward search, we explore in-neighbors in a non-descending temporal order}.

\cblack{\textit{Note that, due to the space limitation, the pseudocode of BiDirSearch can be found in Appendix B online~\cite{appendix}.}}

\begin{theorem}\label{the:tspgconstruct}
Algorithm~\ref{alg:tspgconstruct} needs $\mathcal{O}(m  \cdot d_t^{\theta-1})$ time, where $d_t$ is the largest degree of the vertices in $\gt$, i.e., $d_t=\max_{u\in \mathcal{V}(\gt)}\{\max(|N_{in}(u,\gt)|,|N_{out}(u,\gt)|)\}$ and 
$\theta=\tau_e-\tau_b+1$. 
Its space complexity
is $\mathcal{O}(n+m)$.
\end{theorem}

\section{Experiments}
\label{sec:exp}

\subsection{Experiment Setup}

\myparagraph{Algorithms}
To the best of our knowledge, we are the first to
investigate the temporal simple path graph generation problem.
\cblack{Therefore, we implement and evaluate the following four algorithms in our experiments.
Three baseline methods i.e., \texttt{\epdt}, \texttt{\epes}, \texttt{\eptg}, whose details can be found in Section~\ref{sec:sol_overview_a}. \texttt{\VUG:} Algorithm~\ref{alg:framework} proposed in Section~\ref{sec:framework} with all the optimized strategies.}


\myparagraph{Datasets}
We employ \cblack{10} real-world temporal graphs in our experiments. 
\cblack{Details of these datasets are summarized in TABLE I of Appendix C online~\cite{appendix}.}


\myparagraph{Parameters and workloads}
\cblack{
For each dataset, we generate 1000 random queries with different source vertices $s$, target vertices $t$, time intervals $[\tau_b,\tau_e]$ where $s$ can temporally reach $t$ within $[\tau_b,\tau_e]$ and report the total query time on each dataset.
We conduct the experiment by varying $\tau_e-\tau_b+1$ (i.e., $\theta$),
whose default values are set following existing
works about path problems in temporal graphs (e.g.,~\cite{DBLP:journals/kais/JinCL24,DBLP:journals/pvldb/GouYZY24,DBLP:journals/pvldb/LiHDZ17}).
Details are shown in the last column of TABLE I.}
For those algorithms that cannot finish within 12 hours, we set them as \textbf{INF}.
All the programs are implemented in standard C++. 
All the experiments are performed on a server with Intel(R) Xeon(R) Gold 5218R CPU @ 2.10GHz and 256G RAM.


\begin{figure}[t]
	\centering
	\includegraphics[width=0.75\linewidth]{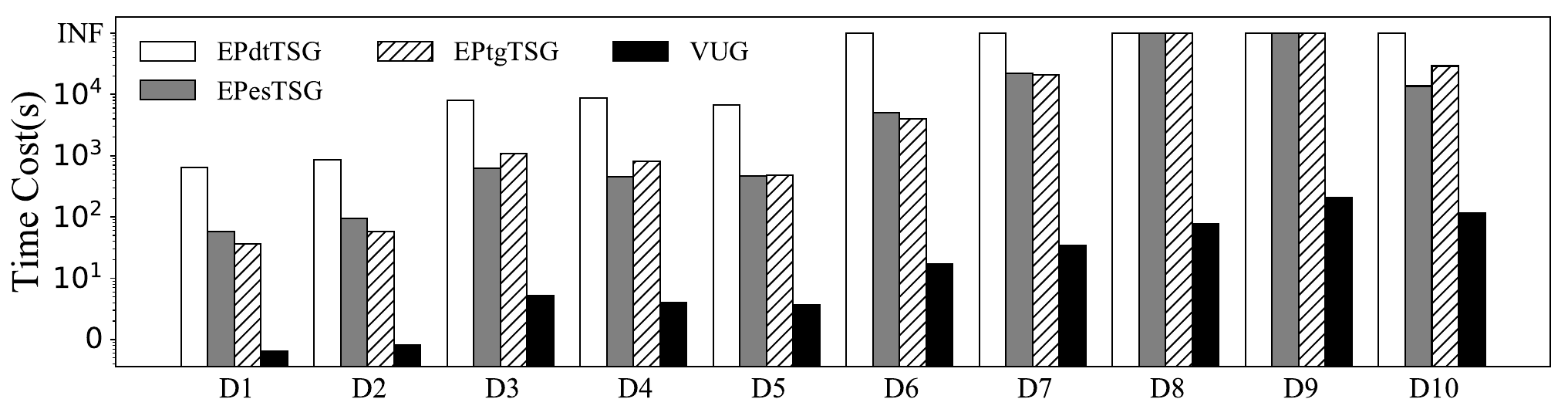}
	\caption{\cblack{Response time on all the datasets}}
	\label{fig:exp1}
\end{figure}

\begin{figure}[t]
    \centering
    \begin{subfigure}{0.18\textwidth}
        \includegraphics[width=\textwidth]{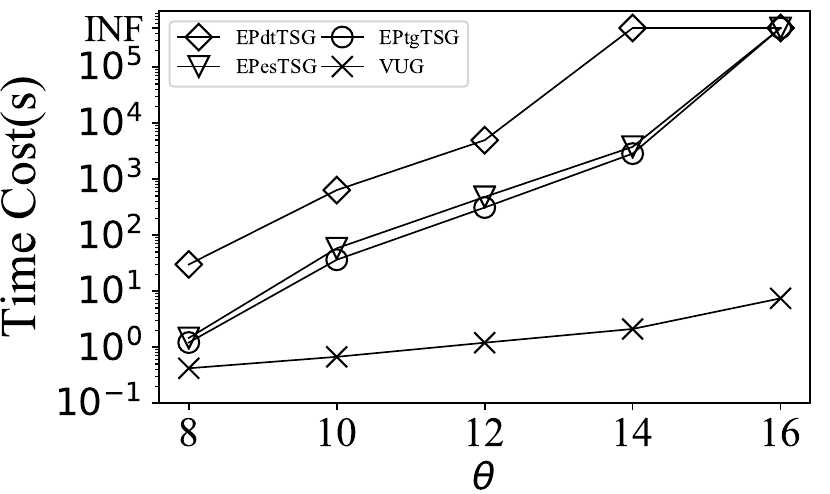}
        \caption{\cblack{D1}}
    \end{subfigure}
    \hspace{2mm}
    \begin{subfigure}{0.18\textwidth}
        \includegraphics[width=\textwidth]{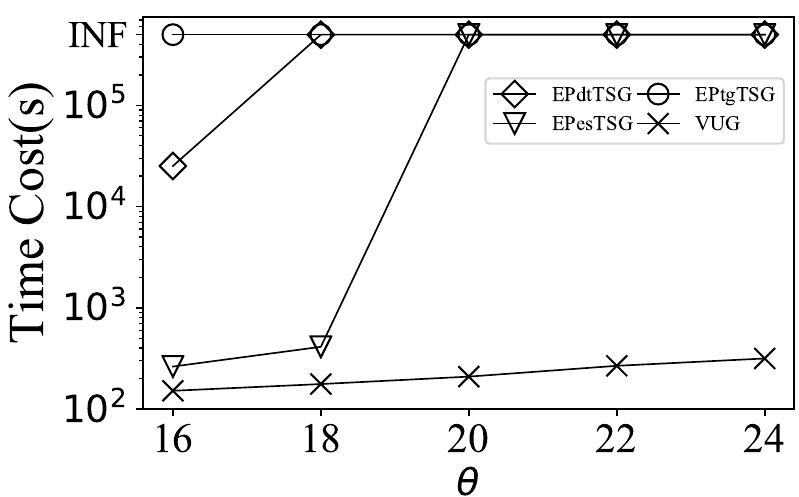}
        \caption{\cblack{D9}}
    \end{subfigure}
    \caption{\cblack{Response time by varying parameter $\theta$}}
    \label{fig:exp5}
\end{figure}

\subsection{Performance Comparison with Baselines}

\myparagraph{Exp-1: Response time on all the datasets}
\cblack{Fig.~\ref{fig:exp1}} reports the total response time of $\epdt$, $\epes$, $\eptg$ and \texttt{\VUG} for 1000 queries over all the datasets under the default settings.
As observed, $\epes$ and $\eptg$ are faster than $\epdt$ due to their optimization for tighter upper-bound graphs, 
\ie integrating the temporal path constraint.
\texttt{\VUG} outperforms the other three algorithms on all the datasets by a significant margin as a result of $i)$ it further excludes many unpromising edges by incorporating the simple path constraint when generating the tight upper-bound graph; and $ii)$ it avoids lots of verifications for the existence of edges in the final result.
\cblack{For example, on the datasets D8 and D9, three baseline methods
cannot finish queries within 12 hours due to the large search space, while \texttt{\VUG} only takes 78 seconds and 208 seconds to return the result, respectively}.

\myparagraph{Exp-2: Response time by varying parameter $\theta$}
\cblack{In Fig.~\ref{fig:exp5}, we report the total response time of $\epdt$, $\epes$, $\eptg$ and \texttt{\VUG} for 1000 queries by varying $\theta$ on the datasets D1 and D9.
Due to the space limitation, results on the other datasets can be found in Appendix C online~\cite{appendix}}.
We can see that the response time of the three baselines grows exponentially as $\theta$ increases, while the response time of \texttt{\VUG} grows modestly, demonstrating its better scalability.
For example, when $\theta$ varies from 8 to 12 on D1, the total response time of $\epdt$, $\epes$, $\eptg$ increases by a factor of 165, 320, 259, respectively, while that of \texttt{\VUG} only increases by a factor of 3.




\begin{figure}[t]
	\centering
	\includegraphics[width=0.75\linewidth]{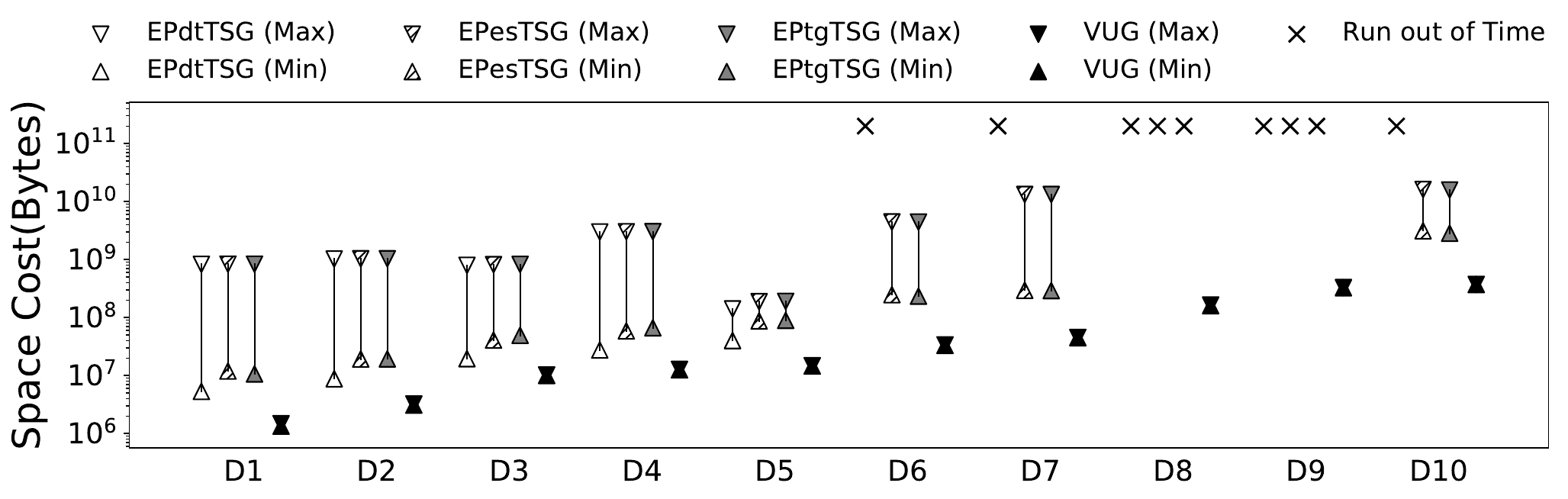}
	\caption{\cblack{Space consumption comparison}}
	\label{fig:exp2}
\end{figure}

\begin{figure}[t]
	\centering
	\includegraphics[width=0.75\linewidth]{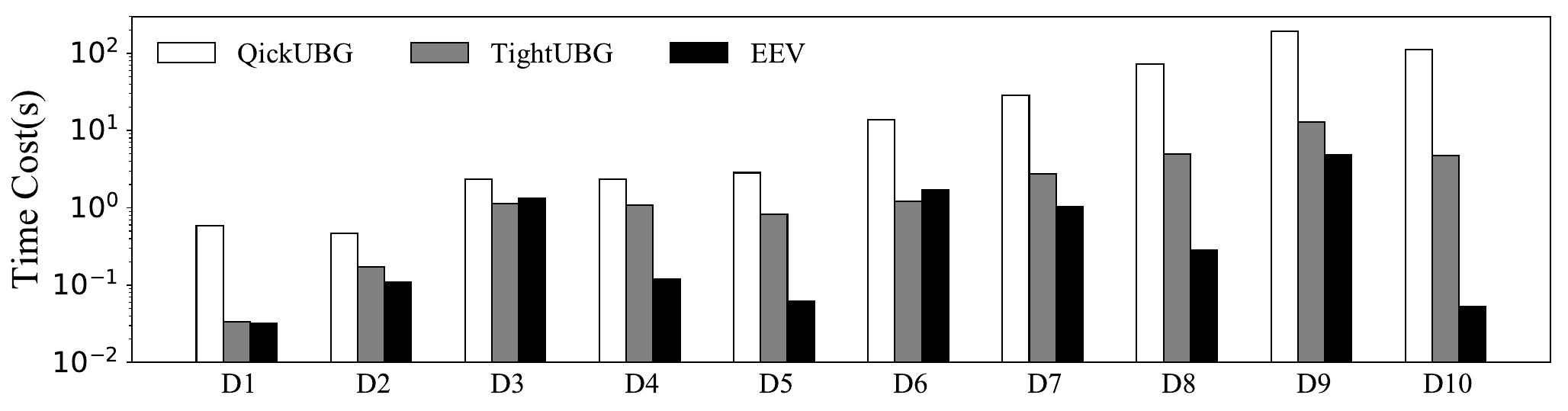}
	\caption{\cblack{Response time of each phase in \texttt{\VUG}}}
	\label{fig:exp3}
\end{figure}

\myparagraph{Exp-3: Space consumption comparison}
\cblack{Fig.~\ref{fig:exp2}} reports the space consumption of \texttt{\VUG} and the three baseline algorithms.
For each algorithm, we report its maximum and minimum space consumption among 1000 queries on each dataset under the default settings.
As observed, \texttt{\VUG} consistently consumes less space compared to baseline algorithms, attributed to the fact that it avoids enumerating and explicitly storing all temporal simple paths.
\cblack{In addtion, \texttt{\VUG} maintains a stable space consumption across different queries, regardless of the changing number of temporal simple paths, whereas the baselines exhibit exponential differences between their maximum and minimum space consumption}.
This supports our theoretical analysis in Theorems~\ref{the:gq}-\ref{the:tspgconstruct}, that is, the space complexity of \texttt{\VUG} is linear in the number of vertices and edges in the datasets.




\subsection{Performance Evaluation of Each Phase in \texttt{\VUG}}
\myparagraph{Exp-4: Response time of each phase in \texttt{\VUG}}
In this experiment, we test the total response time of each phase in \texttt{\VUG} (\ie \quick, \tight and \eev) processing 1000 queries on each dataset.
The results are illustrated in \cblack{Fig.~\ref{fig:exp3}}.
As we can see, 
the response time of \eev is the shortest among those three phases in most datasets.
\cblack{For example, on D10, the running time of \eev accounts for only 0.04$\%$ of the total response time, indicating that the bidirectional DFS, though theoretically exponential, has a limited practical overhead}.


\myparagraph{Exp-5: Evaluation of upper-bound graph generation}
In this experiment, we evaluate the performance of five upper-bound graph generation methods, i.e., $\dt$, $\es$, $\tg$,
\quick and \tight.
\cblack{For each algorithm, 
we define the upper-bound ratio as the percentage of the resulting $\stspgraph$ in the corresponding upper-bound graph, i.e., 
$\frac{\textup{the number of edges in the result graph}~\stspgraph}{\textup{the number of edges in the upper-bound graph obtained by the algorithm}}$,
and then calculate it for 1000 queries on each dataset, whose average upper-bound ratio is reported in
TABLE~\ref{tab:ratio}.}
As observed, the upper-bound ratios achieved by $\dt$ are consistently less than \cblack{0.1$\%$}.
The upper-bound graphs produced by $\es$ and $\tg$ are comparatively tighter,
and \texttt{\VUG} generates the tightest upper-bound graph, achieving an upper-bound ratio exceeding \cblack{90$\%$} on \cblack{8} datasets.

\begin{figure}[t]
	\centering
	\includegraphics[width=0.75\linewidth]{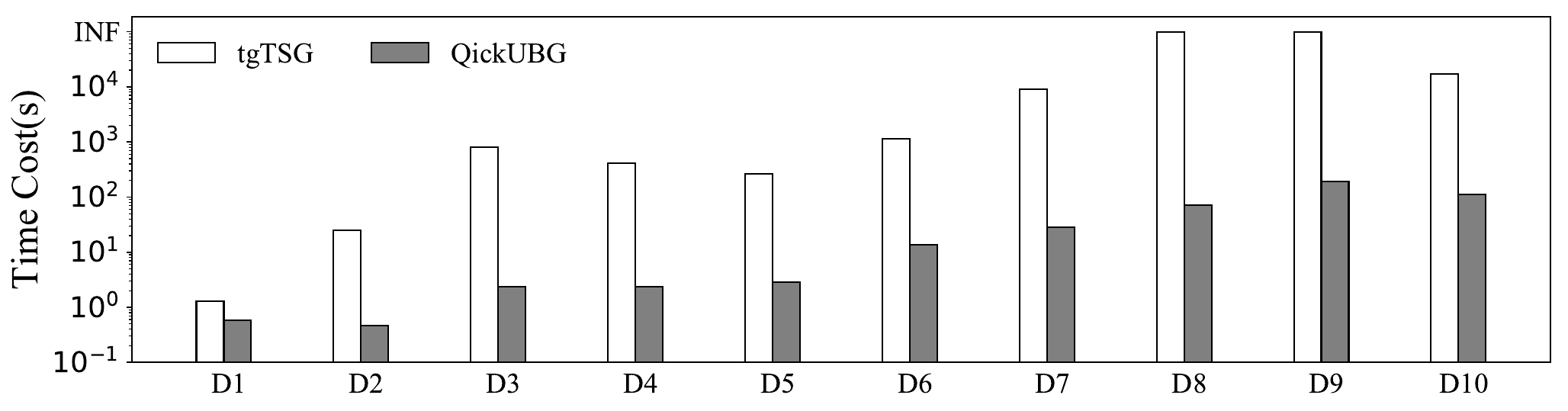}
	\caption{\cblack{Response time of \tg and \quick}}
	\label{fig:exp8}
\end{figure}

\cblack{As discussed in Section~\ref{sec:ubg_one}, while $\tg$ and \quick achieve the same effect in terms of graph reduction, \quick reduces the time complexity of $\tg$ by a factor of $\mathcal{O}(\log n)$.
To further demonstrate the efficiency of \quick, we report the total response time of $\tg$ and \quick for 1000 queries on each dataset. 
Details are shown in Fig.~\ref{fig:exp8}.
\quick is more efficient since it further optimizes the traversal process.
For example, on the dataset D7, $\tg$ takes 2.4 hours to complete the upper-bound graph generation, while \quick only takes 32 seconds.}

\begin{figure}[t]
    \centering
    \begin{subfigure}{0.18\textwidth}
        \includegraphics[width=\textwidth]{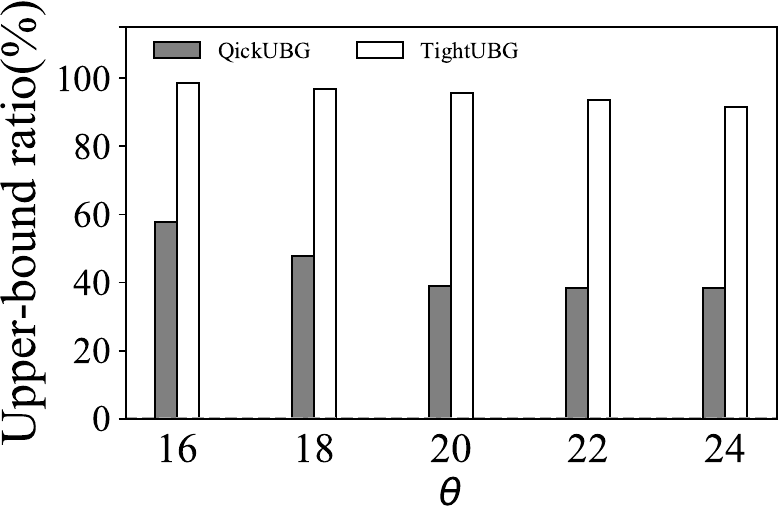}
        \caption{\cblack{D9}}
    \end{subfigure}
    \hspace{2mm}
    \begin{subfigure}{0.18\textwidth}
        \includegraphics[width=\textwidth]{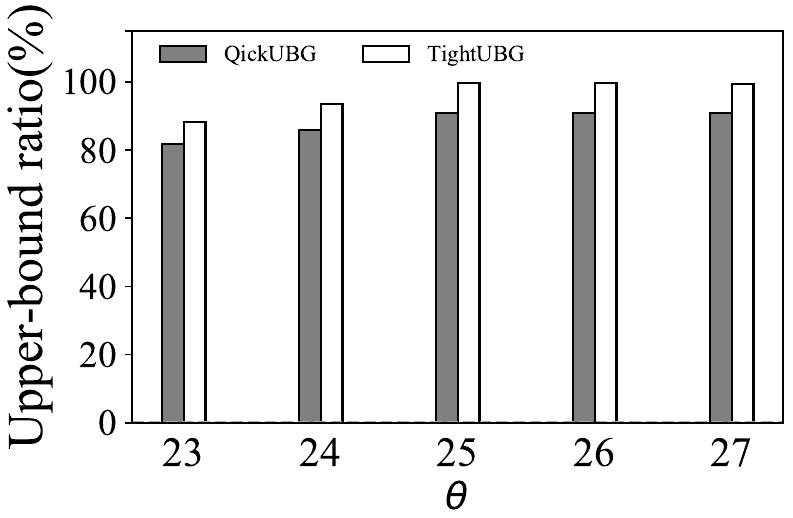}
        \caption{\cblack{D10}}
    \end{subfigure}
    \begin{subfigure}{0.18\textwidth}
        \includegraphics[width=\textwidth]{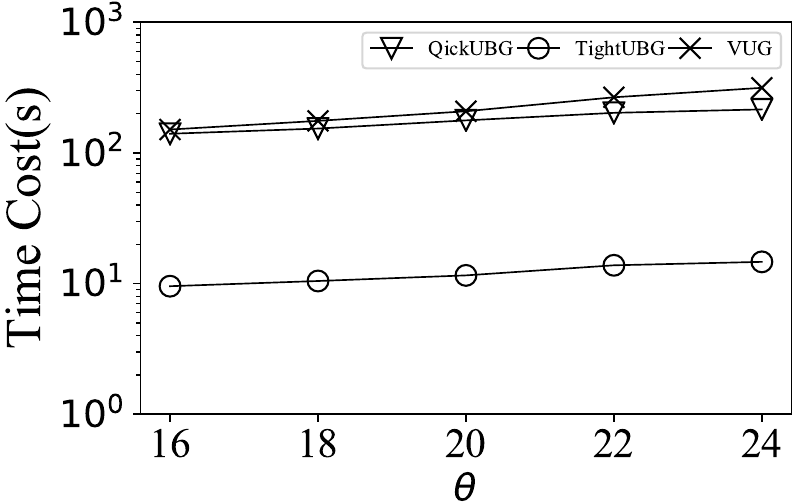}
        \caption{\cblack{D9}}
    \end{subfigure}
    \hspace{2mm}
    \begin{subfigure}{0.18\textwidth}
        \includegraphics[width=\textwidth]{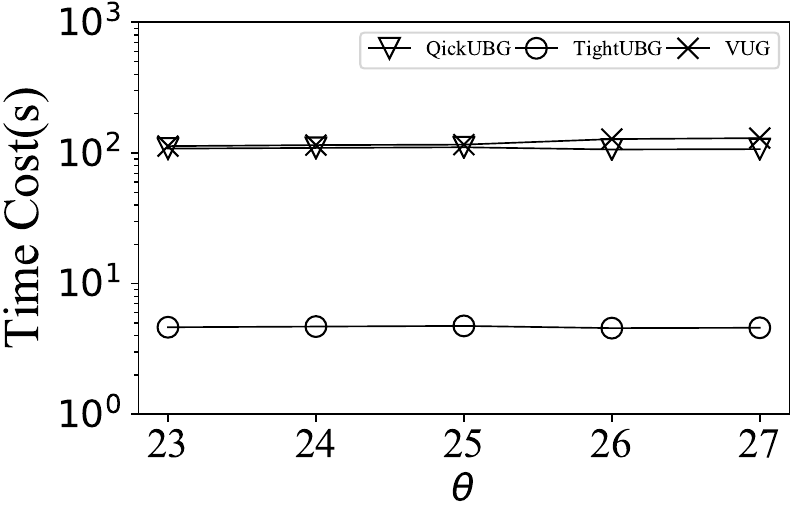}
        \caption{\cblack{D10}}
    \end{subfigure}
    \caption{\cblack{Evaluation of upper-bound graph generation}}
    \label{fig:exp7}
\end{figure}

\cblack{We also evaluate the impact of $\theta$ on upper-bound graph generation.
Fig.~\ref{fig:exp7} reports the response time and the upper-bound ratio for \quick and \tight on the two largest datasets D9 and D10 (results on other datasets can be found in Appendix C online~\cite{appendix}).
As shown, the upper-bound ratio slightly decreases with increasing $\theta$ in most cases due to the introduction of more redundant edges that can not be pruned. 
For example, on D9, when $\theta$ increases from 16 to 24, the ratio decreases from 98.5$\%$ to 91.6$\%$.
However, in some cases, such as $\theta$ from 23 to 25 on D10, the ratio slightly increases from 88.2$\%$ to 99.6$\%$, because certain previously redundant edges can be included in newly established temporal simple paths from $s$ to $t$ within a larger time interval and thus belong to $\stspgraph$. 
Meanwhile, the overhead of \quick and \tight remains stable across various settings of $\theta$, which confirms the scalability.
For example, on the dataset D9, when $\theta$ increases from 16 to 24, the time overhead of \VUG increases from 151 seconds to 315 seconds. Within this, the overhead of \quick ranges from 140 seconds to 215 seconds, and that of \tight ranges from 9 seconds to 14 seconds.
}

\setcounter{table}{1}
\begin{table}[t]
  \centering
  \caption{\cblack{The average upper-bound ratio (\%)}}
  \label{tab:ratio}
  \footnotesize
  \renewcommand\arraystretch{1.25}
  \setlength\tabcolsep{1.2pt}
  \begin{tabular}{lccccccccccc}
    \hline
    \multicolumn{2}{c|}{ } & \textbf{D1} & \textbf{D2} & \textbf{D3} & \textbf{D4} & \textbf{D5} & \textbf{D6} & \textbf{D7} & \textbf{D8} & \cblack{\textbf{D9}} & \cblack{\textbf{D10}}\\
    \hline
    \multicolumn{2}{c|}{\texttt{\dtTSG}} & $<$0.1 & $<$0.1 & $<$0.1 & $<$0.1 & $<$0.1 & - & - & - & \cblack{-} & \cblack{-}\\
    \hline
    \multicolumn{2}{c|}{\texttt{\esTSG}} & 10.4 & 32.1 & $<$0.1 & 0.2 & 13.5 & 9.8 & 15.8 & - & \cblack{-} & \cblack{64.6}\\
    \hline
    \multicolumn{2}{c|}{\texttt{\tgTSG}} & 59.4 & 51.1 & 3.4 & 4.6 & 30.6 & 24.6 & 28.4 & - & \cblack{-} & \cblack{90.9}\\
    \hline
    \multirow{2}{*}{\texttt{\VUG-UB}} & \multicolumn{1}{|c|}{\quick} & 59.4 & 51.1 & 3.4 & 4.6 & 30.6 & 24.6 & 28.4 & 88.9 & \cblack{38.9} & \cblack{90.9}\\
                          \cline{2-12}
                          & \multicolumn{1}{|c|}{\tight} & 94.9& 98.4 & 70.6 & 90.1 & 97.2 & 92.4 & 87.9 & 98.8 & \cblack{95.5} & \cblack{99.6}\\
  \hline
\end{tabular}
\end{table}


\begin{figure}[t]
    \centering
    \begin{subfigure}{0.18\textwidth}
        \includegraphics[width=\textwidth]{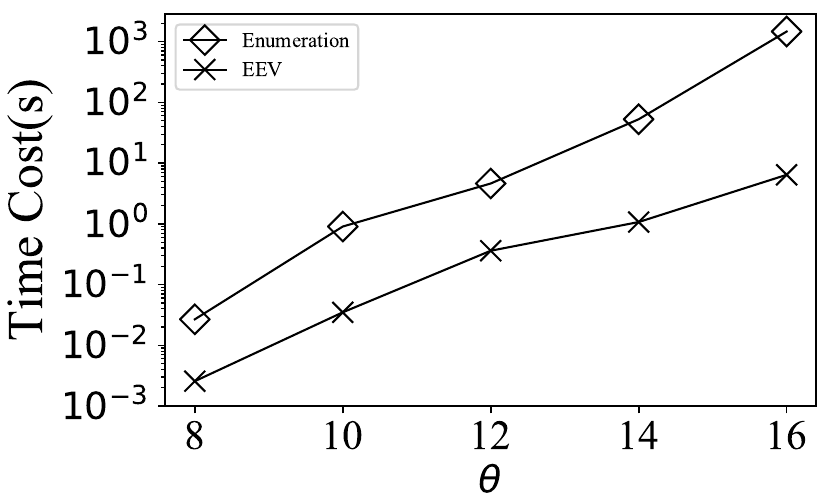}
        \caption{\cblack{D1}}
    \end{subfigure}
    \hspace{2mm}
    \begin{subfigure}{0.18\textwidth}
        \includegraphics[width=\textwidth]{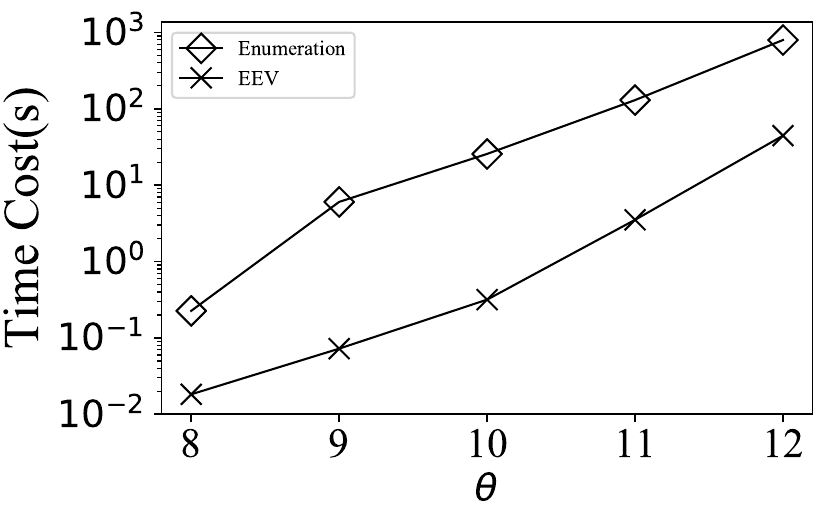}
        \caption{\cblack{D8}}
    \end{subfigure}
    \caption{\cblack{Evaluation of \eev}}
    \label{fig:exp4}
\end{figure}

\begin{figure}[t]
    \centering
    \begin{subfigure}{0.18\textwidth}
        \includegraphics[width=\textwidth]{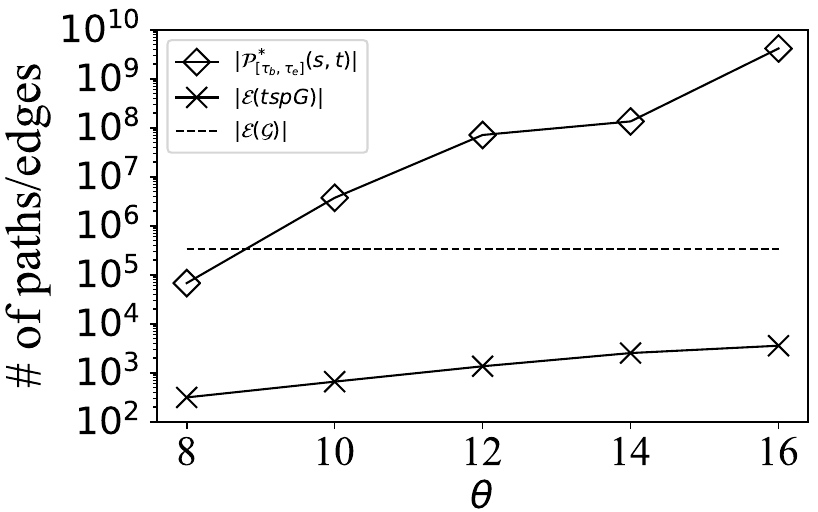}
        \caption{\cblack{D1}}
    \end{subfigure}
    \hspace{2mm}
    \begin{subfigure}{0.18\textwidth}
        \includegraphics[width=\textwidth]{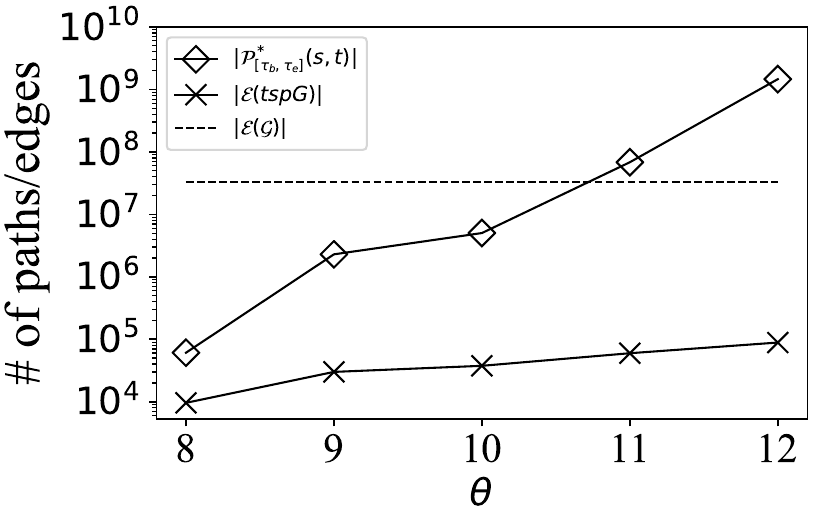}
        \caption{\cblack{D8}}
    \end{subfigure}
    \caption{\cblack{Numbers of paths/edges in $\stspgraph$}}
    \label{fig:exp6}
\end{figure}

\myparagraph{Exp-6: Evaluation of \eev}
We compare \eev with path enumeration method by applying them separately on $\gt$ and generating $\stspgraph$.
\cblack{The total response time for 1000 queries on D1 and D8 is reported in Fig.~\ref{fig:exp4} (results on the other datasets can be found in Appendix C online~\cite{appendix})}.
As observed, \eev accelerates the process of generating $\stspgraph$ by at least an order of magnitude compared to the enumeration method, since based on a tight upper-bound graph, \eev has the potential to reduce repeated searches as opposed to the path enumeration method, thereby accelerating the generation of $\stspgraph$.
\cblack{For example, when $\theta$ is 12 on D8, the enumeration-based method takes 796 seconds, while \eev only takes 44 seconds.}


\begin{figure}[t]
    \centering
    \includegraphics[width=0.8\linewidth]{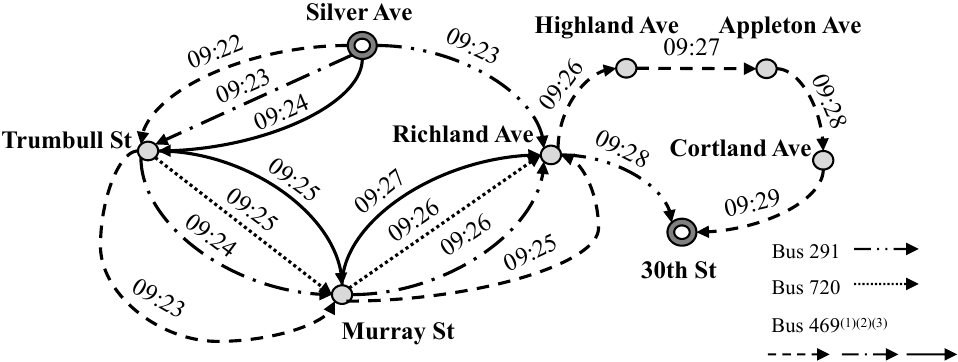}
    \caption{\cblack{Case study on SFMTA}}
    \label{fig:case}
\end{figure}

\cblack{\subsection{Effectiveness Evaluation of $\stspgraph$}}

\cblack{
\myparagraph{{Exp-7: Number of edges/paths in $\stspgraph$}}
In this experiment, we report the number of edges in $\stspgraph$ and the number of temporal simple paths it contains on D1 and D8 by varying $\theta$. 
Results are shown in Fig.~\ref{fig:exp6} (results on all other datasets can be found in Appendix C online~\cite{appendix}).
As we can see, the number of temporal simple paths in $\stspgraph$ far exceeds the number of edges in $\stspgraph$.
For example, when $\theta=10$ on D1, the generated $\stspgraph$ with 659 edges contains more than 3 million temporal simple paths.
This means many edges are shared among multiple paths, 
highlights the effectiveness of representing all paths as a single graph structure.
}

\myparagraph{\cblack{Exp-8: Case study on SFMTA}}
\cblack{
To demonstrate the effectiveness of our model, we collect the GTFS transit data from the San Francisco Municipal Transportation Agency (SFMTA\footnote{https://www.sfmta.com/reports/gtfs-transit-data}), which includes 936,188 scheduled trips across 3,267 stops made by 25,533 buses in a single day. 
We model this data as a temporal directed graph where each edge $(u,v)$ is associated with a timestamp that records the departure time of a bus from a stop $u$ to the next stop $v$~\cite{DBLP:conf/sigmod/WangLYXZ15}.
Given the query (“Silver Ave”, “30th St”, [9:20, 9:30]), Fig.~\ref{fig:case} illustrates the corresponding temporal simple path graph $\stspgraph$ with 8 transit stops (i.e., vertices) and 17 transit trips (i.e., edges). 
This graph provides possible transfer options that allow passengers to reach the destination within the given time interval. For example, if passengers miss buses departing from “Silver Ave” before 9:23, they can still board Bus 469 at 9:24, transfer to Bus 291 at "Richland Ave", and ultimately arrive at “30th St” on time.
This demonstrates the utility of $\stspgraph$ in capturing dynamic transit options and aiding real-time travel planning.
}

\section{Related Work}
\label{sec:rel}

\myparagraph{Simple Path Enumeration}
Graphs are widely used to model relationships between entities across various domains~\cite{DBLP:journals/pvldb/WangWWZQZL24,DBLP:journals/tkde/SunWWCZL24,wang2025effective,DBLP:conf/www/WuWSCWZ24,DBLP:conf/icde/SunWCWZL22}.
Simple path enumeration is one of the fundamental problems in graph analysis and has been extensively studied in~\cite{DBLP:journals/vldb/PengLZZQZ21,DBLP:conf/sigmod/SunCHH21,DBLP:conf/icde/LaiPY0021,DBLP:journals/pvldb/HaoYZ21,DBLP:conf/icde/YuanH0024,DBLP:conf/soda/BirmeleFGMPRS13,DBLP:conf/icde/ZhangYO00Y23,DBLP:journals/pvldb/LiangOZYLT24,DBLP:conf/latin/GrossiMV18,DBLP:conf/wise/LiHYCZYL22,DBLP:conf/iwoca/RizziSS14,DBLP:journals/kais/JinCL24,DBLP:conf/tamc/MutzelO19}.
Peng \etal~\cite{DBLP:journals/vldb/PengLZZQZ21} propose prune-based algorithms that dynamically maintain hop bounds for each vertex, avoiding the exploration of unpromising branches when the remaining hop budget is smaller than a vertex's hop bound.
Sun \etal~\cite{DBLP:conf/sigmod/SunCHH21} introduce index-based algorithms to reduce the search space towards real-time enumeration queries. 
Yuan \etal~\cite{DBLP:conf/icde/YuanH0024} focus on batch hop-constrained $s$-$t$ path enumeration, sharing common computations across queries to boost efficiency.
\cite{DBLP:conf/icde/LaiPY0021} and \cite{DBLP:journals/pvldb/HaoYZ21} further study the implementation on FPGA and distributed environments, respectively.
In addition, Li \etal~\cite{DBLP:conf/wise/LiHYCZYL22} study 
labelled hop-constrained $s$-$t$ simple path enumeration on billion-scale labelled graphs.
Rizzi \etal~\cite{DBLP:conf/iwoca/RizziSS14} propose length-constrained $s$-$t$ simple path enumeration on non-negative weighted graphs, where the total weight of each path does not exceed $\alpha$.
Jin \etal~\cite{DBLP:journals/kais/JinCL24} explore multi-constrained $s$-$t$ path enumeration on temporal graphs.
Mutzel \etal~\cite{DBLP:conf/tamc/MutzelO19} introduce two bicriteria temporal min-cost path problems, focusing on enumerating simple paths that are efficient w.r.t. costs and duration or costs and arrival times.
Although there are many works on the path enumeration, it is still inefficient to extend the existing solutions to our problem.

\myparagraph{Path Graph Queries}
Path graph captures the structural relationships between vertices by characterizing connections through reachable paths.
Existing works~\cite{DBLP:conf/sigmod/WangWKL21,DBLP:conf/dasfaa/LiuGP021,DBLP:journals/pacmmod/CaiLZ023} address path graph queries between two given vertices $s$ and $t$.
Liu \etal~\cite{DBLP:conf/dasfaa/LiuGP021} define the $k$-hop $s$-$t$ subgraph query, which involves all paths from $s$ to $t$ within $k$-hops.
Cai \etal~\cite{DBLP:journals/pacmmod/CaiLZ023} extend this problem and further study hop-constraint $s$-$t$ simple path graphs.
Wang \etal~\cite{DBLP:conf/sigmod/WangWKL21} formalize the $s$-$t$ shortest path graph, containing all shortest paths between $s$ and $t$.
However, these graphs lack temporal information, and a naive extension of their methods to our problem is inefficient.
Some complex path-based graph queries do involve temporal information~\cite{DBLP:journals/dam/ChengGL03,DBLP:conf/esa/BastCEGHRV10,DBLP:conf/adhoc-now/BhadraF03}.
Bhadra \etal~\cite{DBLP:conf/adhoc-now/BhadraF03} compute strongly connected components where any pair of vertices within a component are connected by a temporal path, 
which might be too rigid and inadequate to identify relationships in many real-life situations.
Cheng \etal~\cite{DBLP:journals/dam/ChengGL03} generate the associated influence digraph where each edge represents a non-decreasing path in the original graph.
Bast \etal~\cite{DBLP:conf/esa/BastCEGHRV10} generate a graph that contains a set of optimal connections with multi-criteria shortest paths.
However, the resulting graphs generated in
these works
are not well-suited for illustrating the underlying connections, as they are not subgraphs of the original graph.

\section{Conclusion}
\label{sec:concl}
Temporal graph is an important structure for modeling many real-world applications.
To analyze the relationships between vertices in temporal graphs,
in this paper, we study the problem of generating temporal simple path graph.
To solve the problem, three reasonable baselines are first proposed by explicitly enumerating all temporal simple paths.
To overcome the limitations in baselines and scale for larger networks, 
we further propose an efficient algorithm \texttt{\VUG} with two main components, \ie Upper-bound Graph Generation and Escaped Edges Verification.
Extensive experiments on \cblack{10} real-world temporal graphs show that \texttt{\VUG} significantly outperforms all baselines by at least two orders of magnitude.

\noindent \textbf{Acknowledgments}.
This work was supported by UoW R6288 and ARC DP240101322, DP230101445.

\bibliographystyle{IEEEtran}

\bibliography{ref}

\clearpage
\section*{Appendix A}
\label{appendixA}

\begin{proof}[\textbf{Proof of Theorem~\ref{the:gq}}]
Line 2 iterates through each edge in $\mathcal{E}(\mathcal{G})$ once, therefore, the total time complexity is $\mathcal{O}(m)$.
$\gq$ takes $\mathcal{O}(n+m)$ space, therefore, the total space complexity is $\mathcal{O}(n+m)$.
\end{proof}

\begin{proof}[\textbf{Proof of Theorem~\ref{the:polarity}}]
Lines 1-2 initialize the $\ea{u}$ and $\ld{u}$ for each vertex $u\in \mathcal{V}(\mathcal{G})$, which takes $\mathcal{O}(n)$ time.  
Line 6 scans each edge in $\mathcal{E}(\mathcal{G})$ once, which takes $\mathcal{O}(m)$ time.
Therefore, the total time complexity is $\mathcal{O}(n+m)$.

Throughout the algorithm, we maintain $\ea{u}$, $\ld{u}$ and pointer in $N_{out}(u,\mathcal{G})$ (resp. $N_{in}(u,\mathcal{G})$) for each vertex $u \in \mathcal{V}(\mathcal{G})$; during the traversal, in Line 9, at most one copy of each vertex is in ${Q}$, resulting in a space complexity of $\mathcal{O}(n)$.
\end{proof}

\begin{proof}[\textbf{Proof of Theorem~\ref{the:time-stream}}]
Lines 4 initialize $\tcv{\tau}{s}{u}$ for each $\tau \in \mathcal{T}_{in}(u,\gq)$ of each vertex $u \in \mathcal{V}(\mathcal{G}_q)\setminus \{s,t\}$, which takes $\mathcal{O}(m)$ time.
Lines 5-6 take $\mathcal{O}(n)$ time to initialize the {\upshape completed} indicator and the pointer of each vertex $u \in \mathcal{V}(\mathcal{G}_q)\setminus \{s,t\}$.
For each edge in $\mathcal{E}(\mathcal{G}_q)$, Lines 8-15 take $\mathcal{O}(1)$ time, Lines 17 and 19 take $\mathcal{O}(\theta)$ time as the number of vertices in each entry of $\tcv{\cdot}{s}{\cdot}$ is bounded by $\theta-1$.
Therefore, the total time complexity is $\mathcal{O}(n + \theta \cdot m)$, \cblack{which is polynomial with respect to the input size since $\theta \leq | \mathcal{\Tau}| \ll m$}.

There are $\mathcal{O}(m)$ entries in $\tcv{\cdot}{s}{\cdot}$ and $\tcv{\cdot}{\cdot}{s}$, and each entry has a length bounded by $\theta-1$. 
Throughout the algorithm, we maintain the {\upshape completed} indicator and the pointer for each vertex $u \in \mathcal{V}(\gq)$.
Therefore, the total space complexity is $\mathcal{O}(n + \theta \cdot m)$.
\end{proof}

\begin{proof}[\textbf{Proof of Theorem~\ref{the:ipgconstruct}}]
The pointer initialization in Line 1 takes $\mathcal{O}(n)$ time and the pointer operations in Lines 8-10 take $\mathcal{O}(m)$ time in total.
Line 3 iterates through each edge in $\mathcal{E}(\gq)$ once and the intersection operation in Line 17 is performed at most $m$ times.
Each intersection operation takes $\mathcal{O}(\theta)$ time as the length of each entry in $\tcv{\cdot}{s}{\cdot}$ and $\tcv{\cdot}{\cdot}{s}$ is bounded by $\theta-1$. 
Therefore, the total time complexity is $\mathcal{O}(n + \theta \cdot m)$, \cblack{which is polynomial with respect to the input size, since $\theta \leq |\mathcal{\Tau}| \ll m$}. 
$\gt$ takes $\mathcal{O}(n+m)$ space, therefore, the total space complexity is $\mathcal{O}(n+m)$.
\end{proof}

\begin{proof}[\textbf{Proof of Theorem~\ref{the:tspgconstruct}}]
Line 2 initializes a {\upshape verified} indicator for each edge, therefore takes $\mathcal{O}(m)$ time. Lines 3-5 can be implemented as a traversal in $\gt$ from $s$ (resp. $t$), which takes $\mathcal{O}(m)$ time.
For each unverified edge $e(u,v,\tau)$, the Bidirectional DFS in Line 9 has a depth of $\theta-1$ at most with a time complexity bounded by
$\mathcal{O}(d_t^{\theta-1})$,  Lines 11-18 takes $\mathcal{O}(d_t \cdot \theta)$ time to verify edges in the batch of paths.
Therefore the total time complexity is $\mathcal{O}(m \cdot d_t^{\theta-1})$.

$\stspgraph$ takes $\mathcal{O}(n+m)$ space and the verified indicator for all edges takes $\mathcal{O}(m)$ space.
The space for stacks $\mathcal{S}_v$ and $\mathcal{S}_e$ is in $\mathcal{O}(n)$.
Thus, the total space complexity is $\mathcal{O}(n+m)$.
\end{proof}

\begin{proof}[\textbf{Proof of Lemma~\ref{lem:suf}}]

\myparagraph{Sufficiency}
If there exists a temporal simple path $p_{[\tau_b,\tau_e]}^*(s,t)$ through $e(u,v,\tau)$, there exist two temporal simple paths $p_{[\tau_b,\tau_i]}^*(s,u)$,$p_{[\tau_j,\tau_e]}^*(v,t)$ s.t. $\tau_i<\tau<\tau_j$, and $\mathcal{V}(p_{[\tau_b,\tau_i]}^*(s,u)) \cap \mathcal{V}(p_{[\tau_j,\tau_e]}^*(v,t)) = \emptyset$.
Based on the definition of time-stream common vertices, we have
$\tcv{\tau_i}{s}{u} \subseteq \mathcal{V}(p_{[\tau_b,\tau_i]}^*(s,u))$ and $\tcv{\tau_j}{v}{t} \subseteq \mathcal{V}(p_{[\tau_j,\tau_e]}^*(v,t))$, thus, $\tcv{\tau_i}{s}{u} \cap \tcv{\tau_j}{v}{t} = \emptyset$.

\myparagraph{Necessity}
Consider $e(c,f,4) \in \mathcal{E}(\gq)$ in Fig.~\ref{tg:computing tpgraph}(c) as an counterexample.
There only exist $\tau_i=3$ and $\tau_j=5$ satisfying $ \tau_i<4<\tau_j$, and we have $\tcv{3}{s}{c} \cap \tcv{5}{f}{t} = \emptyset$ as $\tcv{3}{s}{c}=\{b,c\}$ and $\tcv{5}{f}{t}=\{f\}$.
However, there does not exist a temporal simple path $p_{[2,7]}^*(s,t)$ through $e(c,f,4)$.
Therefore, the necessity is not established.
\end{proof}


\begin{proof}[\textbf{Proof of Lemma~\ref{lem:timestamps}}]
To proof Lemma~\ref{lem:timestamps}, we first proof $\mathcal{P}_{[\tau_b,\tau_l]}^*(s,u)$ = $\mathcal{P}_{[\tau_b,\tau]}^*(s,u)$ by contradiction.
Since $\tau_l \leq \tau$, $\mathcal{P}_{[\tau_b,\tau_l]}^*(s,u)$  $ \subseteq$  $\mathcal{P}_{[\tau_b,\tau]}^*(s,u)$.
Suppose $\mathcal{P}_{[\tau_b,\tau_l]}^*(s,u)\neq \mathcal{P}_{[\tau_b,\tau]}^*(s,u)$, this implies that there exists $p_{[\tau_b,\tau]}^*(s,u) \in \mathcal{P}_{[\tau_b,\tau]}^*(s,u)$ such that $p_{[\tau_b,\tau]}^*(s,u) \notin \mathcal{P}_{[\tau_b,\tau_l]}^*(s,u)$. 
It is evident that such $p_{[\tau_b,\tau]}^*(s,u)$ contains an in-coming edge $e(v,u,\tau')$ of $u$ where $\tau_l < \tau' \leq \tau$, 
which contradicts $\tau_{l}$ = $\max\{\tau_{i} | \tau_{i} \in \mathcal{T}_{in}(u,\gq), \tau_{i} \leq \tau\}$.
In conclusion, $\mathcal{P}_{[\tau_b,\tau_l]}^*(s,u)$ = $\mathcal{P}_{[\tau_b,\tau]}^*(s,u)$, and we have $\tcv{\tau}{s}{u}=\tcv{\tau_{l}}{s}{u}$ following the definition of the time-stream common vertices.
We omit the proof for $\tcv{\tau}{u}{t}=\tcv{\tau_{r}}{u}{t}$ as it follows a similar approach.
\end{proof}

\begin{proof}[\textbf{Proof of Lemma~\ref{lem:tcvpath}}]

For each temporal path $p \in \mathcal{P}_{[\tau_b,\tau]}(s,u) $, there exists a corresponding temporal simple path $p^* \in \mathcal{P}_{[\tau_b,\tau]}^*(s,u)$ such that the path $p^*$ contains all edges in the path $p$ except those forming cycles. 
Then, we have $\mathcal{V}(p^*) \subseteq \mathcal{V}(p)$ to derive that  $\mathcal{V}(p^*) \cap \mathcal{V}(p) = \mathcal{V}(p^*)$. 
Thus, $\tcv{\tau}{s}{u}$ = $\bigcap_{p^* \in \mathcal{P}^*_{[\tau_b,\tau]}(s,u) ~s.t.~t \notin \mathcal{V}(p^*)}\mathcal{V}(p^*) \setminus \{s\}$ = 
$\bigcap_{p \in \mathcal{P}_{[\tau_b,\tau]}(s,u) ~s.t.~t \notin \mathcal{V}(p)}\mathcal{V}(p) \setminus \{s\}$.
The proof for $\tcv{\tau}{u}{t}$ follows the same approach.

\end{proof}

\begin{proof}[\textbf{Proof of Lemma~\ref{lem:certain_timestamp}}]
    If $\tcv{\tau_l}{s}{u} \cap \tcv{\tau_r}{v}{t} \ne \emptyset$, \ie there exits $w$ such that $w \in \tcv{\tau_l}{s}{u}$ and $w \in \tcv{\tau_r}{v}{t}$, then based on the definition of time-stream common vertices, $\forall$ $p^* \in \mathcal{P}^*_{[\tau_b,\tau_l]}(s,u) ~s.t.~t \notin \mathcal{V}(p^*)$, $w \in p^*$.
    Since $\forall$ $\tau_b \leq \tau_i<\tau_l$, we have $\mathcal{P}^*_{[\tau_b,\tau_i]}(s,u) \subseteq \mathcal{P}^*_{[\tau_b,\tau_l]}(s,u)$, thus, $\forall$ $p^*_i \in \mathcal{P}^*_{[\tau_b,\tau_i]}(s,u) ~s.t.~t \notin \mathcal{V}(p^*_i)$, $w \in p^*_i$, that is, $w \in \tcv{\tau_i}{s}{u}$.
    Similarly, $\forall$ $\tau_r < \tau_j \leq \tau_e$, $w \in \tcv{\tau_j}{v}{t}$.
    Therefore, $\tcv{\tau_i}{s}{u} \cap \tcv{\tau_j}{v}{t} \ne \emptyset$.
\end{proof}

\begin{proof}[\textbf{Proof of Lemma~\ref{the:ipgcorrectness}}]
    $(\Rightarrow)$
    If an edge $e(u,v,\tau) \in \mathcal{E}(\gq)$, where $u \neq s$ and $v \ne t$, satisfies condition $i)$, based on Lemma~\ref{lem:suf} and Lemma~\ref{lem:certain_timestamp}, such an edge will not be excluded as an unpromising edge. Therefore, it should be included in $\gt$.
    If an edge $e(u,v,\tau) \in \mathcal{E}(\gq)$, where $u = s$ or $v = t$, based on the condition $ii)$ of Lemma~\ref{lem:tpgtotspexistamended}, we can conclude that $e(u,v,\tau)$ belongs to $\stspgraph$.
    Since $\gt$ is the upper-bound graph of $\stspgraph$, $e(u,v,\tau)$ must belong to $\gt$.

    $(\Leftarrow)$
    If an edge $e(u,v,\tau)$ belongs to $\gt$, there does not exist a vertex $w$ appearing in all $p_{[\tau_b,\tau_i]}^*(s,u)$ and $p_{[\tau_j,\tau_e]}^*(v,t)$ where $\tau_i<\tau<\tau_j$.
    Based on the Definiton~\ref{def:common}, it is easy to derive that $e(u,v,\tau)$ satisfies condition $i)$ s.t. $u \ne s$ and $v \ne t$.
    Since $e(u,v,\tau) \in \gt$ must belong to $\gq$, and we have discussed $w \ne s$ and $w \ne t$ for each edge in $\gq$, therefore $u$ can be the source vertex $s$ or $v$ can be the target vertex $t$. 
\end{proof}

\begin{proof}[\textbf{Proof of Lemma~\ref{lem:2hopedges}}]
    Let $\tau_l=\max\{\tau''|\tau'' \in \mathcal{T}_{in}(u, $ $\gq) \land \tau_b \leq \tau'' <\tau\}$, $\tau_r=\min\{\tau''|\tau'' \in \mathcal{T}_{out}(v,\gq) \land \tau < \tau'' \leq \tau_e\}$.
    For condition $i)$, if there exists an edge $e(s, u, \tau') \in \mathcal{E}(\gt)$ such that $\tau_b \leq \tau' \leq \tau_l < \tau$, then we have a temporal simple path $p_{[\tau_b,\tau_l]}^*(s,u) = \langle e(s, u, \tau') \rangle$, therefore $\tcv{\tau_l}{s}{u} = \{u\}$. 
    Since $e(u,v,\tau) \in \gt$, according to Lemma~\ref{the:ipgcorrectness}, 
    $\tcv{\tau_l}{s}{u} \cap \tcv{\tau_r}{v}{t}=\emptyset$, that is, $u \notin \tcv{\tau_r}{v}{t}$. 
    Then, based on the definition of $\tcv{\tau_r}{v}{t}$, there exists a temporal simple path $p_{[\tau_r,\tau_e]}^*(v,t)$ that does not pass through $s$ and $u$.
    Therefore, a temporal simple path $p_{[\tau_b,\tau_e]}^*(s,t)$ that includes $e(u,v,\tau)$ can be formed.
    We omit the proof for condition $ii)$ as it is similar to that of condition $i)$.
\end{proof}

\section*{Appendix B}
\label{appendixB}

\begin{algorithm}[t]  
{
    \SetVline
    \caption{BiDirSearch}\label{alg:bidirsearch}
    \Input{an edge $e(u, v, \tau) \in \mathcal{E}(\gt)$}
    \Output{the existence of $p_{[\tau_b,\tau_e]}^*(s,t)$ through $e(u, v, \tau)$}

    \SetKwFunction{FwdSearch}{\textrm{FwdSearch}}
    \SetKwFunction{BwdSearch}{\textrm{BwdSearch}}

    \SetKwFunction{Search}{\textrm{Search}}
    
    \State{$S_v.push(u)$, $S_v.push(v)$}
    \lIf{$\tau-\tau_b > \tau_e-\tau$}{$dir=f$}
    \lElse{\State{$dir=b$}}
    \State{$F\gets false, B \gets false$}
    \lIf{\Search{$u,v,\tau, dir, F, B$}}{\State{\Return{true}}}
    \lElse{\State{\Return{false}}}

    \SetKwBlock{Procedure}{Procedure \Search{$u,v,\tau, dir, F, B$}:}{end}

    \Procedure{
        \lIf{$dir=f$}{$S_v.push(v)$}
        \lElse{\State{$S_v.push(u)$}}
        \State{$S_e.push(e(u,v,\tau))$}
        \If{$v=t \lor u=s$} {
            \lIf{$v=t$}{\State{$dir=b$, $F \gets true$}}
            \lIf{$u=s$}{\State{$dir=f$, $B \gets true$}}
            \lIf{$F \land B$}{\State{\Return{true}}}
            \State{$e(u_0,v_0,\tau_0) \gets$ the first edge in $S_e$}
            \lIf{\Search{$u_0,v_0,\tau_0, dir, F, B$}}{\State{\Return{true}}}
        }
        
        \If{$dir=f$} {
            \ForEach{$(w_i,\tau_i) \in N_{out}(v, \gt)$}{
                \lIf{$\tau_i \leq \tau$}{\State{\textbf{break}}}
                \lIf{$w_i \in S_v$}{\State{\textbf{continue}}}
                \lIf{\Search{$v,w_i,\tau_i,dir, F, B$}}{\State{\Return{true}}}
            }
        }
        \Else{
            \ForEach{$(w_i,\tau_i) \in N_{in}(u, \gt)$}{
                \lIf{$\tau_i \geq \tau$}{\State{\textbf{break}}}
                \lIf{$w_i \in S_v$}{\State{\textbf{continue}}}
                \lIf{\Search{$w_i,u,\tau_i,dir, F, B$}}{\State{\Return{true}}}
            }
        }

        \State{$S_v.pop()$, $S_e.pop()$}
        \State{\Return{false}}
    }
}
\end{algorithm}

\myparagraph{\underline{BiDirSearch}}
With the proposed optimizations, our implementation of the bidirectional DFS is detailed in Algorithm~\ref{alg:bidirsearch}.
In Line 2, we determine whether to perform a forward or a backward search first.
Then, Line 4 starts the \textit{search} procedure with two flags $F$ (resp. $B$), indicating whether a forward path (resp. backward path) has been found in the current search, initially set to $false$.
The \textit{search} procedure recursively searches for a temporal simple path in its given direction.
Once a forward (resp. backward) path is completed, it sets the corresponding flag $F$ (resp. $B$) and toggles the search direction in Lines 10-11.
If the path in only one direction is completed, the \textit{search} procedure will continue in the other direction and will terminate upon finding paths in both directions in Line 12.
In Lines 16 and 21, the neighbor exploration order is determined based on the current search direction.

\section*{Appendix C}
\label{appendixC}

\setcounter{table}{0}
\begin{table}[t]
  \centering
  \caption{Statistics of datasets}
  \label{tab:dataset}
  \renewcommand\arraystretch{1.2}
  \setlength\tabcolsep{2.5pt}
  \begin{tabular}{lcccccccc}
    \toprule
    \textbf{Dataset}     & $|\mathcal{V}|$     & $|\mathcal{E}|$      & $|\mathcal{T}|$  & $d$ & $\theta$ \\
    \midrule
    D1~(email-Eu-core)   & 1,005  & 332,334  & 803  & 9,782 & 10 \\ 
    D2~(sx-mathoverflow) & 88,581 & 506,550  & 2,350   & 5,931 & 20 \\ 
    D3~(sx-askubuntu)    & 159,316 & 964,437 & 2,613   & 8,729 & 20 \\ 
    D4~(sx-superuser)    & 194,085 & 1,443,339   & 2,773   & 26,996 & 20 \\ 
    D5~(wiki-ru)         & 457,018 & 2,282,055   & 4,715   & 188,103 & 25 \\
    D6~(wiki-de)         & 519,404 & 6,729,794   & 5,599    & 395,780 & 25 \\ 
    D7~(wiki-talk)       & 1,140,149   & 7,833,140   & 2,320   & 264,905 & 20 \\ 
    D8~(flickr)          & 2,302,926   & 33,140,017  & 196   & 34,174 & 10 \\  
    \cblack{D9~(sx-stackoverflow)} & \cblack{6,024,271} & \cblack{63,497,050} & \cblack{2,776} & \cblack{101,663} & \cblack{20} \\ 
    \cblack{D10~(wikipedia)} & \cblack{2,166,670} & \cblack{86,337,879} & \cblack{3,787} & \cblack{218,465} & \cblack{25} \\ 
  \bottomrule
\end{tabular}
\end{table}

\myparagraph{Datasets}
We employ \cblack{10} real-world temporal graphs in our experiments. 
Details of these datasets are summarized in TABLE I,
where 
$d$ is the maximum degree in a dataset.
Among these datasets, D5, D6, D8 \cblack{and D10} are obtained from KONECT\footnote{http://konect.cc/networks/}, 
the other \cblack{six} datasets are obtained from SNAP\footnote{http://snap.stanford.edu}.

\begin{figure*}[t]
    \centering
    \begin{subfigure}{0.18\textwidth}
        \includegraphics[width=\textwidth]{Figure_Experiment/exp_5_2.pdf}
        \caption{\cblack{D1}}
    \end{subfigure}
    \begin{subfigure}{0.18\textwidth}
        \includegraphics[width=\textwidth]{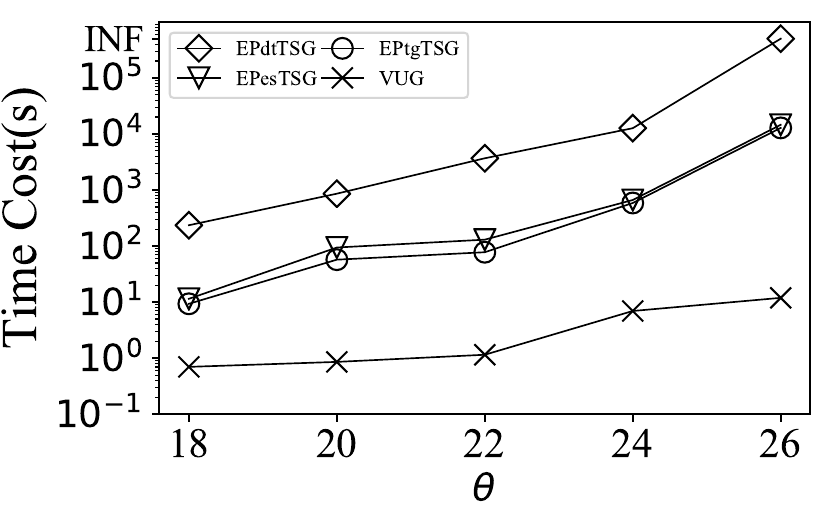}
        \caption{\cblack{D2}}
    \end{subfigure}
    \begin{subfigure}{0.18\textwidth}
        \includegraphics[width=\textwidth]{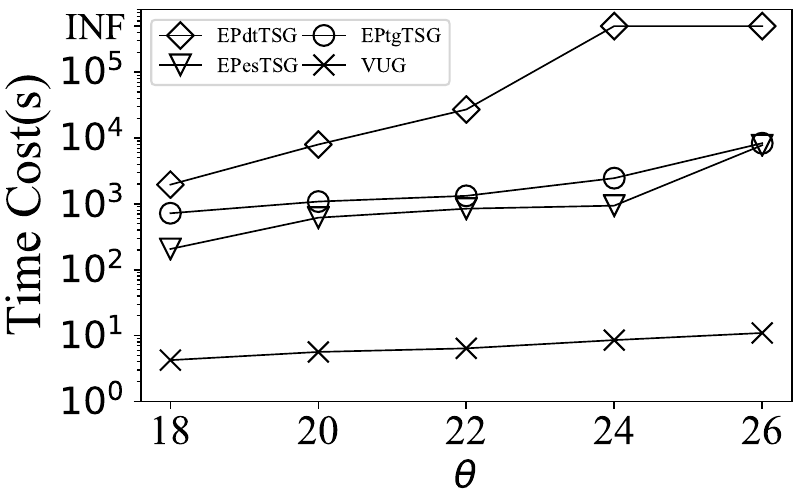}
        \caption{\cblack{D3}}
    \end{subfigure}
    \begin{subfigure}{0.18\textwidth}
        \includegraphics[width=\textwidth]{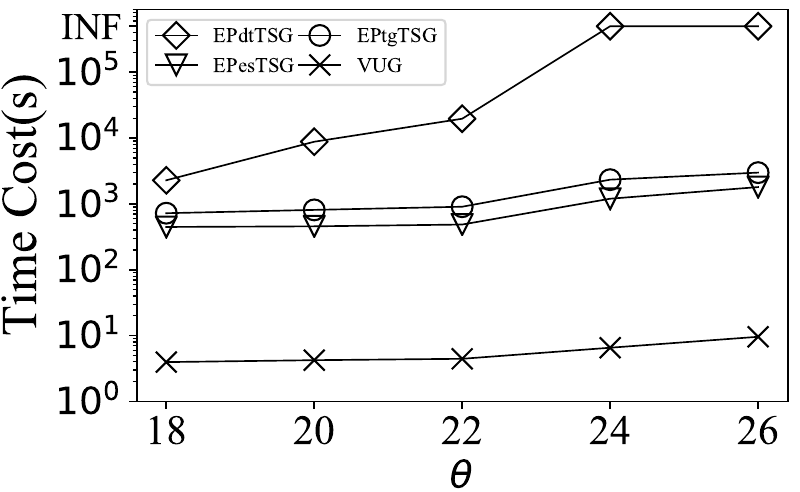}
        \caption{\cblack{D4}}
    \end{subfigure}
    \begin{subfigure}{0.18\textwidth}
        \includegraphics[width=\textwidth]{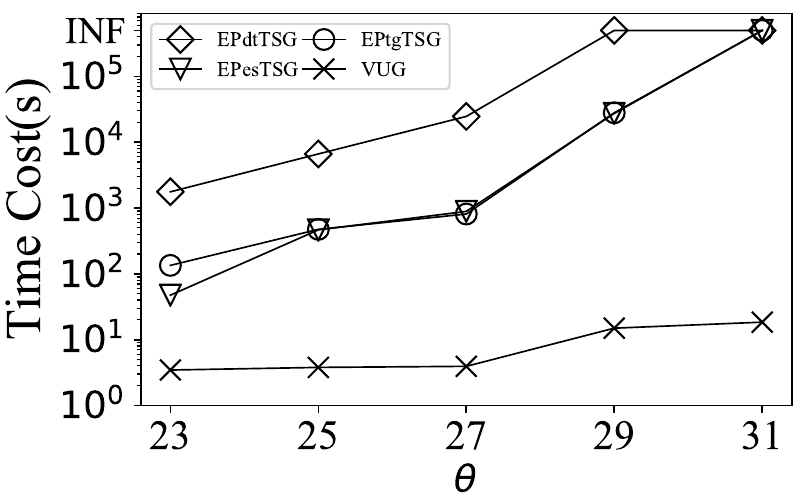}
        \caption{\cblack{D5}}
    \end{subfigure}
    \begin{subfigure}{0.18\textwidth}
        \includegraphics[width=\textwidth]{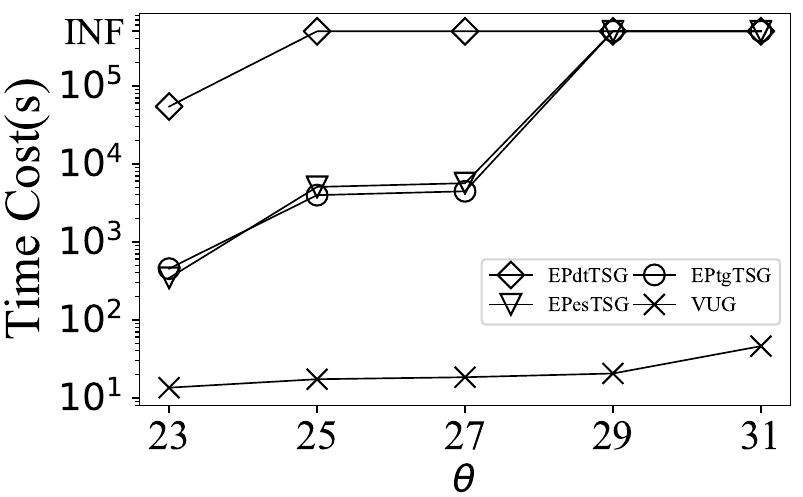}
        \caption{\cblack{D6}}
    \end{subfigure}
    \begin{subfigure}{0.18\textwidth}
        \includegraphics[width=\textwidth]{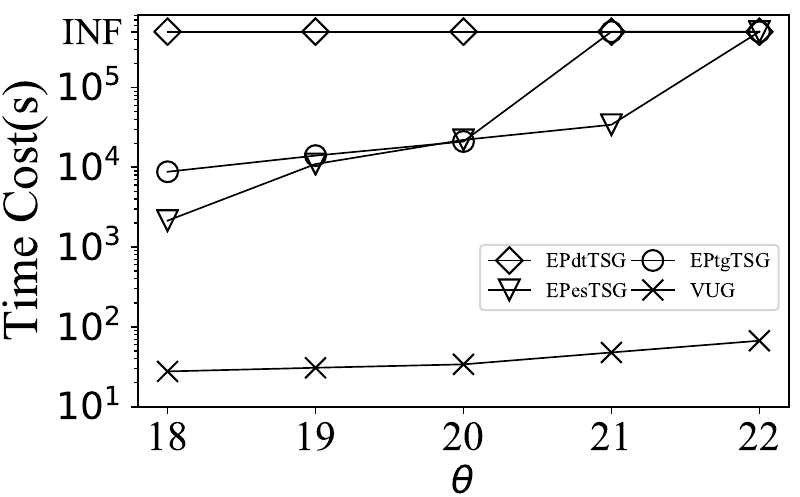}
        \caption{\cblack{D7}}
    \end{subfigure}
    \begin{subfigure}{0.18\textwidth}
        \includegraphics[width=\textwidth]{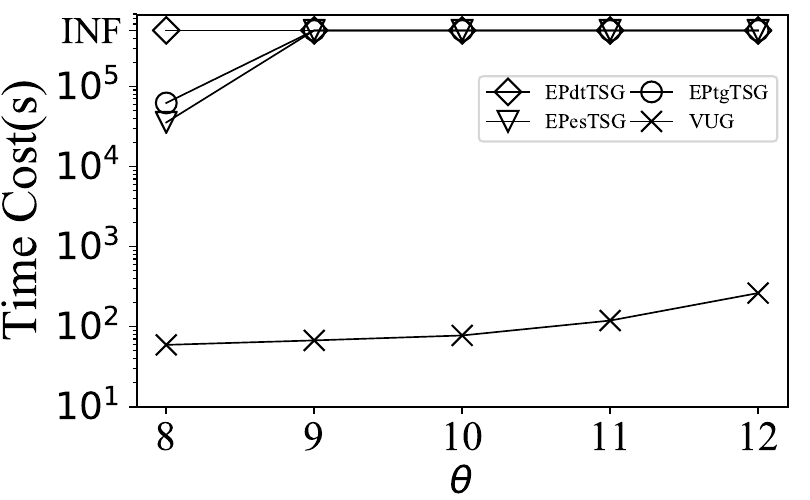}
        \caption{\cblack{D8}}
    \end{subfigure}
    \begin{subfigure}{0.18\textwidth}
        \includegraphics[width=\textwidth]{Figure_Experiment/exp_5_10.pdf}
        \caption{\cblack{D9}}
    \end{subfigure}
    \begin{subfigure}{0.18\textwidth}
        \includegraphics[width=\textwidth]{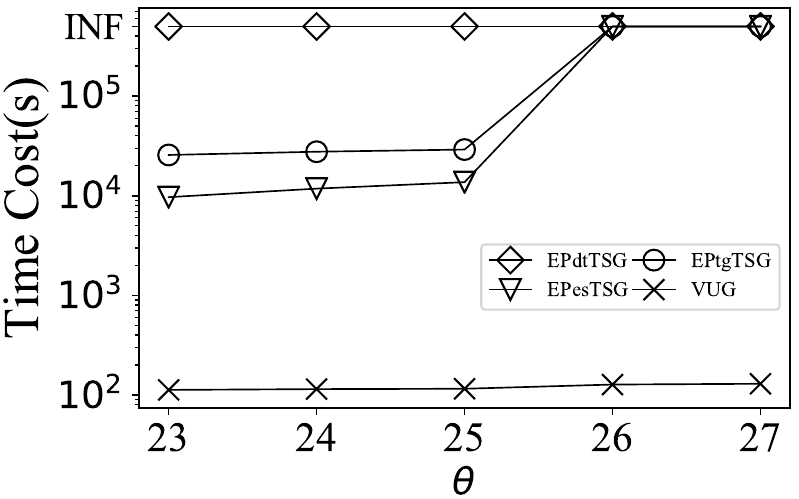}
        \caption{\cblack{D10}}
    \end{subfigure}
    \caption{\cblack{Response time by varying parameter $\theta$ on all other datasets}}
    \label{fig:exp5_more}
\end{figure*}

\myparagraph{\cblack{Supplement to Exp-2: Response time by varying parameter $\theta$ on the other datasets}}
\cblack{In Fig.~\ref{fig:exp5_more}, we report the total response time of \texttt{\VUG} for 1000 queries by varying $\theta$, along with the three baseline algorithms as comparison.
On each dataset, the response time of \VUG grows modestly, which confirms the scalability of our proposed methods. 
For example, on the largest dataset D10, three baseline methods cannot finish queries within 12 hours when $\theta=26$, while the total response time of \VUG only increases by a factor of 1.2 when $\theta$ increases from 23 to 27.
}

\begin{figure*}[t]
    \centering
    \begin{subfigure}{0.18\textwidth}
        \includegraphics[width=\textwidth]{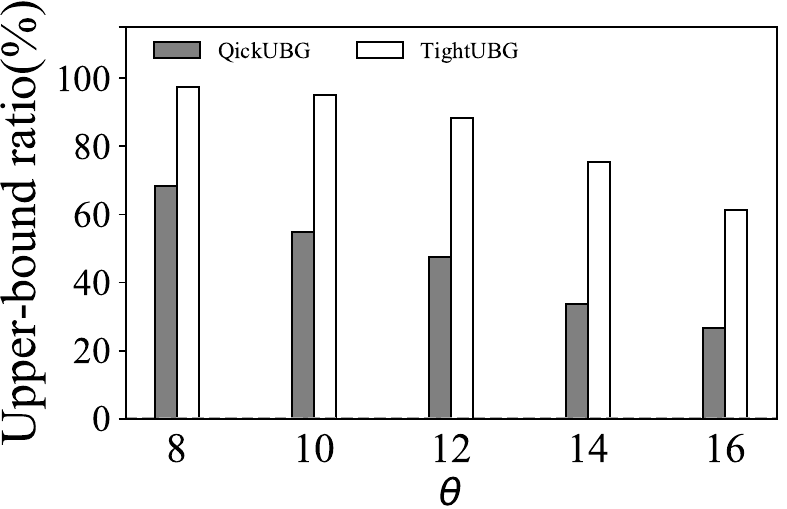}
        \caption{\cblack{D1}}
    \end{subfigure}
    \begin{subfigure}{0.18\textwidth}
        \includegraphics[width=\textwidth]{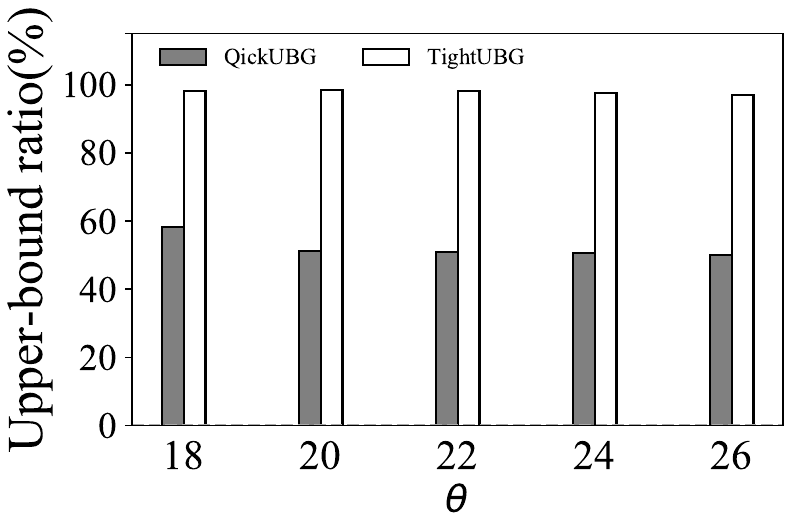}
        \caption{\cblack{D2}}
    \end{subfigure}
    \begin{subfigure}{0.18\textwidth}
        \includegraphics[width=\textwidth]{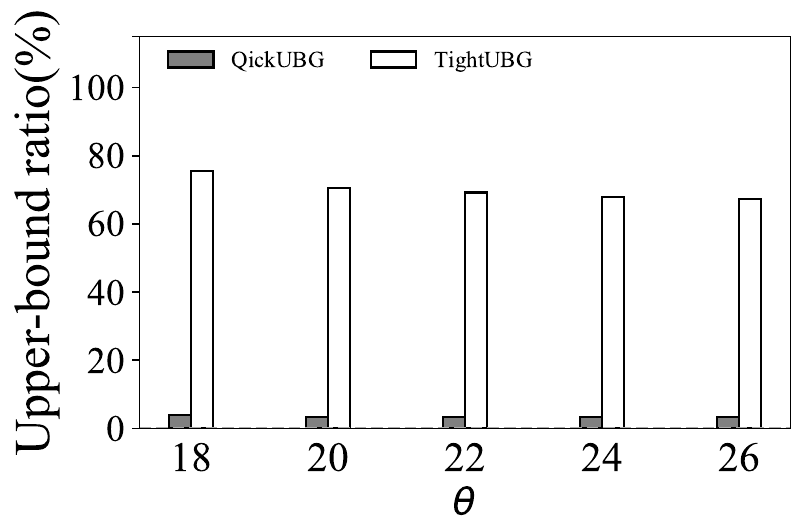}
        \caption{\cblack{D3}}
    \end{subfigure}
    \begin{subfigure}{0.18\textwidth}
        \includegraphics[width=\textwidth]{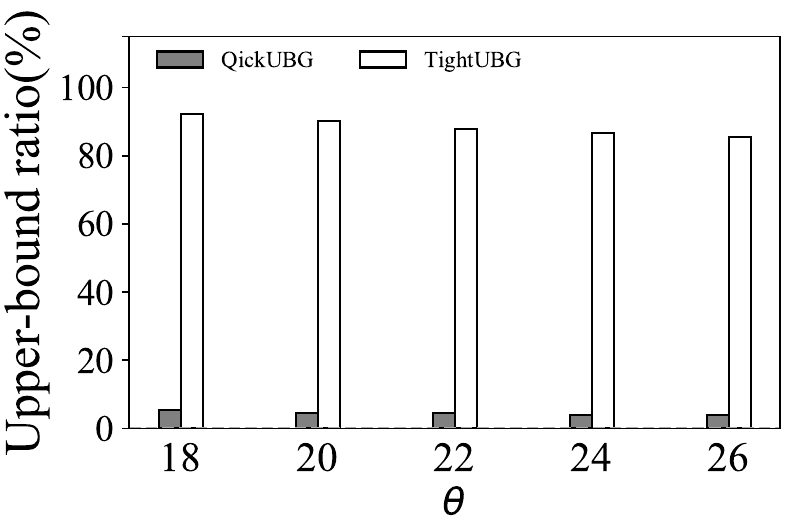}
        \caption{\cblack{D4}}
    \end{subfigure}
    \begin{subfigure}{0.18\textwidth}
        \includegraphics[width=\textwidth]{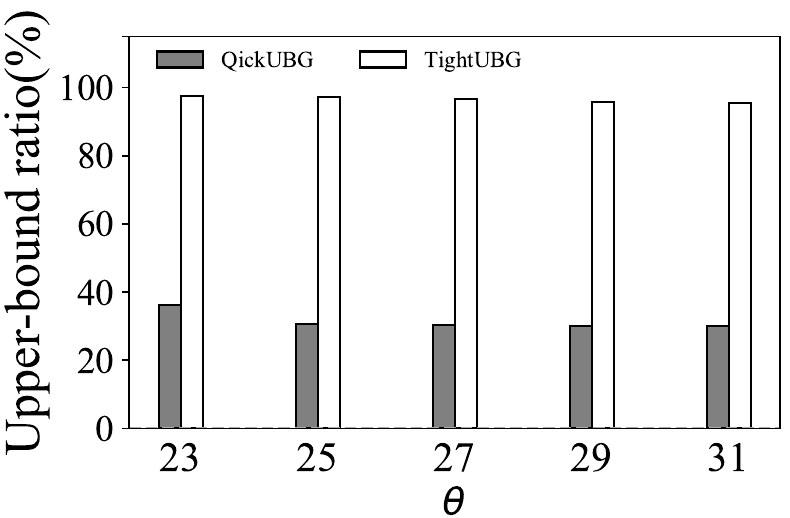}
        \caption{\cblack{D5}}
    \end{subfigure}
    \begin{subfigure}{0.18\textwidth}
        \includegraphics[width=\textwidth]{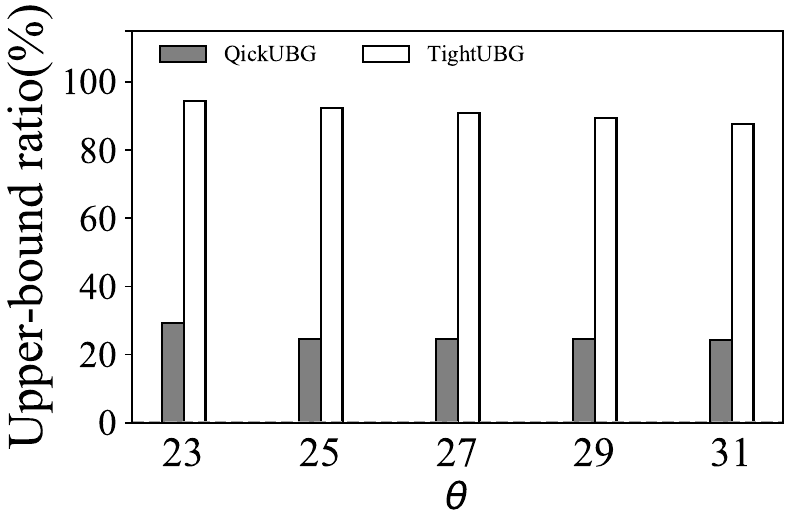}
        \caption{\cblack{D6}}
    \end{subfigure}
    \begin{subfigure}{0.18\textwidth}
        \includegraphics[width=\textwidth]{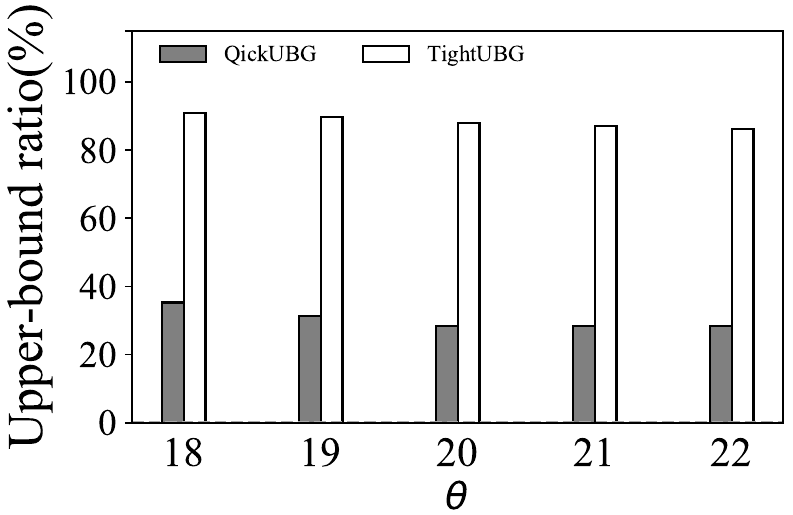}
        \caption{\cblack{D7}}
    \end{subfigure}
     \begin{subfigure}{0.18\textwidth}
        \includegraphics[width=\textwidth]{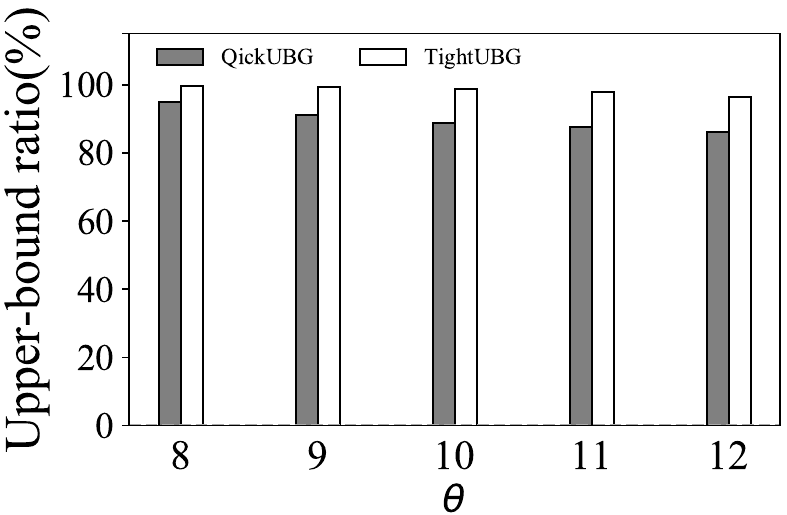}
        \caption{\cblack{D8}}
    \end{subfigure}
    \begin{subfigure}{0.18\textwidth}
        \includegraphics[width=\textwidth]{Figure_Experiment/exp_7_9_1.pdf}
        \caption{\cblack{D9}}
    \end{subfigure}
     \begin{subfigure}{0.18\textwidth}
        \includegraphics[width=\textwidth]{Figure_Experiment/exp_7_10_1.pdf}
        \caption{\cblack{D10}}
    \end{subfigure}
    \begin{subfigure}{0.18\textwidth}
        \includegraphics[width=\textwidth]{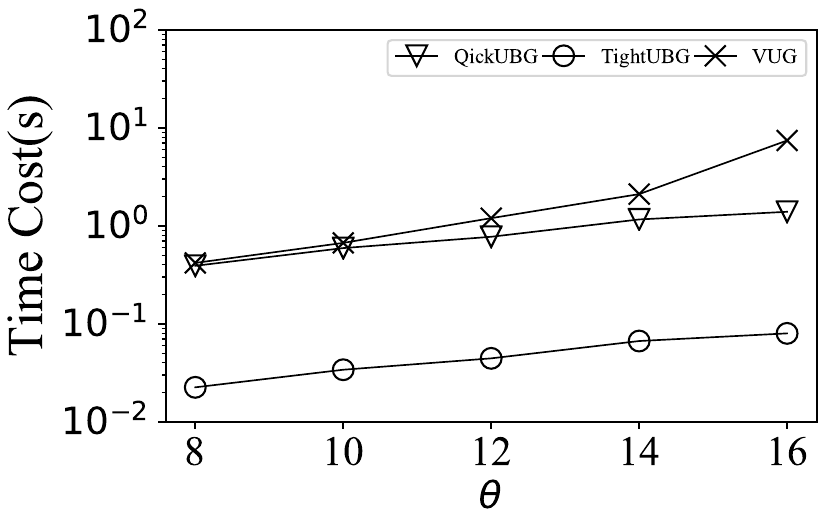}
        \caption{\cblack{D1}}
    \end{subfigure}
    \begin{subfigure}{0.18\textwidth}
        \includegraphics[width=\textwidth]{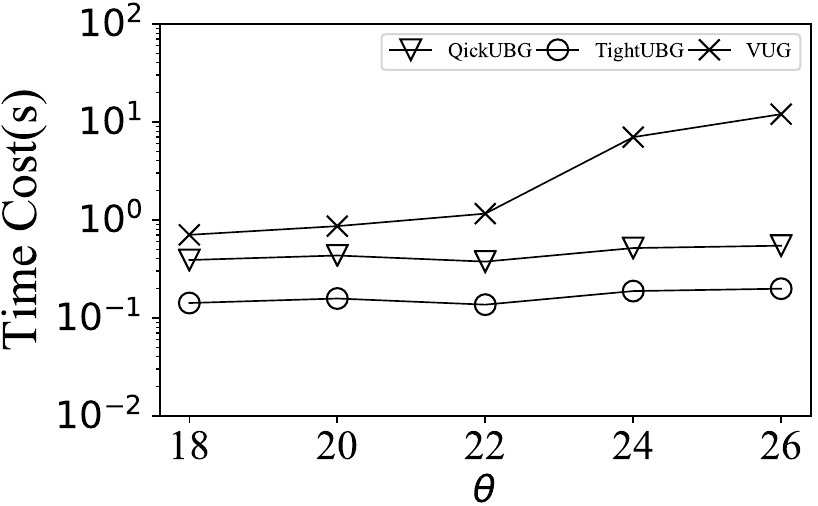}
        \caption{\cblack{D2}}
    \end{subfigure}
    \begin{subfigure}{0.18\textwidth}
        \includegraphics[width=\textwidth]{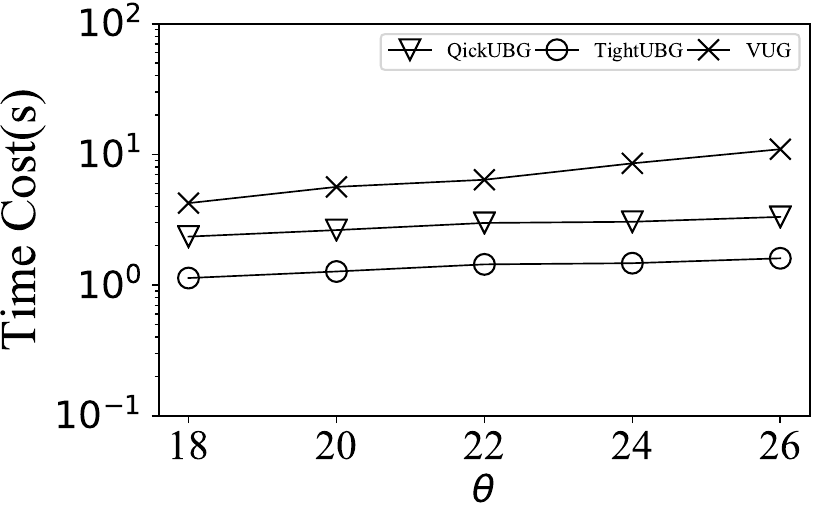}
        \caption{\cblack{D3}}
    \end{subfigure}
    \begin{subfigure}{0.18\textwidth}
        \includegraphics[width=\textwidth]{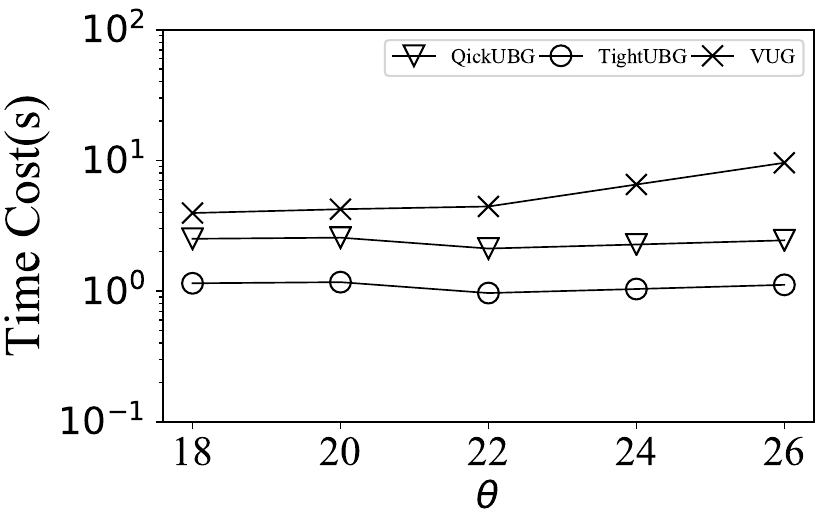}
        \caption{\cblack{D4}}
    \end{subfigure}
    \begin{subfigure}{0.18\textwidth}
        \includegraphics[width=\textwidth]{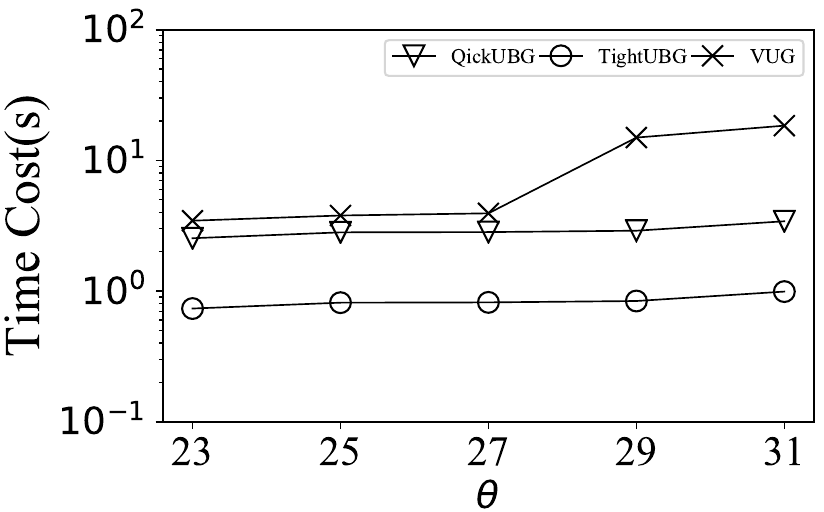}
        \caption{\cblack{D5}}
    \end{subfigure}
    \begin{subfigure}{0.18\textwidth}
        \includegraphics[width=\textwidth]{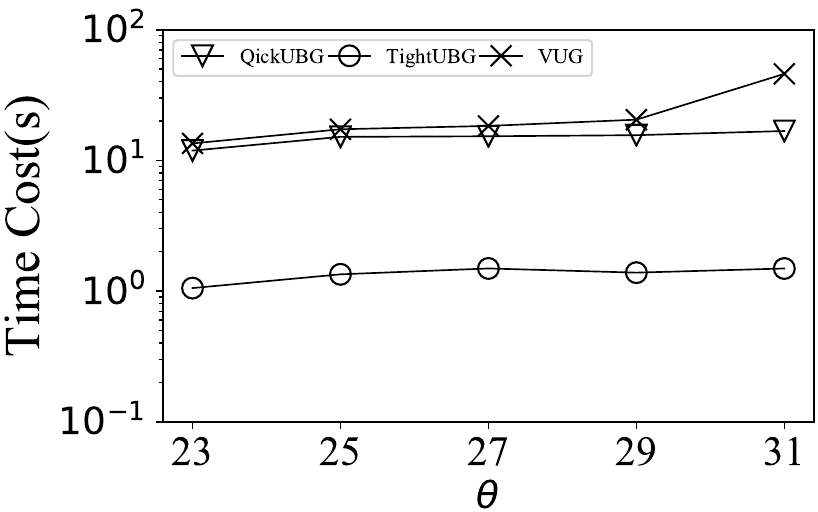}
        \caption{\cblack{D6}}
    \end{subfigure}
    \begin{subfigure}{0.18\textwidth}
        \includegraphics[width=\textwidth]{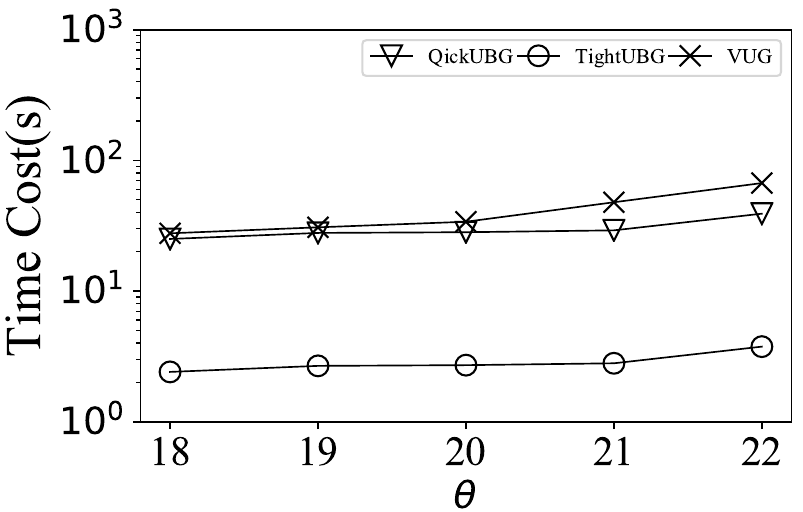}
        \caption{\cblack{D7}}
    \end{subfigure}
    \begin{subfigure}{0.18\textwidth}
        \includegraphics[width=\textwidth]{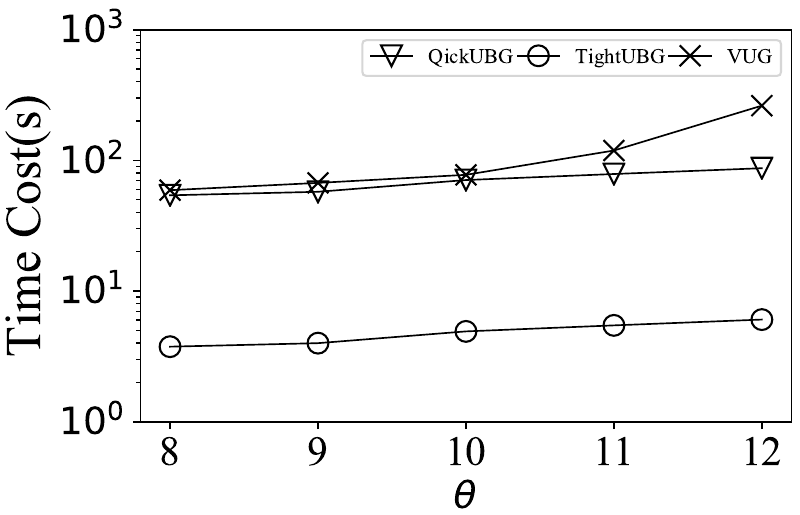}
        \caption{\cblack{D8}}
    \end{subfigure}
    \begin{subfigure}{0.18\textwidth}
        \includegraphics[width=\textwidth]{Figure_Experiment/exp_7_9_2.pdf}
        \caption{\cblack{D9}}
    \end{subfigure}
    \begin{subfigure}{0.18\textwidth}
        \includegraphics[width=\textwidth]{Figure_Experiment/exp_7_10_2.pdf}
        \caption{\cblack{D10}}
    \end{subfigure}
    \caption{\cblack{Evaluation of upper-bound graph generation on all other datasets}}
    \label{fig:exp7_more}
\end{figure*}

\myparagraph{\cblack{Supplement to Exp-5: Evaluation of upper-bound graph generation of $\theta$'s impact on the other datasets}}
\cblack{
Fig.~\ref{fig:exp7_more} reports the response time and the upper-bound ratio for \quick and \tight on each dataset.
As shown, the upper-bound ratio slightly decreases with increasing $\theta$ in most cases, \ie D1 ($97.5\%$-$61.3\%$), D3 ($75.4\%$-$67.3\%$), D4 ($92.2\%$-$85.5\%$), D5 ($97.6\%$-$95.4\%$), D6 ($94.4\%$-$87.6\%$), D7 ($90.8\%$-$86.2\%$), D8 ($99.5\%$-$96.5\%$), D9 ($98.5\%$-$91.6\%$) within presented ranges of $\theta$.
However, in some cases, the ratio slightly increases, \ie D2 ($98.2\%$-$98.4\%$) as $\theta$ increases from 18 to 20 and D10 ($88.2\%$-$99.6\%$) as $\theta$ increases from 23 to 25.
Meanwhile, the overhead of \quick and \tight remains stable across various settings of $\theta$ on all the datasets.
For example, on the largest dataset D10, when $\theta$ increases from 23 to 27, the time overhead of \VUG increases from 113 seconds to 130 seconds. Within this, the overhead of \quick ranges from 105 seconds to 107 seconds, and that of \tight ranges from 4.3 seconds to 4.6 seconds.
}

\begin{figure*}[t]
    \centering
    \begin{subfigure}{0.18\textwidth}
        \includegraphics[width=\textwidth]{Figure_Experiment/exp_4_2.pdf}
        \caption{\cblack{D1}}
    \end{subfigure}
    \begin{subfigure}{0.18\textwidth}
        \includegraphics[width=\textwidth]{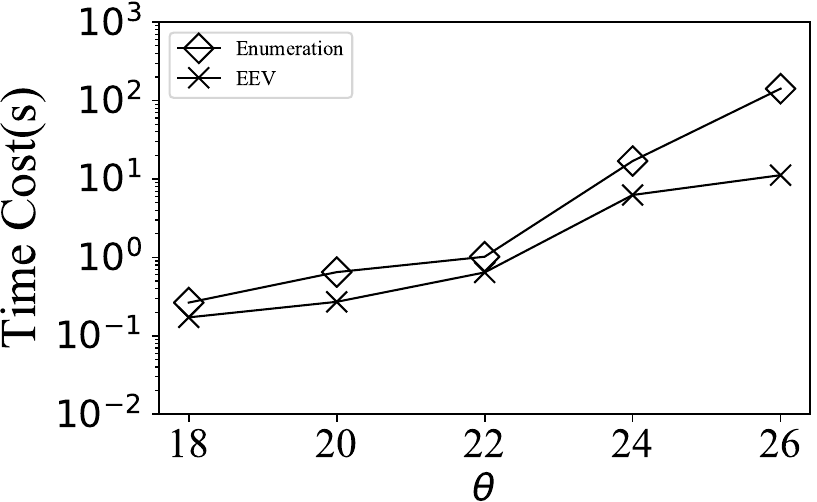}
        \caption{\cblack{D2}}
    \end{subfigure}
    \begin{subfigure}{0.18\textwidth}
        \includegraphics[width=\textwidth]{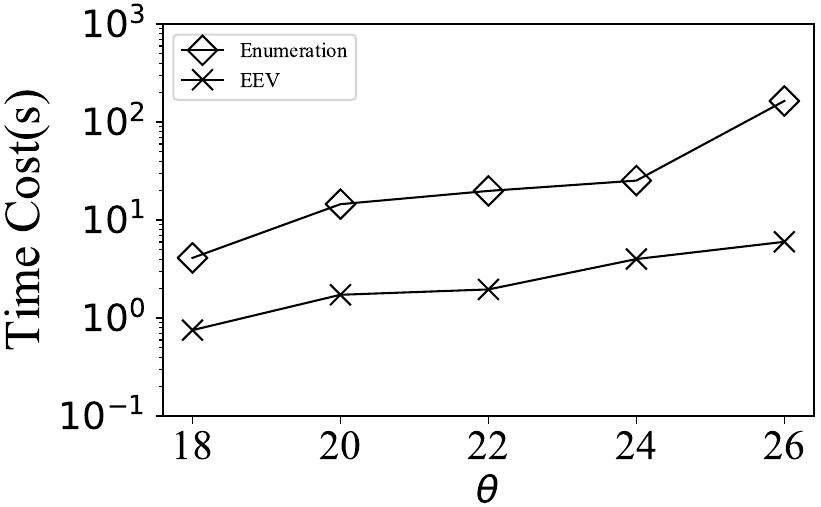}
        \caption{\cblack{D3}}
    \end{subfigure}
    \begin{subfigure}{0.18\textwidth}
        \includegraphics[width=\textwidth]{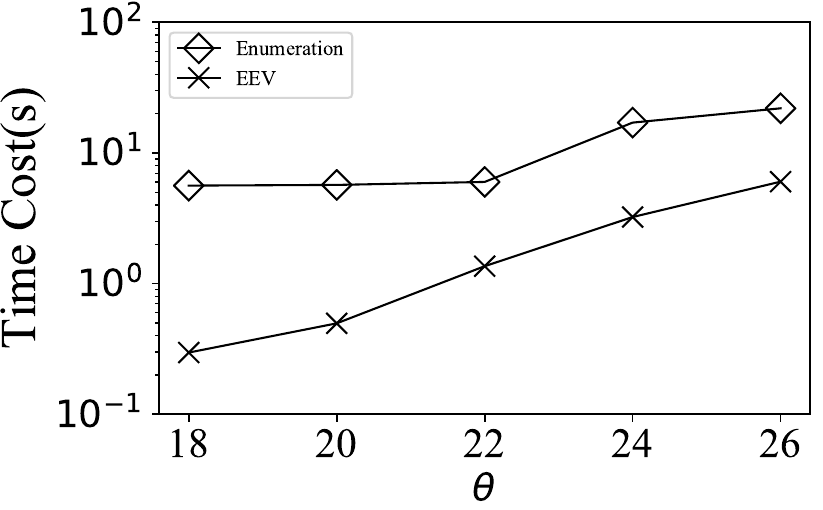}
        \caption{\cblack{D4}}
    \end{subfigure}
    \begin{subfigure}{0.18\textwidth}
        \includegraphics[width=\textwidth]{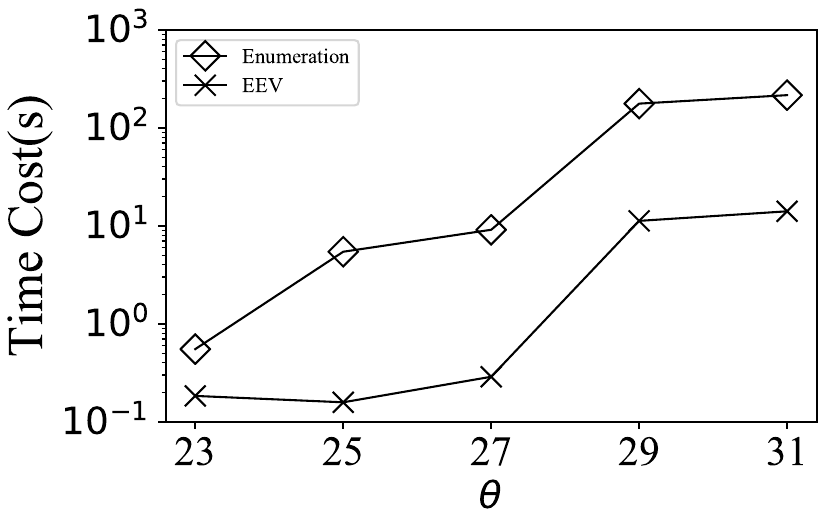}
        \caption{\cblack{D5}}
    \end{subfigure}
    \begin{subfigure}{0.18\textwidth}
        \includegraphics[width=\textwidth]{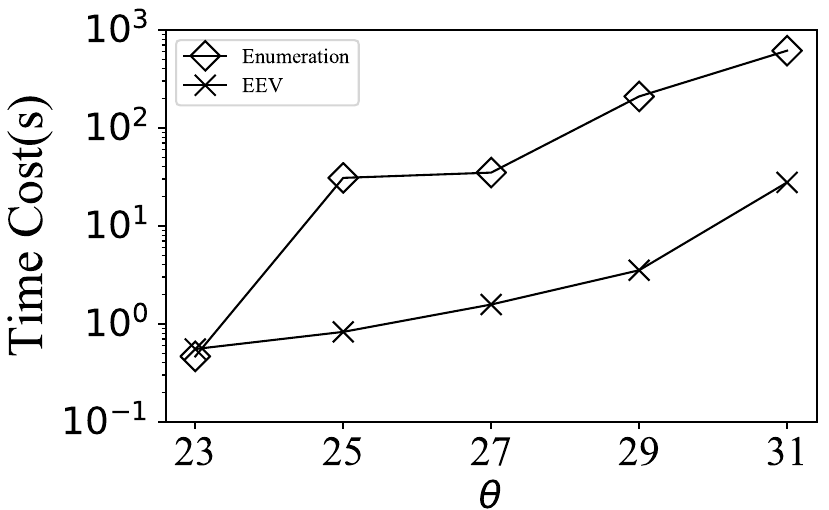}
        \caption{\cblack{D6}}
    \end{subfigure}
    \begin{subfigure}{0.18\textwidth}
        \includegraphics[width=\textwidth]{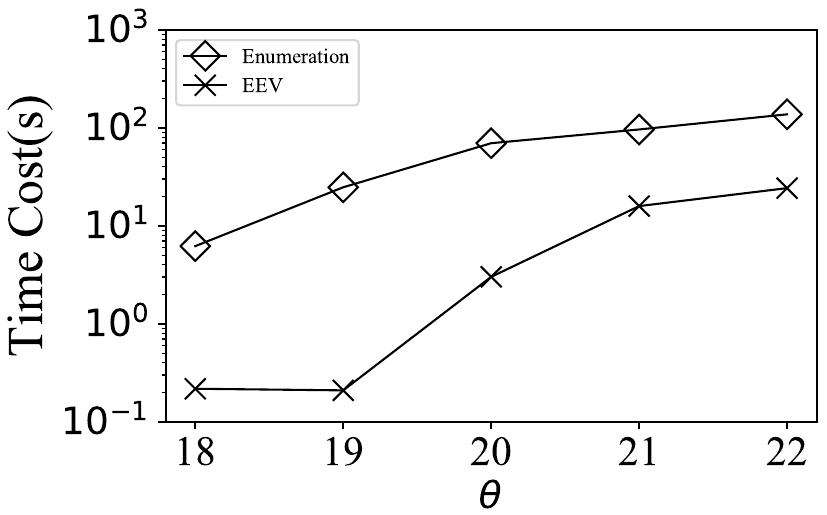}
        \caption{\cblack{D7}}
    \end{subfigure}
    \begin{subfigure}{0.18\textwidth}
        \includegraphics[width=\textwidth]{Figure_Experiment/exp_4_9.pdf}
        \caption{\cblack{D8}}
    \end{subfigure}
    \begin{subfigure}{0.18\textwidth}
        \includegraphics[width=\textwidth]{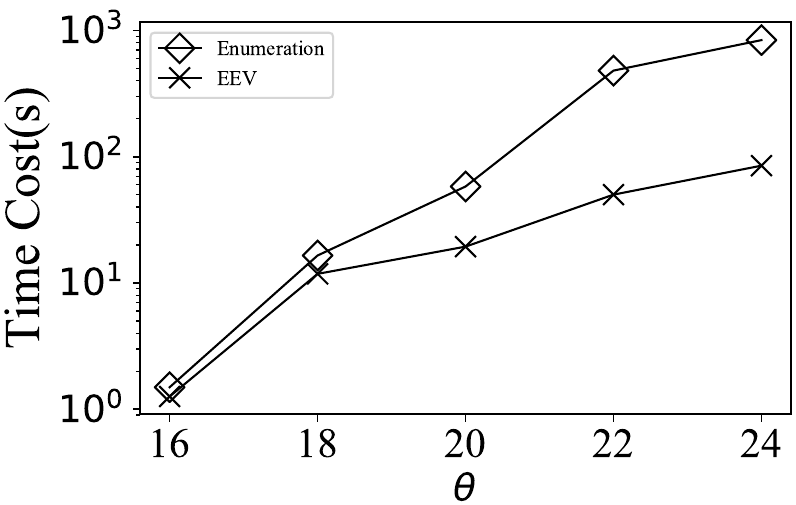}
        \caption{\cblack{D9}}
    \end{subfigure}
    \begin{subfigure}{0.18\textwidth}
        \includegraphics[width=\textwidth]{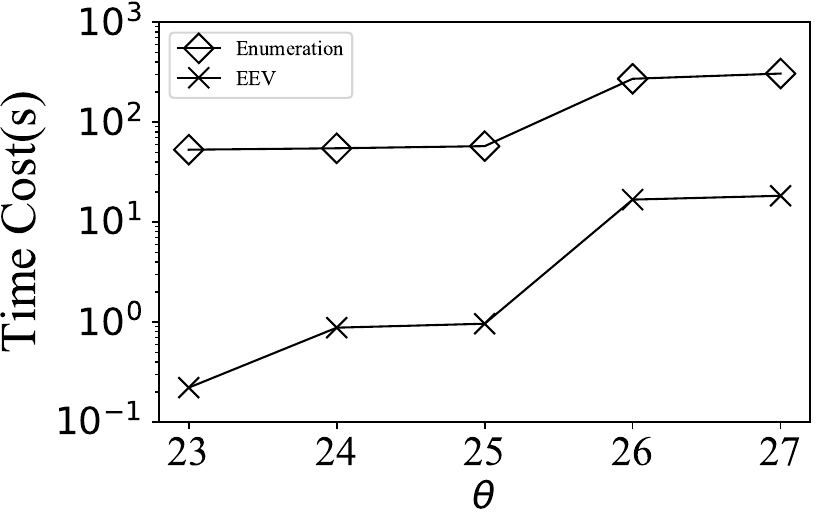}
        \caption{\cblack{D10}}
    \end{subfigure}
    \caption{\cblack{Evaluation of \eev on all other datasets}}
    \label{fig:exp4_more}
\end{figure*}


\myparagraph{\cblack{Supplement to Exp-6: Evaluation of \eev on the other datasets}}
We compare \eev with path enumeration method by applying them separately on $\gt$ and generating $\stspgraph$.
\cblack{Fig.~\ref{fig:exp4_more} reports the total response time of both methods for 1000 queries on each dataset.
As observed, \eev accelerates the process of generating $\stspgraph$ by at least an order of magnitude compared to path enumeration method in the most datasets across various setting of $\theta$.
For example, on the largest dataset D10, when $\theta$ is 23, enumerating paths to generate $\stspgraph$ takes 53 seconds, while \eev only takes 0.2 seconds and when $\theta$ is 27, enumeration method takes 878 seconds, while \eev only takes 18 seconds.}

\begin{figure*}[t]
    \centering
    \begin{subfigure}{0.18\textwidth}
        \includegraphics[width=\textwidth]{Figure_Experiment/exp_6_2.pdf}
        \caption{\cblack{D1}}
    \end{subfigure}
    \begin{subfigure}{0.18\textwidth}
        \includegraphics[width=\textwidth]{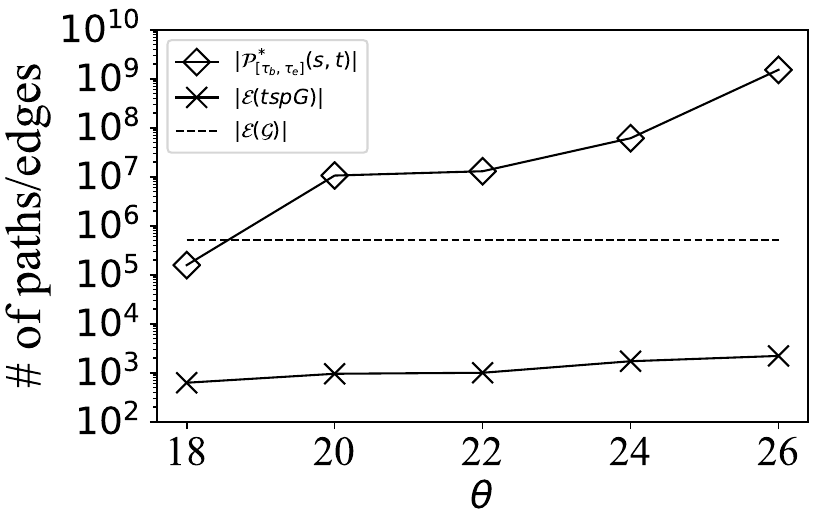}
        \caption{\cblack{D2}}
    \end{subfigure}
     \begin{subfigure}{0.18\textwidth}
        \includegraphics[width=\textwidth]{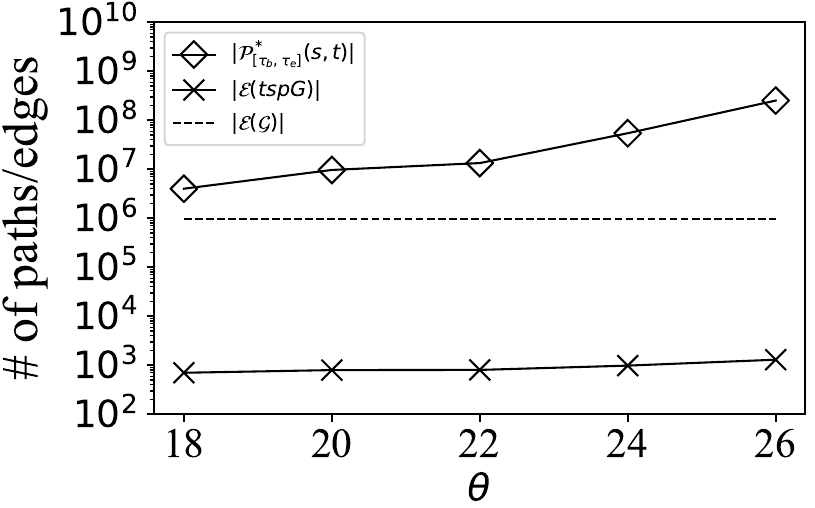}
        \caption{\cblack{D3}}
    \end{subfigure}
    \begin{subfigure}{0.18\textwidth}
        \includegraphics[width=\textwidth]{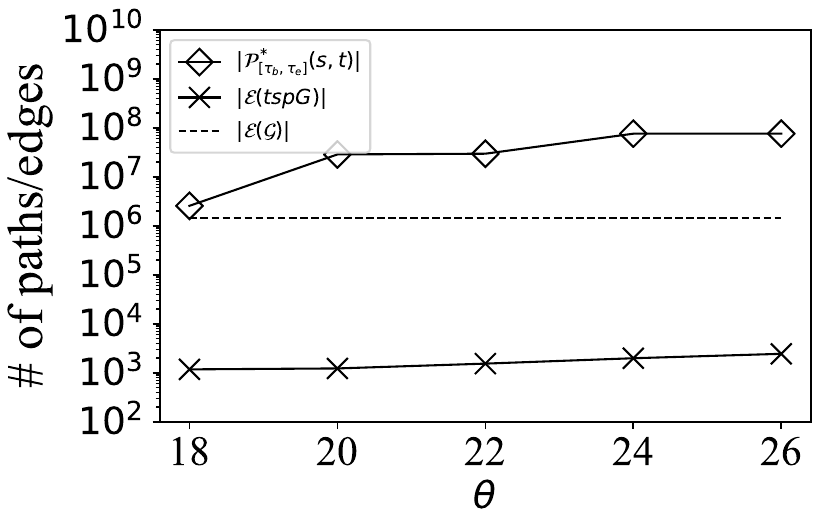}
        \caption{\cblack{D4}}
    \end{subfigure}
     \begin{subfigure}{0.18\textwidth}
        \includegraphics[width=\textwidth]{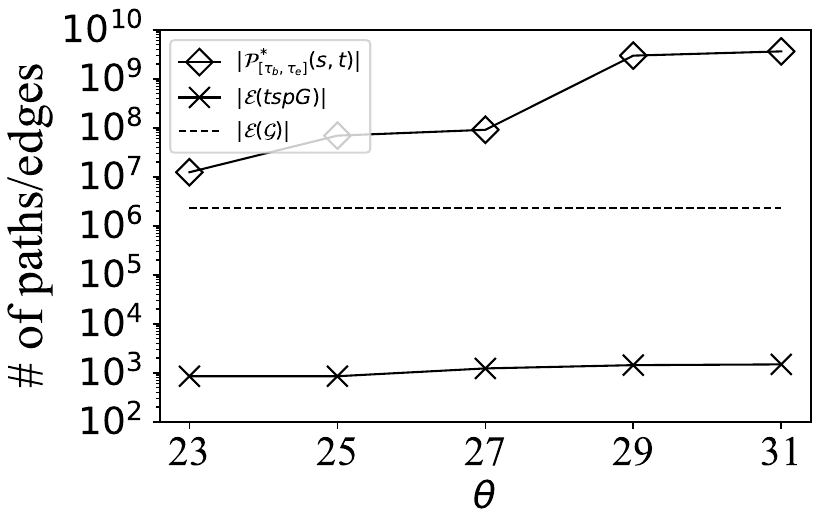}
        \caption{\cblack{D5}}
    \end{subfigure}
    \begin{subfigure}{0.18\textwidth}
        \includegraphics[width=\textwidth]{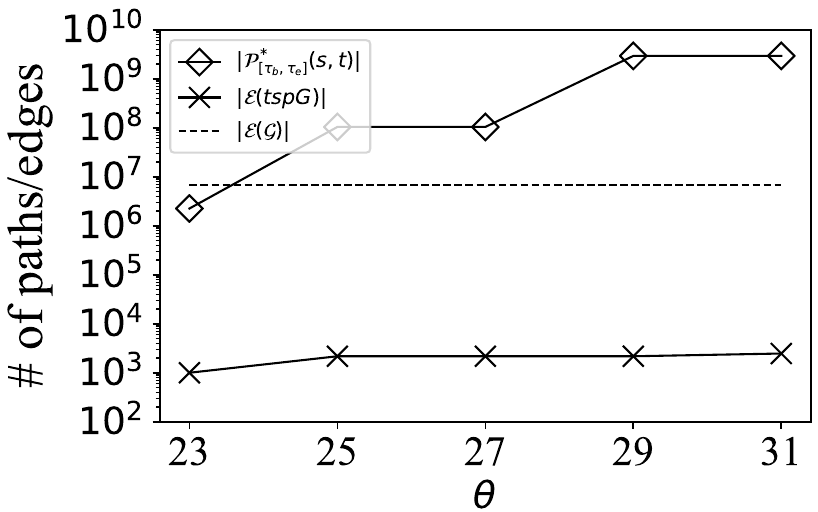}
        \caption{\cblack{D6}}
    \end{subfigure}
     \begin{subfigure}{0.18\textwidth}
        \includegraphics[width=\textwidth]{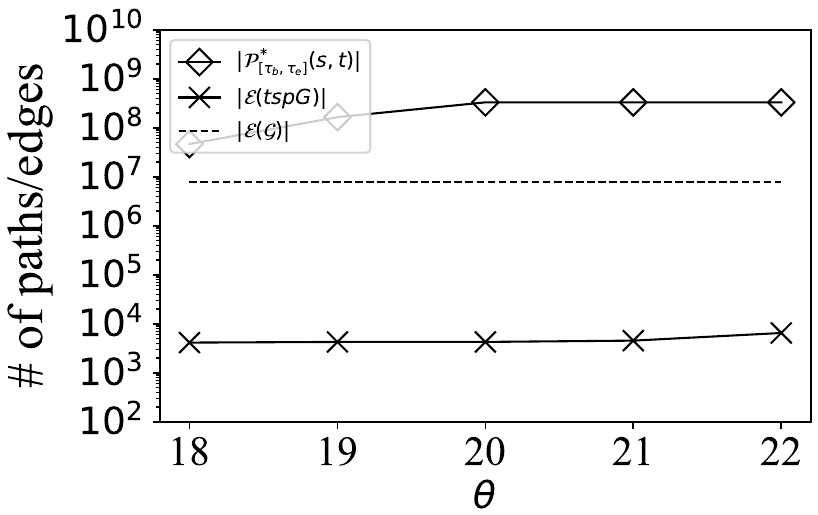}
        \caption{\cblack{D7}}
    \end{subfigure}
    \begin{subfigure}{0.18\textwidth}
        \includegraphics[width=\textwidth]{Figure_Experiment/exp_6_9.pdf}
        \caption{\cblack{D8}}
    \end{subfigure}
     \begin{subfigure}{0.18\textwidth}
        \includegraphics[width=\textwidth]{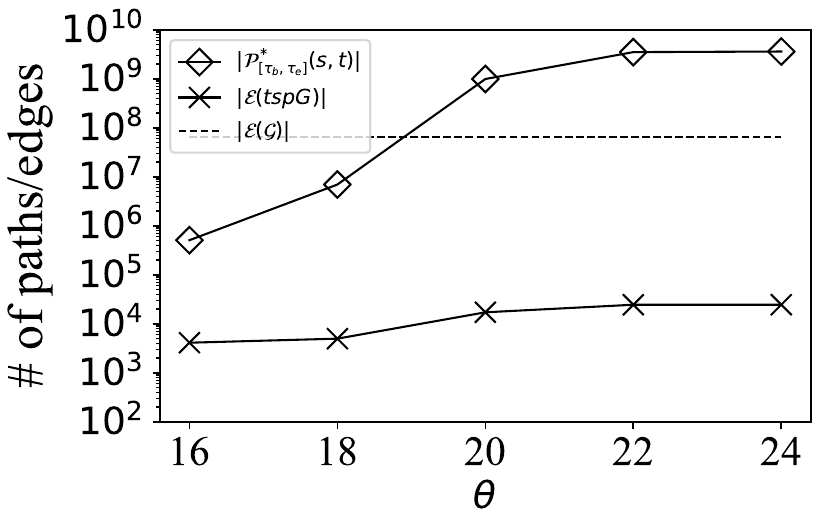}
        \caption{\cblack{D9}}
    \end{subfigure}
    \begin{subfigure}{0.18\textwidth}
        \includegraphics[width=\textwidth]{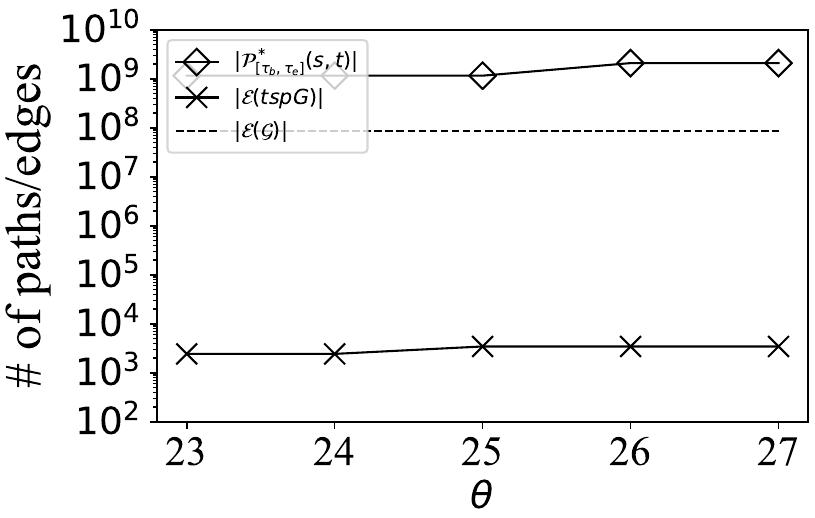}
        \caption{\cblack{D10}}
    \end{subfigure}
    \caption{\cblack{Numbers of paths/edges in $\stspgraph$ on all other datasets}}
    \label{fig:exp6_more}
\end{figure*}

\myparagraph{\cblack{Supplement to Exp-7: Number of edges in $\stspgraph$ on the other datasets}}
\cblack{In this experiment, we report the number of edges in $\stspgraph$ and the number of temporal simple paths it contains on each dataset by varying $\theta$.
Results are shown in Fig.~\ref{fig:exp6_more}.
As we can see, the number of temporal simple paths in $\stspgraph$ far exceeds the number of edges in $\stspgraph$ on all the datasets.
For example, when $\theta=25$ on the largest dataset D10, the generated $\stspgraph$ with 3442 edges contains more than 1.1 billion temporal simple paths.
}

\end{document}